\documentclass[journal]{IEEEtran}

\usepackage{hyperref}

\usepackage{amsmath, amsfonts, amssymb}
\usepackage{algorithm, algorithmic}
\usepackage{graphicx, comment}
\usepackage{color,subfigure,cite,times}

\usepackage{url}
\usepackage{epsfig}
\usepackage{graphics,xspace}
\usepackage{multirow}
\usepackage{multicol}

\def\tran{\mathrm{T}}

\def\a{{\mathbf{a}}}

\def\q{{\mathbf{q}}}

\def\x{{\mathbf{x}}}

\def\0{{\mathbf{0}}}
\def\1{{\mathbf{1}}}
\def\RR{{\mathbb{R}}}
\def\NN{{\mathbb{N}}}

\def\prob{{\mathsf{Prob}}}
\def\EE{{\mathsf{E}}}

\def\lmat{\left(\begin{matrix}}
\def\rmat{\end{matrix}\right)}
\def\eqref#1{(\ref{#1})}

\newtheorem{definition}{Definition}
\newtheorem{lemma}{Lemma}

\newtheorem{proposition}{Proposition}

\def\out{{\mathrm{out}}}

\def\BibTeX{{\rm B\kern-.05em{\sc i\kern-.025em b}\kern-.08em
    T\kern-.1667em\lower.7ex\hbox{E}\kern-.125emX}}

\usepackage{setspace}
\usepackage{framed}

\newcommand{\be}{\begin{equation}}
\newcommand{\ee}{\end{equation}}

\definecolor{BrightBlue}{rgb}{0,0,1}

\def\av{{\mathbf{a}(t)}}
\newcommand{\avm}[1]{\ensuremath{a_{#1}(t)}}
\def\sv{{\mathbf{x}(t)}}
\newcommand{\svm}[1]{\ensuremath{\x_{#1}(t)}}
\newcommand{\ve}[1]{\mathbf{#1}}
\newcommand{\avg}[1]{\overline{#1}}
\newcommand{\one}[1]{{#1}}
\newcommand{\two}[1]{{#1}^{\text{inter}}}

\def\iin{{\text{in}}}
\def\out{{\text{out}}}
\def\qed{{$\blacksquare$}}

\def\im{{\mathcal{A}}} 
\def\om{{\mathcal{B}}} 

\def\xncone{{\ensuremath{x_{\text{NC1}}^{[c]}}}}
\def\xnctwo{{\ensuremath{x_{\text{NC2}}^{[c]}}}}
\def\xdxone{{\ensuremath{x_{\text{DX1}}^{[c]}}}}
\def\xdxtwo{{\ensuremath{x_{\text{DX2}}^{[c]}}}}
\def\xpm{{\ensuremath{x_{\text{PM}}^{[c]}}}}
\def\xrc{{\ensuremath{x_{\text{RC}}^{[c]}}}}
\def\xcx{{\ensuremath{x_{\text{CX}}^{[c]}}}}
\def\SCH{{\ensuremath{\mathsf{SCH}}}}

\def\tran{{\mathsf{T}}}

\def\qinter{{\ensuremath{\ve{q}^{\text{inter}}}}}
\def\Qinter{{\ensuremath{\ve{Q}^{\text{inter}}}}}
\newcommand{\qkinter}[1]{q_{#1}^{\text{inter}}}
\newcommand{\Qkinter}[1]{Q_{#1}^{\text{inter}}}
\newcommand{\Nna}[1]{\ensuremath{N_{\mathsf{NA},#1}}}
\newcommand{\bin}[2]{\ensuremath{\beta_{#1,#2}^{\text{in}}}}
\newcommand{\bine}[2]{\ensuremath{\beta_{#1,#2}^{\text{in}}}}
\newcommand{\bout}[2]{\ensuremath{\beta_{#1,#2}^{\text{out}}}}
\def\Bin{{\ensuremath{\mathcal{B}^\text{in}}}}
\def\Bout{{\ensuremath{\mathcal{B}^\text{out}}}}
\def\Bini{{\ensuremath{\mathcal{B}^{\text{in},(i)}}}}
\def\Bouti{{\ensuremath{\mathcal{B}^{\text{out},(i)}}}}
\def\Bine{{\ensuremath{\mathcal{B}^\text{in}}}}

\DeclareMathOperator*{\argmax}{arg\,max}

\def\eqdef{{\stackrel{\triangle}{=}}}

\newcommand{\SAin}[1]{\mathcal{I}_{#1}}
\newcommand{\SAout}[1]{\mathcal{O}_{#1}}

\newtheorem{claim}{Claim}

\newcount{\myi}
\newcommand{\Repeat}[2]{%
    \myi=0
    \loop
        \ifnum\myi<#2
        #1
        \advance\myi by 1
    \repeat
}

\definecolor{BrickRed}{rgb}{1,0,0}

\begin{document}

\title{Robust And Optimal Opportunistic Scheduling For Downlink 2-Flow Network Coding With Varying Channel Quality and Rate Adaptation}

\author{\IEEEauthorblockN{Wei-Cheng Kuo, Chih-Chun Wang}\thanks{This work was supported in parts by NSF grants CCF-0845968, CNS-0905331, and CCF-1422997. Part of the results were presented in the 2014 INFOCOM.}
\IEEEauthorblockA{\{wkuo, chihw\}@purdue.edu\\
School of Electrical and Computer Engineering, Purdue University, USA}
}

\maketitle

\begin{abstract}
This paper considers the downlink traffic from a base station to two different clients. When assuming infinite backlog, it is known that {\em inter-session} network coding (INC) can significantly increase the throughput of each flow. However, the corresponding scheduling solution (when assuming dynamic arrivals instead and requiring bounded delay) is still nascent.

For the 2-flow downlink scenario, we propose the first opportunistic INC + scheduling solution that is provably optimal for time-varying channels, i.e., the corresponding stability region matches the optimal Shannon capacity.
Specifically, we first introduce a new {\em binary INC} operation, which is distinctly different from the traditional wisdom of XORing two overheard packets. We then develop a {\em queue-length-based} scheduling scheme, which, with the help of the new INC operation,  can robustly and optimally adapt to time-varying channel quality. We then show that the proposed algorithm can be easily extended for rate adaptation and it again robustly achieves the optimal throughput.
 A
byproduct of our results is a scheduling scheme for stochastic
processing networks (SPNs) with {\em random departure}, which relaxes the assumption of {\em deterministic
departure} in the existing results. The new SPN scheduler could thus further broaden the applications of SPN scheduling
to other real-world scenarios.
\end{abstract}

\section{Introduction\label{sec:introduction}}

Since 2000, NC has emerged as a promising technique in communication networks. The seminal work by \cite{LiYeung03} shows linear intra-session NC achieves the min-cut/max-flow capacity of single-session multi-cast networks. The natural connection of intra-session NC and the {\em maximum flow} allows the use of  back-pressure (BP) algorithms to stabilize intra-session NC traffic, see \cite{ho2009dynamic} and the references therein.

However, when there are multiple coexisting sessions, the benefits of {\em inter-session} network coding (INC) are far from fully utilized. The COPE architecture \cite{KattiRahulHuKatabiMedardCrowcroft06} demonstrated that a simple INC scheme can provide 40\%--200\% throughput improvement when compared to the existing TCP/IP architecture in a testbed environment. Several analytical attempts have been made to characterize the INC capacity (or provably achievable throughput) for various small network topologies \cite{GeorgiadisTassiulas13,Wang10b,WangJournalCOPE,EryilmazLunSwapna11}.

\begin{figure}
\centering
\subfigure[INC using only 3 operations\label{fig:BPEC_eg}]{
  \includegraphics[width=3.5cm]{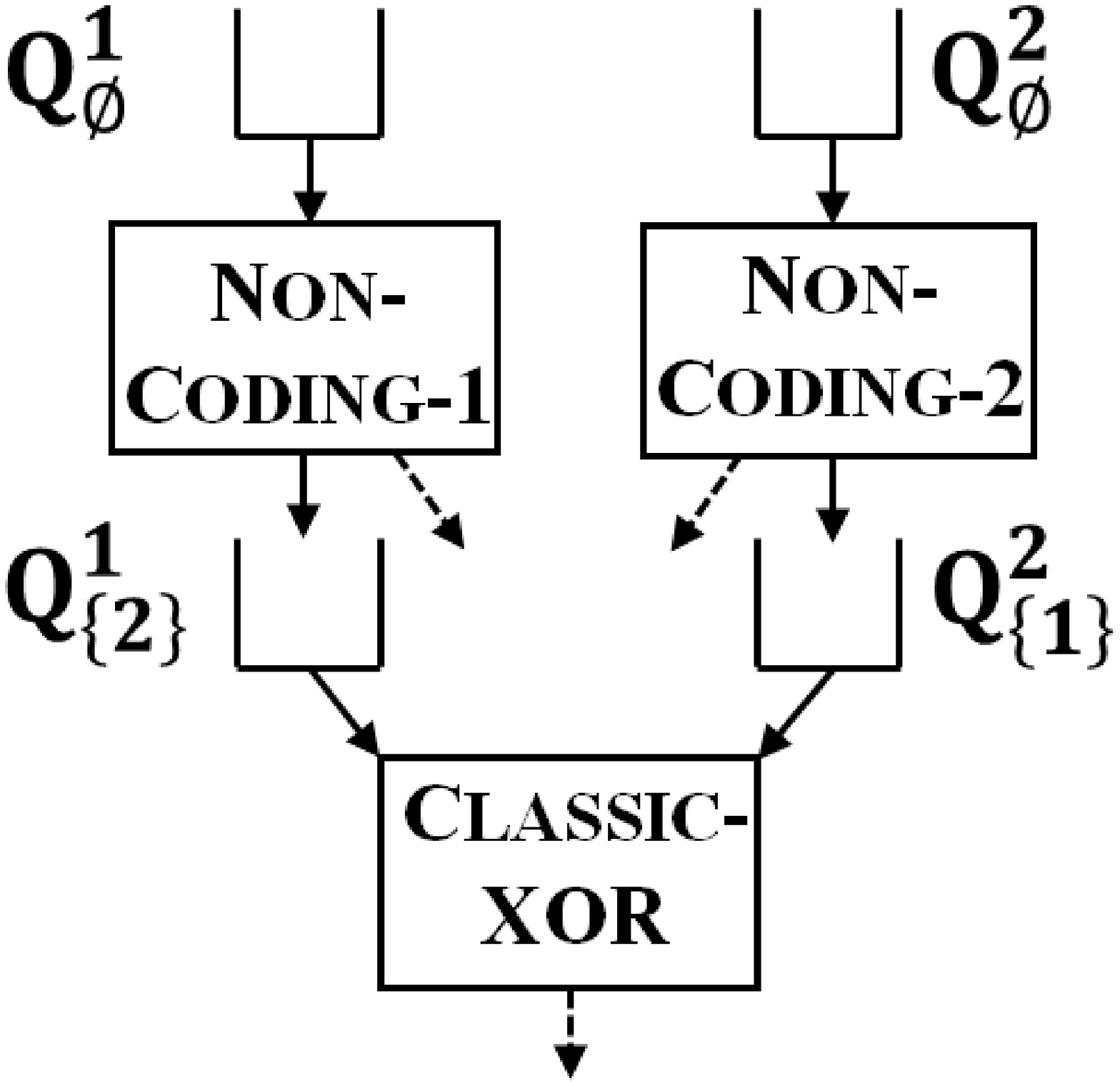}}
\subfigure[INC using only 5 operations\label{fig:5-type}]{
  \includegraphics[width=4.5cm]{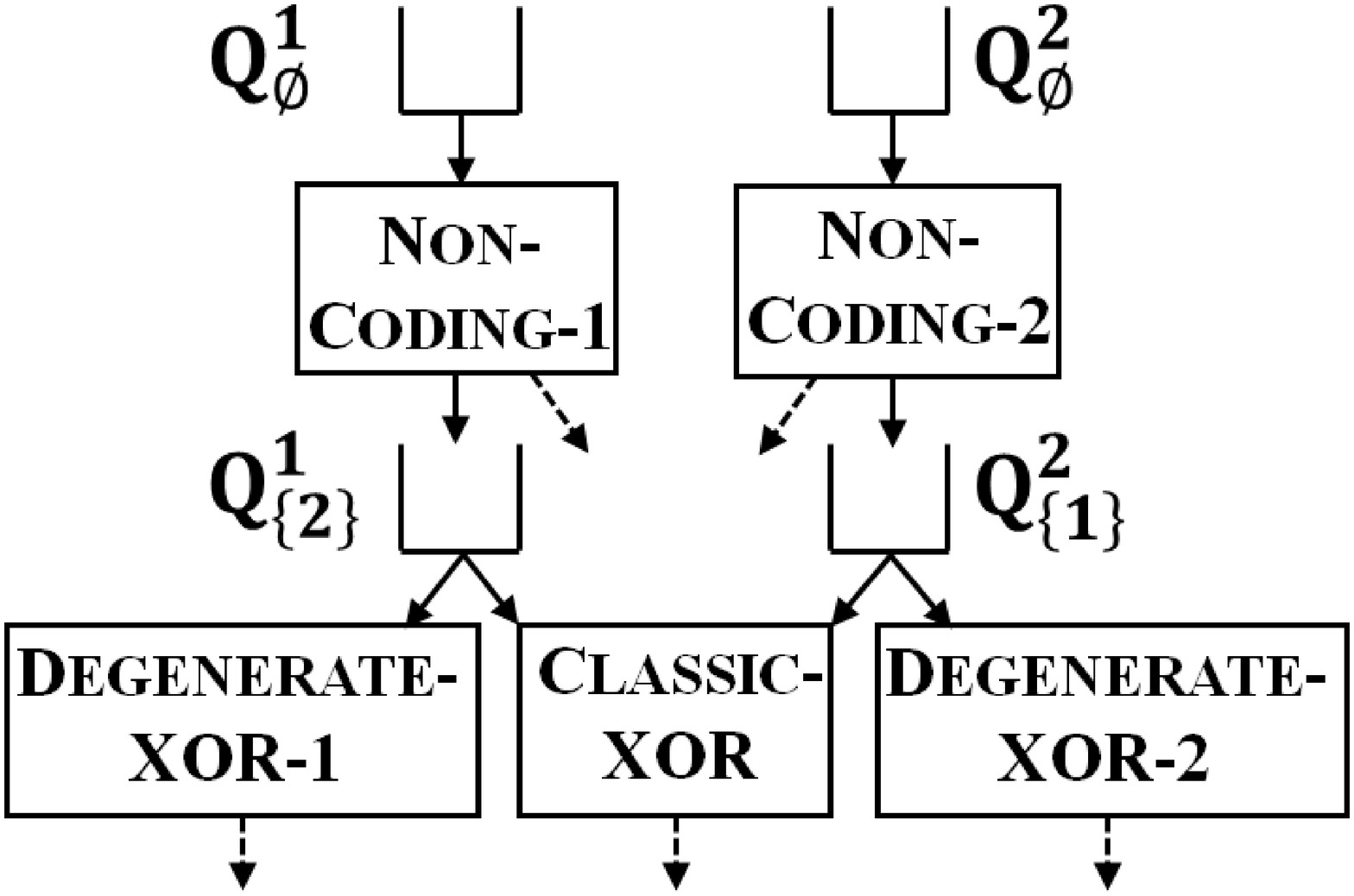}}  \caption{The virtual networks of two INC schemes.}
\vspace{-.5cm}
 \end{figure}
However, unlike the case of intra-session NC, there is no direct analogy from INC to the commodity flow. As a result, it is much more challenging to derive BP-based scheduling for INC traffic. We use the following example to illustrate this point. Consider a single source $s$ and two destinations $d_1$ and $d_2$. Source $s$ would like to send to $d_1$ the $X_i$ packets, $i=1,2,\cdots$; and send to $d_2$ the  $Y_j$ packets, $j=1,2,\cdots$. The simplest INC scheme consists of three operations. OP1: Send uncodedly those $X_i$ that have not been heard by any of $\{d_1,d_2\}$. OP2: Send uncodedly  those $Y_j$ that have not been heard by any of $\{d_1,d_2\}$.  OP3: Send a linear sum $[X_i+Y_j]$ where $X_i$ has been overheard by $d_2$ but not by $d_1$ and $Y_j$ has been overheard by $d_1$ but not by $d_2$. For future reference, we denote OP1 to OP3 by  {\sc Non-Coding-1},  {\sc Non-Coding-2}, and {\sc Classic-XOR}, respectively.

OP1 to OP3 can also be represented by the virtual network (vr-network) in Fig.~\ref{fig:BPEC_eg}. Namely, any newly arrived $X_i$ and $Y_j$ virtual packets\footnote{We often use ``virtual packets'' to refer to the packets (jobs) inside the vr-network.} (vr-packets) that have not been heard by any of $\{d_1,d_2\}$ are stored in queues $Q^1_\emptyset$ and  $Q^2_\emptyset$, respectively. The superscript $k\in\{1,2\}$ indicates that the queue is for the session-$k$ packets. The subscript $\emptyset$ indicates that those packets have not been heard by any of $\{d_1,d_2\}$.  {\sc Non-Coding-1} then takes one $X_i$ vr-packet from $Q^1_\emptyset$ and send it uncodedly. If such $X_i$ is heard by $d_1$, then the vr-packet leaves the vr-network, which is described by the dotted arrow emanating from the {\sc Non-Coding-1} block. If $X_i$ is overheard by $d_2$ but not $d_1$, then we place it in queue $Q^1_{\{2\}}$, the queue for the overheard session-1 packets. {\sc Non-Coding-2} in Fig.~\ref{fig:BPEC_eg} can be interpreted symmetrically.  {\sc Classic-XOR} operation takes an $X_i$ from $Q^{1}_{\{2\}}$ and a $Y_j$ from $Q^2_{\{1\}}$ and sends $[X_i+Y_j]$. If $d_1$ receives $[X_i+Y_j]$, then $X_i$ is removed from $Q^{1}_{\{2\}}$ and leaves the vr-network. If $d_2$ receives $[X_i+Y_j]$, then $Y_j$ is removed from $Q^{2}_{\{1\}}$ and leaves the vr-network. The transition probability (of the edges) of the vr-network can be computed by analyzing the corresponding random reception events when transmitting the packet physically.

It is known \cite{Paschos12} that with dynamic packet arrivals, any INC scheme that (i) uses only these three operations and (ii) attains bounded decoding delay with arrival rates $(R_1,R_2)$ can always be converted to a scheduling solution that stabilizes the vr-network with arrival rates $(R_1,R_2)$, and vice versa. {\em The INC design problem is thus converted to a scheduling problem on the vr-network.} To distinguish the above INC design for dynamical arrivals (the concept of the stability region) from the INC design assuming infinite backlog and decoding delay (the concept of the Shannon capacity), we term the former {\em the dynamic INC} design problem and the latter {\em the block-code INC} design problem.

\begin{figure}
\centering
  \includegraphics[width=6cm]{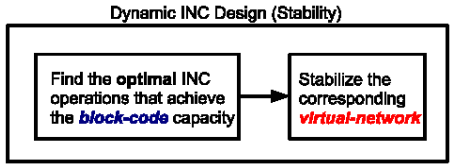}\\
  \caption{The two components of optimal dynamic INC design.}\label{fig:break-down}
  \vspace{-.3cm}
\end{figure}
The above vr-network representation also allows us to divide the optimal dynamic INC design problem into solving the following two major challenges separately. {\bf Challenge~1}: The example in Fig.~\ref{fig:BPEC_eg} focuses on dynamic INC schemes using only 3 possible operations. Obviously, the more INC operations one can choose from, the larger the degree of design freedom, and the higher the achievable throughput. {\em The goal is thus to find a (small) finite set of INC operations that can provably maximize the ``block-code'' achievable throughput}.  {\bf Challenge~2}: Suppose that we have found a set of INC operations that achieves the block-code capacity. However, it does not mean that such a set of INC operations always leads to a dynamic INC design since we still need to consider the delay/stability requirements. Specifically, once the optimal set of INC operations is decided, we can derive the corresponding vr-network. {\em The goal is then to devise a stabilizing scheduling policy for the vr-network, which leads to an equivalent representation of the optimal dynamic INC solution.} See Fig.~\ref{fig:break-down} for the illustration of these two separate tasks.

Both tasks turn out to be highly non-trivial and optimal dynamic INC solution \cite{GeorgiadisTassiulas13,Paschos12,AthanasiadouGeorgiadis13} has been designed only for the scenario of fixed channel quality. Specifically, \cite{GeorgiadisTassiulas09} answers Challenge~1 and shows that for fixed channel quality,  the 3 INC operations in Fig.~\ref{fig:BPEC_eg} plus 2 additional {\sc Degenerate-XOR} operations, see Fig.~\ref{fig:5-type} and Section~\ref{subsubsec:degenerate-XOR}, can achieve the block-code INC capacity. One difficulty of resolving Challenge~2 is that an INC operation may involve multiple queues simultaneously, e.g., {\sc Classic-XOR} can only be scheduled when {\em both} $Q^1_{\{2\}}$ and $Q^2_{\{1\}}$ are non-empty. This is in sharp contrast with the traditional BP solutions in which each queue can act independently.\footnote{To be more precise, a critical assumption in [Section II C.1 \cite{tassiulas1992stability}] is that if two queues $Q_1$ and $Q_2$ can be {\em activated} at the same time, then we can also choose to activate exactly one of the queues if desired. This is unfortunately not the case in the vr-network. E.g., {\sc Classic-XOR} activates both $Q^1_{\{2\}}$ and $Q^2_{\{1\}}$ but no coding operation in Fig.~\ref{fig:BPEC_eg} activates only one of $Q^1_{\{2\}}$ and $Q^2_{\{1\}}$.} For the vr-network in Fig.~\ref{fig:5-type},
\cite{GeorgiadisTassiulas13} circumvents this problem by designing a fixed priority rule that gives strict precedence to the {\sc Classic-XOR} operation. Alternatively, \cite{Paschos12} derives a BP scheduling scheme by noticing that the vr-network in Fig.~\ref{fig:5-type} can be decoupled into two vr-subnetworks (one for each data session) so that the queues in each of the  vr-subnetworks can be activated independently and the traditional BP results follow.

However, the channel quality varies over time for practical wireless downlink scenarios. Therefore, one should  opportunistically choose the most favorable users as receivers, the so-called {\em opportunistic scheduling} technique. Nonetheless, recently \cite{wang2012linear} shows that when allowing opportunistic coding+scheduling for time-varying channels, the 5 operations in Fig.~\ref{fig:5-type} no longer achieve the block-code capacity. The existing dynamic INC design in \cite{GeorgiadisTassiulas13,Paschos12} are thus strictly suboptimal for time-varying channels since they are based on a suboptimal set of INC operations (recall Fig.~\ref{fig:break-down}).

In this work, we propose a new optimal dynamic INC design for 2-flow downlink traffic with time-varying packet erasure channels. Our detailed contributions are summarized as follows.

{\em Contribution 1:} We introduce a new pair of INC operations such that (i) The underlying concept is distinctly different from the traditional wisdom of XORing two overheard packets; (ii) The overall scheme uses only the ultra-low-complexity binary XOR operation; 
and (iii) The new set of INC operations is guaranteed to achieve the block-code-based Shannon capacity.

{\em Contribution 2:} The introduction of new INC operations leads to a new vr-network that is different from Fig.~\ref{fig:5-type} and for which the existing ``{\em vr-network decoupling} + {\em BP}'' approach in \cite{Paschos12} no longer holds. To answer Challenge~2 of the optimal dynamic INC design, we generalize the results of Stochastic Processing Networks (SPNs) \cite{jiang2009stable,huang2011utility} and successfully apply it to the new vr-network. The end result is an opportunistic, dynamic INC solution
that is completely {\em queue-length-based} and can robustly adapt to time-varying channels while achieving the largest possible stability region.

{\em Contribution 3:} The proposed solution can also be readily generalized for rate-adaptation. Through numerical experiments, we have shown that a simple extension of the proposed scheme can opportunistically and optimally choose the order of modulation and the rate of the error correcting codes used for each packet transmission while achieving the optimal stability region, i.e., equal to the Shannon capacity.

{\em Contribution 4:} A byproduct of our results is a scheduling scheme for SPNs with {\em random departure}. The new results relax the previous assumption of {\em deterministic departure}, a major limitation of the existing SPN model, by considering stochastic packet departure behavior. The new scheduling solution could thus further broaden the applications of SPN scheduling to other real-world scenarios.

The rest of the paper is organized as follows. Section~\ref{sec:existing} discusses the existing results on INC design and on SPN scheduling. Sections~\ref{sec:new-coding} and~\ref{sec:new_scheduling} propose a new INC operation and a new SPN scheduling solution, respectively. Section~\ref{sec:combined} elaborates how to combine the new INC operation and the new SPN scheduling to derive the optimal dynamic INC solution. Section~\ref{sec:simulation} contains the simulation results and Section~\ref{sec:conclusion} concludes the paper. Most of the proofs are left in the appendices to improve the readability of the main findings.

\section{Problem Formulation and Existing Results\label{sec:existing}}

In this section, we will introduce the problem formulation and then discuss the latest results on the block-code LNC literature (related to solving Challenge~1) and on the SPN scheduling work (related to solving Challenge~2).

\def\cq{{\mathsf{cq}}}
\def\bigcq{{\mathsf{CQ}}}
\def\stdef{\stackrel{\Delta}{=}}
\def\vp{{\vec{p}}}
\def\rcpt{{\mathsf{rcpt}}}

\subsection{Problem Formulation --- The Broadcast Erasure Channel\label{subsec:setting}}
\begin{figure}
\centering
  \includegraphics[width=5cm]{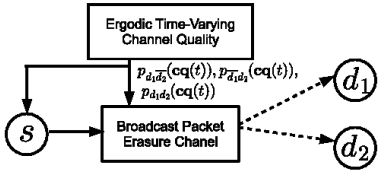}\\
  \caption{The time-varying broadcast packet erasure channel.}\label{fig:VCS}
\end{figure}

We model the 1-base-station/2-client downlink traffic as a broadcast packet erasure channel. See Fig.~\ref{fig:VCS} for illustration. The detailed model description is as follows. Consider the following slotted transmission system.

{\em Dynamic Arrival:} In the beginning of every time $t$, there are $A_1(t)$ session-1 packets and $A_2(t)$ session-2 packets arriving at source $s$. We assume that $A_1(t)$ and $A_2(t)$ are i.i.d.\ integer-valued random variables with mean $(\EE\{A_1(t)\},\EE\{A_2(t)\})=(R_1,R_2)$ and bounded support. Recall that $X_i$ and $Y_j$, $i,j\in\NN$, denote the session-1 and session-2 packets, respectively.

{\em Time-Varying Channel:}
We model the time-varying channel quality by a random process $\cq(t)$, which, as will be elaborated shortly after, decides the reception probability of the {\em broadcast packet erasure channel.}  In our stability proofs, we assume $\cq(t)$ is i.i.d. On the other hand, our numerical experiments show that the proposed scheme achieves the optimal stability region for any ergodic $\cq(t)$, say $\cq(t)$ being periodic. 

\par Let $\bigcq$ denote the support of $\cq(t)$ and we assume $|\bigcq|$ is finite. For any $c\in\bigcq$, we use $f_c$ to denote the expected/long-term frequency of $\cq(t)=c$. Without loss of generality, assume $f_c>0$ for all $c\in\bigcq$.  Obviously $\sum_{c\in\bigcq}f_c=1$ since the total frequency is 1.

{\em Broadcast Packet Erasure Channel:}
For each time slot $t$, source $s$ can transmit one packet, which will be received by a random subset of destinations $\{d_1,d_2\}$. Specifically, there are 4 possible {\em reception status} $\{\overline{d_1d_2},d_1\overline{d_2},\overline{d_1}d_2,d_1d_2\}$, e.g., the reception status $\rcpt=d_1\overline{d_2}$ means that the packet is received by $d_1$ but not $d_2$. The reception status probabilities can be described jointly by a vector $\vec{p}\stdef(p_{\overline{d_1d_2}}, p_{d_1\overline{d_2}},p_{\overline{d_1}d_2},p_{d_1d_2})$. For example, $\vp=(0,0.5,0.5,0)$ means that every time we transmit a packet, with 0.5 probability it will be received by $d_1$ only and with 0.5 probability it will be received by $d_2$ only. It will never be received by $d_1$ and $d_2$ simultaneously. In contrast, if we have $\vp=(0,0,0,1)$, then it means that the packet is always received by $d_1$ and $d_2$ simultaneously. Since our model allows arbitrary joint probability vector $\vec{p}$, it captures the scenarios in which the erasure events of $d_1$ and $d_2$ are dependent, e.g., when the erasures at $d_1$ and $d_2$ are caused by a common (random) external interference source.

{\em Opportunistic INC:}
Since the reception probability is decided by the channel quality, we write $\vp(\cq(t))$ as a function of $\cq(t)$ at time $t$. In the beginning of time $t$, we assume that $s$ is aware of the channel quality $\cq(t)$ (and thus knows $\vp(\cq(t))$) so that $s$ can opportunistically decide how to encode the packet for the current time slot. See Fig.~\ref{fig:VCS}. This opportunistic setting thus models the use of cognitive radio at source $s$.

{\em ACKnowledgement:} In the end of time $t$, both $d_1$ and $d_2$ will report back to $s$ whether they have received the transmitted packet or not. This models the use of ACK.

\subsection{Existing Results on Block INC Design}
References \cite{wang2012linear,WangHan14} focus on the above setting but consider the infinite backlog block-code design instead of dynamic arrivals. Two findings of \cite{wang2012linear,WangHan14} are summarized here.

\subsubsection{The 5 INC operations in Fig.~\ref{fig:5-type} are no longer optimal for time-varying channels\label{subsubsec:degenerate-XOR}}

In Section~\ref{sec:introduction}, we have detailed 3 INC operations: {\sc Non-Coding-1}, {\sc Non-Coding-2}, and {\sc Classic-XOR}. Two additional INC operations are introduced in \cite{GeorgiadisTassiulas09}: {\sc Degenerate-XOR-1} and {\sc Degenerate-XOR-2} as illustrated in Fig.~\ref{fig:5-type}. Specifically, {\sc Degenerate-XOR-1} is designed to handle the degenerate case in which $Q^1_{\{2\}}$ is non empty but  $Q^2_{\{1\}}=\emptyset$. Namely, there is at least one $X_i$ packet overheard by $d_2$ but there is no $Y_j$ packet overheard by $d_1$. Not having such $Y_j$ implies that one cannot send $[X_i+Y_j]$ (the {\sc Classic-XOR} operation). An alternative is thus to send the overheard $X_i$ uncodedly (as if sending $[X_i+0]$). We term this operation {\sc Degenerate-XOR-1}. One can see from Fig.~\ref{fig:5-type} that {\sc Degenerate-XOR-1} takes a vr-packet from $Q^1_{\{2\}}$ as input. If $d_1$ receives it, the vr-packet will leave the vr-network. {\sc Degenerate-XOR-2} is the symmetric version of {\sc Degenerate-XOR-1}.

We use the following example to illustrate the sub-optimality of the above 5 operations. Suppose $s$ has an $X$ packet for $d_1$ and a $Y$ packet for $d_2$ and consider a duration of 2 time slots. Also suppose that $s$ knows beforehand that the time-varying channel will have (i) $\vp=(0,0.5,0.5,0)$ for slot~1; and (ii) $\vp=(0,0,0,1)$ for slot~2. The goal is to transmit as many packets in 2 time slots as possible.

{\em Solution~1: INC based on the 5 operations in Fig.~\ref{fig:5-type}.}  In the beginning of time 1, both $Q^1_{\{2\}}$ and $Q^2_{\{1\}}$ are empty. Therefore, we can only choose either {\sc Non-Coding-1} or {\sc Non-Coding-2}. Since the setting is symmetric, without loss of generality we assume that we choose {\sc Non-Coding-1} and thus send $X$ uncodedly. Since $\vp=(0,0.5,0.5,0)$ in slot~1, there are only two cases to consider. Case 1: $X$ is received only by $d_1$. In this case, we can send $Y$ in the second time slot, which is guaranteed to arrive at $d_2$ since $\vp=(0,0,0,1)$ in slot~2. The total sum rate is sending 2 packets ($X$ and $Y$) in 2 time slots.
Case 2: $X$ is received only by $d_2$. In this case, $Q^1_{\{2\}}$ contains one packet $X$, and $Q^2_\emptyset$ contains one packet $Y$, and all the other queues in Fig.~\ref{fig:5-type} are empty. We can thus choose either {\sc Non-Coding-2} or {\sc Degenerate-XOR-1} for slot~2. Regardless of which coding operation we choose, slot~2 will then deliver 1 packet to either $d_2$ or $d_1$, depending on the INC operation we choose. Since no packet is delivered in slot~1, the total sum rate is 1 packet in 2 time slots. Since both cases have probability 0.5, the expected sum rate is $2\cdot 0.5 + 1\cdot 0.5=1.5$ packets in 2 time slots.

{\em An optimal solution:} We can achieve strictly better throughput by introducing new INC operations. Specifically, in slot~1, we send the linear sum $[X+Y]$ {\em even though neither $X$ nor $Y$ has ever been transmitted}, a distinct departure from the existing 5-operation-based solutions.

Again consider two cases:  Case 1: $[X+Y]$ is received only by $d_1$. In this case, we let $s$ send $Y$ uncodedly in slot~2. Since $\vp=(0,0,0,1)$ in slot~2, the packet $Y$ will be received by both $d_1$ and $d_2$. $d_2$ is happy since it has now received the desired $Y$ packet. $d_1$ can use $Y$ together with the $[X+Y]$ packet received in slot~1 to decode its desired $X$ packet. Therefore, we deliver 2 packets ($X$ and $Y$) in 2 time slots. Case 2: $[X+Y]$ is received only by $d_2$. In this case, we let\footnote{ACK is critical in this scheme. I.e., $s$ needs to  know whether it is $d_1$ or $d_2$ who has received $[X+Y]$ in slot~1 before deciding whether to send $Y$ or $X$ in slot~2. } $s$ send $X$ uncodedly in slot~2. By the symmetric argument of Case~1, we deliver 2 packets ($X$ and $Y$) in 2 time slots.
As a result, the sum-rate of the new solution is 2 packets in 2 slots, a 33\% improvement over the existing solution.

{\em Remark:} This example focuses on a 2-time-slot duration due to the simplicity of the analysis. It is worth noting that the throughput improvement persists even for infinitely many time slots. See the simulations results in Section~\ref{sec:simulation}.

\subsubsection{\cite{wang2012linear,WangHan14} also derive the block-code capacity region}

We summarize the high-level description of \cite{WangHan14}: 

\begin{proposition}\label{prop:recite}[Propositions~1 and 3, \cite{WangHan14}] For the block-code setting, 
a rate vector $(R_1,R_2)$ can be achieved if and only if the corresponding linear programming (LP) problem is feasible. Given any $(R_1,R_2)$, the LP problem of interest involves $18\cdot|\bigcq|+7$ non-negative variables and $|\bigcq|+16$ (in-)equalities and can be explicitly computed.\label{prop:WangISIT12}
\end{proposition}

Our goal is to design a dynamic INC scheme, of which the stability region matches the block-code capacity region in Proposition~\ref{prop:recite}.

\subsection{Stochastic Processing Networks (SPNs)}
The main tool that we use to stabilize the vr-network is scheduling for the stochastic processing networks (SPNs). In the following, we will discuss the basic definitions and existing results on SPNs.

\subsubsection{The Main Feature of SPNs}
The SPN is a generalization of the store-and-forward networks. In an SPN, a packet can not be transmitted directly from one queue to another queue through links. Instead, it must first be processed by a unit called ``Service Activity'' (SA). The SA first collects a certain amount of packets from one or more queues (named the {\em input queues}), jointly processes/consumes these packets, generates a new set of packets, and finally redistributes them to another set of queues (named the {\em output queues}). The number of consumed packets may be different than the number of generated packets. There is one critical rule for an SPN: {\em An SA can be activated only when all its input queues can provide enough amount of packets for the SA to process.} This rule captures directly the INC behavior and thus makes INC a natural application of SPNs. Other applications of SPNs include the video streaming problem \cite{amini2006adaptive} and  the Map-\&-Reduce scheduling problem\cite{zaharia2008improving}.

\subsubsection{SPNs with Deterministic Departure\label{subsubsec:SPN-deterministic}}
All the existing SPN scheduling solutions \cite{jiang2009stable,huang2011utility} assume a special class of SPNs, which we call SPNs with deterministic departure. We elaborate the detailed definition in the following.

Consider a time-slotted system with i.i.d. channel quality $\cq(t)$. An SPN consists of three components: the input activities (IAs), the service activities (SAs), and the queues. We suppose that there are $K$ queues, $M$  IAs, and $N$ SAs in the SPN.

{\em Input Activities:} Each IA represents a session (or a flow) of packets. Specifically, each IA injects a deterministic number of packets to a deterministic set of queues when activated. That is, when an IA $m$ is activated, it it injects a deterministic number of $\alpha_{k,m}$ packets to queue $k$ for a group of different $k$. Let $\im\in \RR^{K*M}$ be the ``input matrix" with the $(k,m)$-th entry equals to $\alpha_{k,m}$, for all $m$ and $k$. At each time $t$, a random subset of IAs will be activated. Equivalently, we define $\av\stdef(a_1(t),a_2(t),\cdots, a_M(t))\in \{0,1\}^M$ as the random ``arrival vector" at time $t$. If $\avm{m}=1$, then IA $m$ is activated at time $t$. We assume that the random vector  $\av$ is i.i.d.\ over time with the average rate vector $\ve{R} = \EE\{\av\}$. In our setting, the $\im$ matrix is a fixed (deterministic) system parameter and all the randomness of IAs lies in $\av$.

{\em Service Activities:} For each service activity SA $n$, we define the {\em input queues} of SA $n$ as the queues which are required to provide specified amounts of packets when SA $n$ is activated. Let $\SAin{n}$ denote the collection of the input queues of SA $n$. Similarly, we define the {\em output queues} of SA $n$ as the queues which will possibly receive packets when SA $n$ is activated, and let $\SAout{n}$ be the collection of the output queues of SA $n$. That is, when SA $n$ is activated, it takes packets from queues in $\SAin{n}$, and sends packets to queues in $\SAout{n}$. We assume that $\cq(t)$ does not change $\SAin{n}$ and $\SAout{n}$.

Let $\bin{k}{n}(c)$ be the number of packets from queue $k\in \SAin{n}$ that will be consumed by SA $n$ if SA $n$ is activated under channel quality $\cq(t)=c$. Specifically, $\bin{k}{n}(c)\geq 0$ if queue $k$ is the input queue of SA $n$ (i.e. $k\in\SAin{n}$), and we set $\bin{k}{n}(c)=0$ otherwise. Similarly, let $\bout{k}{n}(c)$ be the number of packets received by queue $k$ if SA $n$ is activated under channel quality $\cq(t)=c$. Specifically, $\bout{k}{n}(c)\geq 0$ if queue $k \in\SAout{n}$, and $\bout{k}{n}(c)=0$ otherwise. Let $\Bin(c)\in\RR^{K*N}$ be the {\em input service matrix} under channel quality $c$ with the $(k,n)$-entry equals to $\bin{k}{n}(c)$, and let $\Bout(c)\in\RR^{K*N}$ be the {\em output service matrix} under channel quality $c$ with the $(k,n)$-entry equals to $\bout{k}{n}(c)$. For simplicity, we sometimes write $\Bin$ and $\Bout$ instead of $\Bin(c)$ and $\Bout(c)$. In the deterministic SPN setting, the matrices $\Bin(c)$ and $\Bout(c)$  are deterministic. The only random part is the arrival vector $\ve{a}(t)$ and the channel quality $\cq(t)$.

At the beginning of each time $t$, the SPN scheduler is made aware of the current channel quality $\cq(t)$ and can choose to ``activate'' a subset of the SAs.   Let $\sv\in \{0,1\}^N$ be the ``service vector" at time $t$. If the $n$-th coordinate $\svm{n}=1$, then it implies that we choose to activate  SA $n$  at time $t$. Note that for some applications we may need to impose the condition that some of the SAs cannot be scheduled in the same time slot. To model this {\em interference constraint}, we require $\sv$ to be chosen from a pre-defined set of binary vectors $\mathfrak{X}$. Define $\Lambda$ to be the convex hull of $\mathfrak{X}$ and let $\Lambda^\circ$ be the interior of $\Lambda$.

{\em Acyclicness of The Underlying SPN:} The input/outuput queues $\SAin{n}$ and $\SAout{n}$ of the SAs can be used to plot the corresponding SPN. We assume that the SPN is acyclic. 

\par {\em Existing results on the stability region of deterministic SPNs:} Recall that $f_c$ is the relative frequency of $\cq(t)=c$ and all the vectors are row vectors. We then have the following proposition.
\begin{definition}
For the deterministic SPNs, an arrival rate vector $\ve{R}$ is ``feasible" if there exist $\ve{s}_c\in \Lambda$ for all $c\in\bigcq$ such that
\begin{align}
\im\cdot \ve{R}^\tran+\sum_{c\in\bigcq}f_c\cdot \Bout(c)\cdot \ve{s}_c^\tran=\sum_{c\in\bigcq}f_c\cdot \Bin(c)\cdot \ve{s}_c^\tran \label{eq:balance}
\end{align}where $(\vec{v})^\tran$ is the transpose of the row vector $\vec{v}$. A rate vector $\ve{R}$ is ``strictly feasible" if there exist $\ve{s}_c\in \Lambda^\circ$ for all $c\in\bigcq$  such that \eqref{eq:balance} holds.
\end{definition}

\par Eq.~\eqref{eq:balance} can be viewed as a flow conservation law of the deterministic SPN, for which the left-hand side describes the packets injected to queues 1 to $k$ and the right-hand side corresponds to the packets leaving the queues.

\begin{proposition}\label{prop:ach}[A combination of \cite{jiang2009stable,huang2011utility}]
For deterministic SPNs, only feasible $\ve{R}$ can possibly be stabilized. Moreover, there exists an SPN scheduler that can stabilize all $\ve{R}$ that are strictly feasible.
\end{proposition}

The achievability part for SPNs with deterministic departure (Proposition~\ref{prop:ach}) is proven by the Deficit Max-Weight (DMW) algorithm in  \cite{jiang2009stable} and by the Perturb Max-Weight (PMW) algorithm in \cite{huang2011utility}. In the following, we briefly explain the existing DMW algorithm \cite{jiang2009stable}.

\subsubsection{The Deficit Maximum Weight (DMW) Scheduling}

In the DMW algorithm\cite{jiang2009stable} for SPNs with deterministic departure, each queue $k$ maintains a real-valued counter $q_k(t)$, called the virtual queue length. Initially, $q_k(1)$ is set to 0. For comparison, the actual queue length is denoted by $Q_k(t)$ instead.

The key feature of a DMW algorithm is that it makes a back-pressure-based decision based on the virtual queue-lengths, not on the actual queue lengths. Specifically, for each time $t$, we compute the ``preferred\footnote{As we can see later, sometimes we may not be able to execute/schedule the preferred service activities chosen by \eqref{eq:DMW-schedule}. This is the reason why we only call the $\x^*(t)$ vector in \eqref{eq:DMW-schedule} a preferred choice, instead of a scheduling choice.} service vector"  by
\begin{align}
\ve{x}^*(t) = \arg\max_{\ve{x}\in \mathfrak{X}} \ve{d}^\tran(t)\cdot \ve{x}, \label{eq:DMW-schedule}
\end{align}
where $\ve{d}(t)$ is the back pressure vector defined as $\ve{d}(t)=\left(\Bin(\cq(t))-\Bout(\cq(t) \right)^\tran \ve{q}(t)$, and $\ve{q}(t)$ is the vector of the virtual queue lengths.
After computing the preferred SA vector $\ve{x}^*(t)$, we update $\ve{q}(t)$ according to the following flow conservation law:
\begin{align}
\ve{q}(t+1) =& \ve{q}(t)+\im\cdot \ve{a}(t)\nonumber \\
&+\left(\Bout(\cq(t))-\Bin(\cq(t))\right)\cdot \ve{x}^*(t). \label{eq:DMW-update}
\end{align}

Unlike the actual queue lengths $Q_k(t)$, which is always $\geq 0$, the virtual queue length $\ve{q}(t)$ can be smaller than 0 when updated via \eqref{eq:DMW-update}. That is, we do not need to take the projection to positive numbers when computing $\ve{q}(t)$.

It is worth emphasizing that the actual queue length still has to follow the SPN rule. That is, suppose SA $n$ is the preferred service activity according to \eqref{eq:DMW-schedule} but for at least one of its input queues, say queue $k$,  the actual queue length $Q_k(t)$ is smaller than $\bin{k}{n}(\cq(t))$, the number of packets that are supposed to leave queue $k$. According to the model of SPN, we cannot schedule the preferred SA $n$ due to the lack of enough packets in queue $k$. When this scenario happens, DMW simply skips activating SA $n$ for this particular time slot, the system remains idle, and the actual queue length $Q_k(t+1)=Q_k(t)$. On the other hand, even though the system stays idle, the virtual queue length $\q(t)$ is still updated by \eqref{eq:DMW-update}. The above DMW algorithm is used to prove Propisition~\ref{prop:ach} in \cite{jiang2009stable}.

\subsubsection{Open Problems for SPNs with Random Departure} \label{sec:SPN_random}
Although the SPN with deterministic departure is relatively well understood, those SPN scheduling results cannot be applied to the INC vr-network. The reason is as follows. When a packet is broadcast by the base station, it can arrive at a random subset of receivers with certain probability distributions. Therefore, the vr-packets move among the vr-queues according to some probability distribution. This is not compatible with the deterministic departure SPN model, in which when an SA is activated we know {\em deterministically} $\bin{k}{n}(c)$ and $\bout{k}{n}(c)$, the service rates when the channel quality is $\cq(t)=c$. We call the SPN model that allows random
$\bin{k}{n}(c)$ and $\bout{k}{n}(c)$ the SPN with random departure.

\begin{figure}
  \center
  \includegraphics[width=6cm]{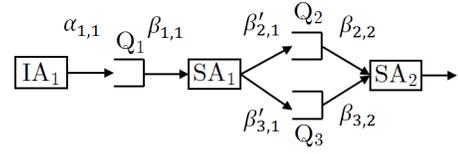}\\
  \caption{An SPN with random departure.}\label{fig:SPN_eg}
  \vspace{-.5cm}
\end{figure}

SPNs with random departure provide a unique challenge for the scheduling design. \cite{jiang2009stable} provides the following example illustrating this issue. Fig.~\ref{fig:SPN_eg} describes an SPN with 6 transition edges. We assume IA1 is activated at every time slot and $\alpha_{1,1}=\bin{1}{1}=\bin{2}{2}=\bin{3}{2}=1$. Namely, for every time $t$, $\alpha_{1,1}=1$ packet will enter $Q_1$; in every time slot if we activate SA1, $\bin{1}{1}=1$ packet will leave $Q_1$; if we activate SA2, $\bin{2}{2}=1$ packet will leave $Q_2$ and $\bin{3}{2}=1$ packet will leave $Q_3$. We assume these 4 transitions are deterministic but the two transitions $\text{SA1}\rightarrow Q_2$ and $\text{SA1}\rightarrow Q_3$ are random. Specifically, we assume that there are two possible values of the pair $(\bout{2}{1},\bout{3}{1})$: $(\bout{2}{1},\bout{3}{1})= (1,0)$ with probability $0.5$ and $(\bout{2}{1},\bout{3}{1})= (0,1)$ with probability $0.5$. That is, whenever SA1 is activated, it takes a packet from $Q_1$, and with probability $0.5$ this packet goes to $Q_2$. Otherwise, this packet goes to $Q_3$. The random departure of SA1 implies that the queue length difference $|Q_2|-|Q_3|$ forms a binary random walk. Note that SA2 has no impact on $|Q_2|-|Q_3|$ since it always takes 1 packet from each of the queues. The analysis of the random walk  shows that $|Q_2|-|Q_3|$ goes unbounded with rate $\sqrt{t}$. And hence there is no scheduling algorithm which can stabilize both $|Q_2|$ and $|Q_3|$ simultaneously even though this example satisfies the flow-conservation law in \eqref{eq:balance} in the sense of expectation. 

\section{The Proposed New INC Solution} \label{sec:new-coding}
In Section~\ref{sec:existing}, we discuss the limitations of the existing works on the INC block code design and on the schedulers for SPNs, separately. In this section, we describe our new low-complexity binary INC scheme that achieves the block code capacity. In Section~\ref{sec:new_scheduling}, we present our new scheduler design for the SPN with random departure. In Section~\ref{sec:combined}, we will combine the proposed solutions to form the optimal dynamic INC design, see Fig.~\ref{fig:break-down}. For the new block code design in this section, we first describe the encoding steps and then discuss the decoding steps and buffer management.
\begin{figure}
\centering
  \includegraphics[width=7cm]{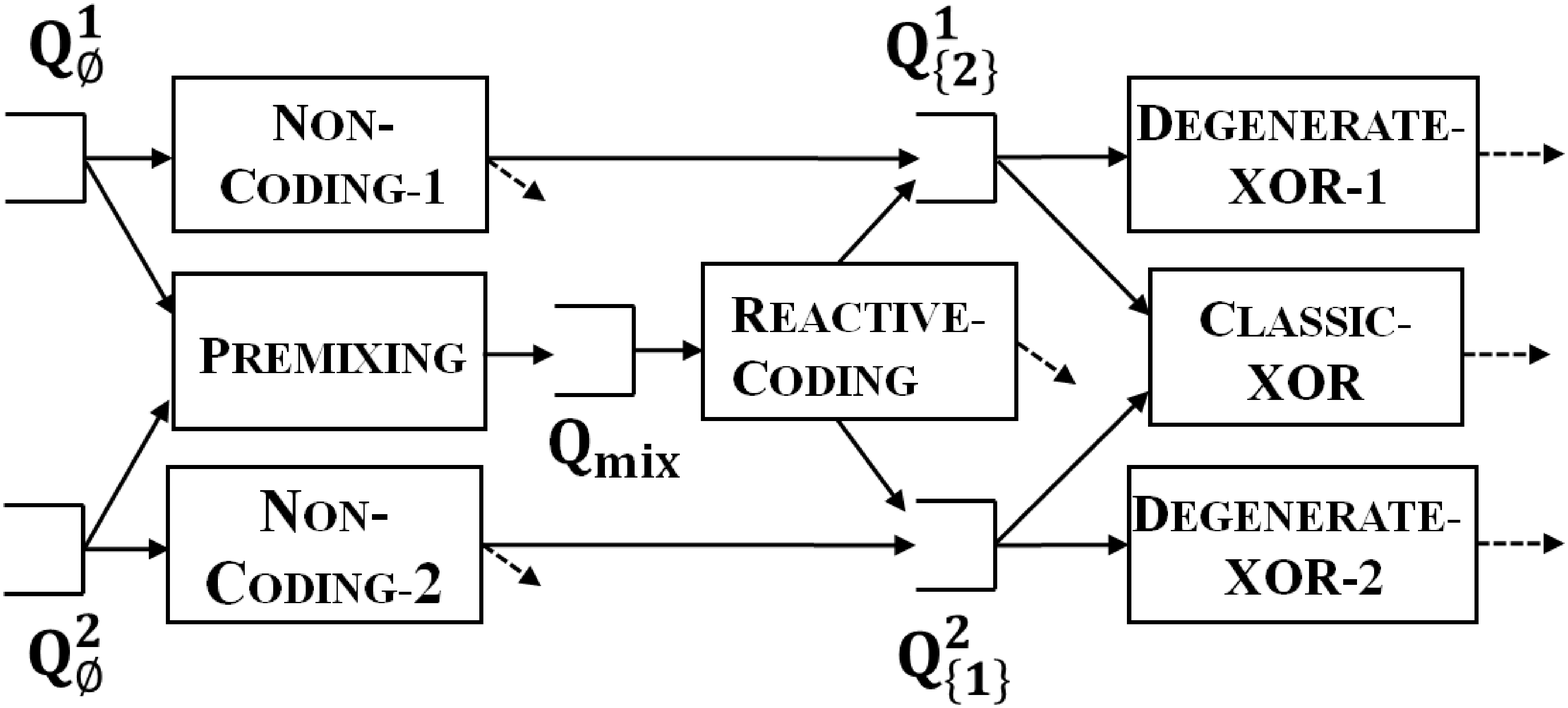}\\
  \caption{The virtual network of the proposed new INC solution.}\label{fig:7-type}
  \vspace{-.3cm}
\end{figure}

\subsection{Encoding}
The proposed new INC solution is described as follows. We build upon the existing 5 operations,
{\sc Non-Coding-1}, {\sc Non-Coding-2}, {\sc Classic-XOR}, {\sc Degenerate-XOR-1}, and {\sc Degenerate-XOR-2}. See Fig.~\ref{fig:5-type} and the discussion in Sections~\ref{sec:introduction} and \ref{subsubsec:degenerate-XOR}. In addition, we add 2 more operations, termed {\sc Premixing} and {\sc Reactive-Coding}, respectively, and 1 new virtual queue, termed $Q_\text{mix}$. We plot the vr-network of the new scheme in Fig.~\ref{fig:7-type}. From Fig.~\ref{fig:7-type}, we can clearly see that {\sc Premixing} involves both $Q^1_\emptyset$ and $Q^2_\emptyset$ as input and outputs to $Q_\text{mix}$. {\sc Reactive-Coding} involves $Q_\text{mix}$ as input and outputs to $Q^1_{\{2\}}$ or $Q^2_{\{1\}}$ or simply lets the vr-packet leave the vr-network (described by the dotted arrow). For every time instant, we can choose one of the 7 operations and the goal  is to stabilize the vr-network. In the following, we describe in details how these two INC operations work and how to integrate them with the other 5 operations. Our description contains 4 parts.

{\em Part I:}
The two operations, {\sc Non-Coding-1} and {\sc Non-Coding-2}, remain the same.

{\em Part II:} We now describe the new operation {\sc Premixing}. We can choose {\sc Premixing} only if both $Q^1_\emptyset$ and $Q^2_\emptyset$ are non-empty. Namely, there are $X_i$ packets and $Y_j$ packets that have not been heard by any of $d_1$ and $d_2$. Whenever we schedule {\sc Premixing}, we choose one $X_i$ from $Q^1_\emptyset$ and one $Y_j$ from $Q^2_\emptyset$ and send $[X_i+Y_j]$. If neither $d_1$ nor $d_2$ receives it, both $X_i$ and $Y_j$ remain in their original queues.

If at least one of $\{d_1,d_2\}$ receives it, we do the following. We remove {\em both} $X_i$ and $Y_j$ from their individual queues. We insert a tuple $(\rcpt; X_i,Y_j)$ into $Q_{\text{mix}}$. That is, unlike the other queues for which each entry is a single vr-packet, each entry of $Q_{\text{mix}}$ is a tuple.

The first coordinate of $(\rcpt; X_i,Y_j)$ is $\rcpt$, the reception status of $[X_i+Y_j]$. For example, if $[X_i+Y_j]$ was received by $d_2$ but not by $d_1$, then we set/record $\rcpt=\overline{d_1}d_2$; If $[X_i+Y_j]$ was received by both $d_1$ and $d_2$, then $\rcpt=d_1d_2$. The second and third coordinates store the participating packets $X_i$ and $Y_j$ separately. The reason why we do not store the linear sum directly is due to the new {\sc Reactive-Coding} operation.

{\em Part III:} We now describe the new operation {\sc Reactive-Coding}.
For any time $t$, we can choose {\sc Reactive-Coding} only if there is at least one tuple $(\rcpt; X_i,Y_j)$ in $Q_\text{mix}$. Choose one tuple from $Q_\text{mix}$ and denote it by $(\rcpt^*; X_i^*, Y_j^*)$. We now describe the encoding part of {\sc Reactive-Coding}.

Whenever we schedule {\sc Reactive-Coding},  if $\rcpt^*=d_1\overline{d_2}$, send $Y_j^*$. If $\rcpt^*=\overline{d_1}d_2$, send $X_i^*$. If $\rcpt^*={d_1}d_2$, send $X_i^*$. One can see that the coding operation depends on the reception status $\rcpt^*$ when $[X_i^*+Y_j^*]$ was first transmitted. This is why it is named {\sc Reactive-Coding}.

\begin{table}
\caption{A summary of the {\sc Reactive-Coding} operation\label{tab:reactive}}
\centering
 \includegraphics[width=8.5cm]{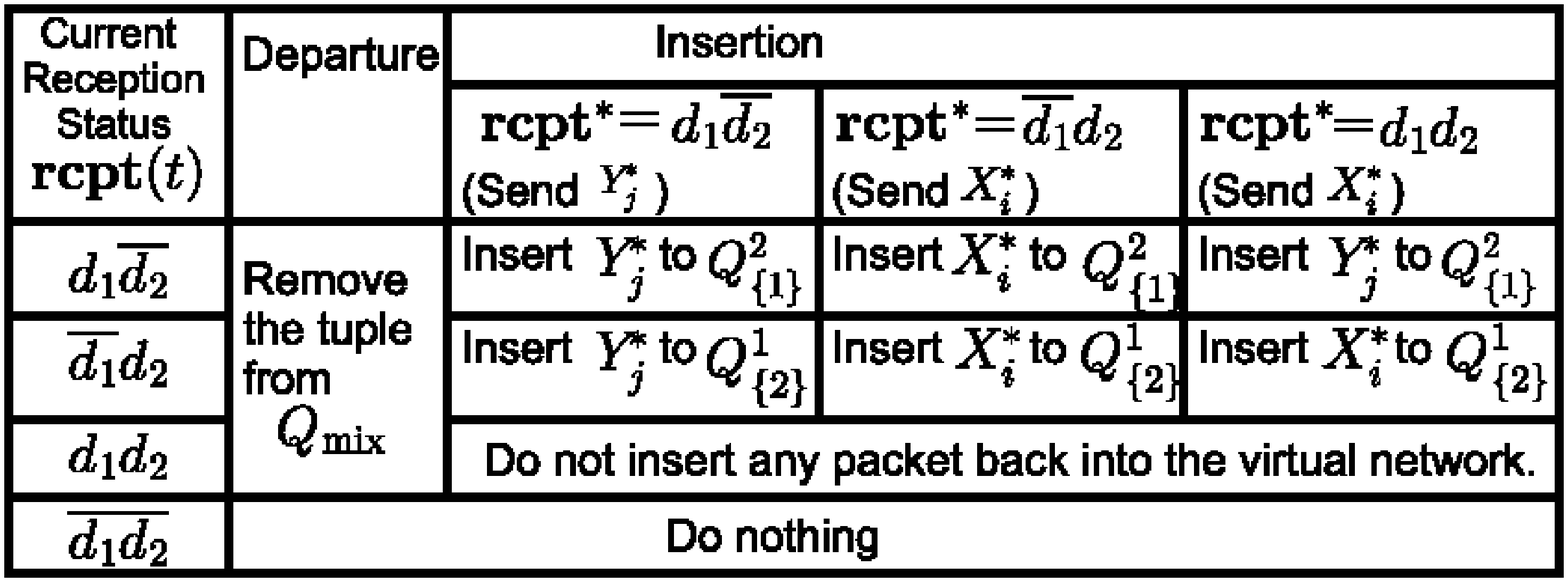}
 \vspace{-.0cm}
\end{table}

The movement of the vr-packets depends on the current reception status of time $t$, denoted by $\rcpt(t)$, and also on the old reception status $\rcpt^*$ when the sum $[X_i^*+Y_j^*]$ was originally transmitted. The detailed movement rules are described in Table~\ref{tab:reactive}. The way to interpret the table is as follows. For example, when $\rcpt(t)=\overline{d_1d_2}$, i.e., neither $d_1$ nor $d_2$ receives the current transmission, then we do nothing, i.e., keep the tuple inside $Q_{\text{mix}}$. On the other hand, we remove the tuple from $Q_\text{mix}$ whenever $\rcpt(t)\in\{d_1\overline{d_2},\overline{d_1}d_2,d_1d_2\}$. If $\rcpt(t)=d_1d_2$, then we remove the tuple but do not insert any vr-packet back to the vr-network, see the second last row of Table~\ref{tab:reactive}. The tuple essentially leaves the vr-network in this case. If $\rcpt(t)=d_1\overline{d_2}$ and $\rcpt^*=d_1d_2$, then we remove the tuple from $Q_\text{mix}$ and insert $Y_j^*$ to $Q^2_{\{1\}}$. The rest of the combinations can be read from Table~\ref{tab:reactive} in the same way. One can verify that the optimal INC example introduced in Section~\ref{subsubsec:degenerate-XOR} is a direct application of the {\sc Premixing} and {\sc Reactive-Coding} operations.

Before we continue describing the slight modification to {\sc Classic-XOR}, {\sc Degenerate-XOR-1}, and {\sc Degenerate-XOR-2}, we briefly explain why the combination of {\sc Premixing} and {\sc Reactive-Coding} works.  To facilitate discussion, we call the time slot in which we use {\sc Premixing} to transmit $[X_i^*+Y_j^*]$ ``slot~1'' and the time slot in which we use {\sc Reactive-Coding} ``slot~2,'' even though the coding operations {\sc Premixing} and {\sc Reactive-Coding} may not be scheduled in two adjacent time slots. Using this notation, if  $\rcpt^*=d_1\overline{d_2}$ and $\rcpt(t)=d_1d_2$, then it means that $d_1$ receives $[X_i^*+Y_j^*]$ and $Y_j^*$ in slots~1 and 2, respectively and $d_2$ receives $Y_j^*$ in slot 2. In this case, $d_1$ can decode the desired $X_i^*$ and $d_2$ directly receives the desired $Y_j^*$. We now consider the perspective of the vr-network. Table~\ref{tab:reactive} shows that the tuple will be removed from $Q_\text{mix}$ and leave the vr-network. Therefore, no queue in the vr-network stores any of $X_i^*$ and $Y_j^*$. This correctly reflects the fact that both $X_i^*$ and $Y_j^*$ have been received by their intended destinations.

Another example is when $\rcpt^*=\overline{d_1}{d_2}$ and $\rcpt(t)=d_1\overline{d_2}$. In this case,  $d_2$ receives $[X_i^*+Y_j^*]$ in slot 1 and $d_1$ receives $X_i^*$ in slot~2.  From the vr-network's perspective, the movement rule (see Table~\ref{tab:reactive}) removes the tuple from $Q_\text{mix}$ and insert an $X_i^*$ packet to $Q^2_{\{1\}}$. Since a vr-packet is removed from a session-1 queue\footnote{$Q_\text{mix}$ is regarded as both a session-1 and a session-2 queue simultaneously.} $Q_\text{mix}$ and inserted to a session-2 queue $Q^2_{\{1\}}$,  the total number of vr-packets in the session-1 queue decreases by 1. This correctly reflects the fact that $d_1$ has received 1 desired packet $X_i^*$ in slot~2.

An astute reader may wonder why in this example we can put $X_i^*$, a session-1 packet, into a session-2 queue $Q^2_{\{1\}}$. The reason is that whenever $d_2$ receives $X_i^*$ in the future, it can recover its desired $Y_j^*$ by subtracting $X_i^*$ from the linear sum $[X_i^*+Y_j^*]$ it received in slot 1 (recall that $\rcpt^*=d_1\overline{d_2}$.) Therefore, $X_i^*$ is now information-equivalent to $Y_j^*$, a session-2 packet. Moreover, $d_1$ has received $X_i^*$. Therefore, in terms of the information it carries, $X_i^*$ is no different than a session-2 packet that has been overheard   by $d_1$. As a result, it is fit to put $X_i^*$ in $Q^2_{\{1\}}$.

{\em Part IV:} We now describe the slight modification to {\sc Classic-XOR}, {\sc Degenerate-XOR-1}, and {\sc Degenerate-XOR-2}. A unique feature of the new scheme is that some packets in $Q^2_{\{1\}}$ may be an $X_i^*$ packet that is inserted by {\sc Reactive-Coding} when $\rcpt^*=\overline{d_1}{d_2}$ and $\rcpt(t)=d_1\overline{d_2}$. (Also some $Q^1_{\{2\}}$ packets may be $Y_j^*$.) However, in our previous discussion, we have shown that those $X_i^*$ in $Q^2_{\{1\}}$ is information-equivalent to a $Y_j^*$ packet overheard by $d_1$. Therefore, in the {\sc Classic-XOR} operation, we should not insist on sending $[X_i+Y_j]$ but can also send $[P_1+P_2]$ as long as $P_1$ is from $Q^1_{\{2\}}$ and $P_2$ is from $Q^2_{\{1\}}$. The same relaxation must be applied to {\sc Degenerate-XOR-1} and {\sc Degenerate-XOR-2} operations. Other than this slight relaxation, the three operations work in the same way as previously described in Sections~\ref{sec:introduction} and \ref{subsubsec:degenerate-XOR}.

As will be seen in Proposition~\ref{prop:combine-capacity} of Section~\ref{sec:combined}, the two new operations {\sc Premixing} and {\sc Reactive-Coding} allow us to achieve the linear block-code capacity for any time-varying channels. We conclude this section by listing in Table~\ref{tab:transition} the transition probabilities of half of the edges of the vr-network of Fig.~\ref{fig:7-type}. For example, when we schedule {\sc Premixing}, we remove a packet from $Q^1_\emptyset$ if at least one of $\{d_1,d_2\}$ receives it. As a result, the transition probability along the $Q^1_\emptyset\rightarrow${\sc Premixing} edge is $p_{d_1\vee d_2}\stdef p_{d_1\overline{d_2}}+p_{\overline{d_1}d_2}+p_{d_1d_2}$. All the other transition probabilities in Table~\ref{tab:transition} can be derived similarly. The transition probability of the other half of the edges can be derived by symmetry.

\begin{table}
\caption{A summary of the transition probability of the virtual network in Fig.~\ref{fig:7-type}, where $p_{d_1\vee d_2}\stdef p_{d_1\overline{d_2}}+p_{\overline{d_1}d_2}+p_{d_1d_2}$; $p_{d_1}\stdef p_{d_1\overline{d_2}}+p_{d_1{d_2}}$; {\sc NC1} stands for {\sc Non-Coding-1}; {\sc CX} stands for {\sc Classic-XOR}; {\sc DX1} stands for {\sc Degenerate-XOR-1}; {\sc PM} stands for {\sc Premixing}; {\sc RC} stands for {\sc Reactive-Coding}. 
\label{tab:transition}}
\centering
\begin{tabular}{cl|cl}
\hline
\hline
Edge & Trans.\ Prob. & Edge & Trans.\ Prob.\\
\hline
$Q^1_\emptyset\rightarrow${\sc NC1} & $p_{d_1\vee d_2}$ & $Q^1_\emptyset\rightarrow${\sc PM} & $p_{d_1\vee d_2}$ \\
{\sc NC1}$\rightarrow Q^1_{\{2\}}$ & $p_{\overline{d_1}d_2}$ & {\sc PM}$\rightarrow Q_\text{mix}$& $p_{d_1\vee d_2}$ \\
$Q^1_{\{2\}}\rightarrow${\sc DX1} & $p_{d_1}$ & $Q_\text{mix}\rightarrow${\sc RC} & $p_{d_1\vee d_2}$ \\
$Q^1_{\{2\}}\rightarrow${\sc CX} & $p_{d_1}$ & {\sc RC}$\rightarrow Q^1_{\{2\}}$& $p_{\overline{d_1}d_2}$\\
\hline
\hline
 \end{tabular}
\end{table}

\subsection{Decoding and Buffer Management at Receivers}

\par It is worth emphasizing that the vr-network is a conceptual tool used by the source $s$ to decide what to transmit in each time slot. As a result, for the encoding purposes $s$ only needs to store in its memory/buffer all the packets that currently participate in the vr-network. This automatically implies that as long as the queues in the vr-network are stabilized, the actual memory usage at the source is also stabilized. However, for the $1$-to-$2$ access point network to be stable, one needs to ensure that the memory usage for the two receivers is stabilized as well. In this subsection we discuss the decoding operations and the memory usage at the receivers.

\par It is clear that each receiver needs to store some packets for the decoding purposes. A very commonly used assumption in the Shannon-capacity literature is to assume that the receivers store all the overheard packets in order to decode the possible XORed packets sent from the source. No packets will ever be removed from the buffer under such a policy. Obviously, such an infinite-buffer scheme is highly impractical. 

\par In the existing INC scheduling works \cite{KattiRahulHuKatabiMedardCrowcroft06,GeorgiadisTassiulas13,AthanasiadouGeorgiadis13,Paschos12}, another commonly used buffer management scheme is the following. For any time $t$, define $i^*$ (resp. $j^*$) as the smallest $i$ (resp. $j$) such that $d_1$ (resp. $d_2$) has not decoded $X_i$ (resp. $Y_j$) in the end of time $t$. Then each receiver can simply remove any $X_i$ and $Y_j$ in the buffer for those $i<i^*$ and $j<j^*$. The reason is that those $X_i$ and $Y_j$ has already been known by their intended receivers, will not participate in any future transmission, and thus can be removed from the receive buffer without any impact to future decoding.  

\par On the other hand, under such a buffer management scheme, the receivers may use significantly more memory than that of the source, which was observed in our numerical experiments. The reason is as follows. Suppose $d_1$ has decoded $X_1$, $X_3$, $X_4$,..., $X_8$, and $X_{10}$ and suppose $d_2$ has decoded $Y_1$ to $Y_4$ and $Y_6$ to $Y_{10}$. In this case $i^*=2$ and $j^*=5$. The aforementioned scheme will keep all $X_2$ to $X_{10}$ in the buffer of $d_2$ and all $Y_5$ to $Y_{10}$ in the buffer of $d_1$. But it turns out that the source is interested in only sending 3 more packets $X_2$, $X_9$, and $Y_5$. This apparent waste of memory is due to the fact that having 3 more packets to send does not mean that we only need to store $X_2$, $X_9$ and $Y_5$ in the buffer of the {\em receivers}. For the decoding purposes, we need to store extra ``overheard" packets that can facilitate decoding in the future. But on the other hand, the above buffer management scheme is too conservative and very inefficient since it does not trace the actual overhearing status of each packet and only use the simplest $i^*$ and $j^*$ pair to decide whether to prune the packets in the buffers of the receivers. 

\par In contrast with the above buffer management scheme used in \cite{KattiRahulHuKatabiMedardCrowcroft06,GeorgiadisTassiulas13,AthanasiadouGeorgiadis13,Paschos12}, our vr-network scheme admits the following efficient decoding operations and buffer management solution. In the following, we describe the decoding and buffer management at $d_1$. The operations at $d_2$ can be done symmetrically. Our description consists of two parts. We first describe how to perform decoding at $d_1$ and which packets need to be stored in $d_1$'s buffer, while assuming that any packets that have been stored in the buffer will never be expunged. In the second part, we describe how to prune the memory usage without affecting the decoding operations.

{\bf Upon $d_1$ receiving a packet:} Case 1: If the received packet is generated by {\sc Non-Coding-1}, then such a packet must be $X_i$ for some $i$. We thus pass such an $X_i$ to the upper layer; Case 2: If the received packet is generated by {\sc Non-Coding-2}, then such a packet must be $Y_j$ for some $j$. We store $Y_j$ in the buffer of $d_1$; Case 3: If the received packet is generated by {\sc Premixing}, then such a packet must be $[X_i+Y_j]$. We store the linear sum $[X_i+Y_j]$ in the buffer. Case 4: If the received packet is generated by {\sc Reactive Coding }, then such a packet can be either $X_i^*$ or $Y_j^*$, see Table~\ref{tab:reactive} for detailed descriptions of {\sc Reactive-Coding}. 

\par We have two sub-cases in this scenario. Case 4.1: If the packet is $X_i^*$, we pass such an $X_i^*$ to the upper layer. Then $d_1$ examines whether it has stored $[X_i^*+Y_j^*]$ in its buffer. If so, use $X_i^*$ to decode $Y_j^*$ and insert $Y_j^*$ to the buffer. If not, store a separate copy of $X_i^*$ in the buffer even though one copy of $X_i^*$ has already been passed to the upper layer. Case 4.2: If the packet is $Y_j^*$, then by Table~\ref{tab:reactive}, it is clear that $d_1$ must have received the linear sum $[X_i^*+Y_j^*]$ in the corresponding {\sc Premixing} operation in the past. Therefore, $[X_i^*+Y_j^*]$ must be in the buffer of $d_1$ already. We can thus use $Y_j^*$ and $[X_i^*+Y_j^*]$ to decode the desired $X_i^*$. Receiver $d_1$ then passes the decoded $X_i^*$ to the upper layer and stores $Y_j^*$ in its buffer.

\par Case 5: If the received packet is generated by {\sc Degenerate XOR-1}, then such a packet can be either $X_i$ or $Y_j$, where $Y_j$ are those packets in $Q^1_{\{2\}}$ but coming from {\sc Reactive Coding}, see Fig.~\ref{fig:7-type}. Case 5.1: If the packet is $X_i$, we pass such an $X_i$ to the upper layer. Case 5.2: If the packet is $Y_j$, then from Table~\ref{tab:reactive}, it must be corresponding to the intersection of the row of $\rcpt=\overline{d_1}d_2$ and the column of $\rcpt^*=d_1\overline{d_2}$. As a result, $d_1$ must have received the corresponding $[X_i+Y_j]$ in the {\sc Premixing} operation. By Case 3, the linear sum has been stored in the buffer, and $d_1$ can thus use the received $Y_j$ to decode the desired $X_i$. After decoding, $X_i$ is passed to the upper layer. 

\par Case 6: the received packet is generated by {\sc Degenerate XOR-2}. Consider two subcases. Case 6.1: the received packet is $X_i$. It is clear from Fig.~\ref{fig:7-type} that such $X_i$ must come from {\sc Reactive-Coding} since any packet from $Q_\emptyset^2$ to $Q_{\{1\}}^2$ must be a $Y_j$ packet. By Table~\ref{tab:reactive} and the row corresponding to $\rcpt=d_1\overline{d_2}$,  any $X_i\in Q^2_{\{1\}} $ that came from {\sc Reactive-Coding} must correspond to the column of $\rcpt^*=\overline{d_1}d_2$. By the second half of Case 4.1, such $X_i\in Q^2_{\{1\}}$ must be in the buffer of $d_1$. As a result, $d_1$ can simply ignore any $X_i$ packet it receives from {\sc Degenerate XOR-2}. Case 6.2: the received packet is $Y_j$. By the discussion of Case 2, if the $Y_j\in Q^2_{\{1\}}$ came from {\sc Non-Coding-2}, then it must be in the buffer of $d_1$ already. As a result, $d_1$ can simply ignore those $Y_j$ packets. If the $Y_j\in Q^2_{\{1\}}$ came from {\sc Reactive-Coding}, then by Table~\ref{tab:reactive} and the row corresponding to $\rcpt=d_1\overline{d_2}$, those $Y_j\in Q^2_{\{1\}}$ must correspond to the column of either $\rcpt^*=d_1\overline{d_2}$ or $\rcpt^*=d_1d_2$. By the first half of Case 4.1 and by Case 4.2, such $Y_j\in Q^2_{\{1\}}$ must be in the buffer of $d_1$ already. Again, $d_1$ can simply ignore those $Y_j$ packets. From the discussion of Cases 6.1 and 6.2, {\em any packet generated by {\sc Degenerate XOR-2} is already known to $d_1$, and nothing needs to be done in this case.}\footnote{The discussion of Cases 5 and 6 echoes our arguments in the end of [Section~\ref{subsec:encoding}: Encoding] that any packet in $Q^2_{\{1\}}$ (which can be either $X_i$ or $Y_j$)  is information-equivalent to a session-2 packet that has been overheard by $d_1$.}

\par  Case 7: the received packet is generated by {\sc Classic-XOR}. Since we have shown in Case~6 that any packet in $Q^2_{\{1\}}$ is already known to $d_1$, receiver $d_1$ can simply subtract the $Q^2_{\{1\}}$ packet from the linear sum received in Case 7. As a result, from $d_1$'s perspective, it is no different than directly receiving a $Q^1_{\{2\}}$ packet, i.e., Case 5. As a result, $d_1$ will repeat the decoding  operation and buffer management in the same way as in Case 5. 

{\bf Periodically pruning the memory:} In the above discussion, we elaborate which packets $d_1$ should store in its buffer and how to use them for decoding, while assuming no packet will ever be removed from the buffer. In the following, we discuss how to remove packets from the buffer of $d_1$.

\par We first notice that by the discussion of Cases 1 to 7, the uncoded packets in the buffer of $d_1$, i.e., those of the form of either $X_i$ or $Y_j$, are used  for decoding  {\em only  in the scenario of Case 7}. Namely, they are used to remove the $Q^2_{\{1\}}$ packet participating in the linear sum of {\sc Classic-XOR}. As a result, periodically we let the source $s$ send to $d_1$ the list\footnote{Only the packet IDs are sent, not the payload. Therefore the overhead of sending the list is small. Moreover, we only need to send the ``incremental changes" of the list and $d_1$ can update the list by itself. In this way, the overhead of sending the list can be made negligible.} of all packets in $Q^2_{\{1\}}$ of the vr-network. After receiving the list, $d_1$ simply removes from its buffer any uncoded packets $X_i$ and/or $Y_j$ that are no longer in $Q^2_{\{1\}}$.

\par We then notice that by the discussion of Cases 1 to 7, the linear sum $[X_i+Y_j]$ in the buffer of $d_1$ is only used in one of the following two scenarios: (i) To decode $Y_j$ in Case 4.1 or to decode $X_i$ in Case 4.2; and (ii) To decode $X_i$ in Case 5.2. As a result, the $[X_i+Y_j]$ in the buffer is ``useful" only if one of the following two conditions are satisfied: (a) The corresponding tuple $(\rcpt,X_i,Y_j)$ is still in the $Q_\text{mix}$ of the vr-network, which corresponds to the scenarios of Cases 4.1 and 4.2; and (b) If the participating $Y_j$ is still in the $Q^1_{\{2\}}$ of the vr-network. By the above observation, periodically we let the source $s$ send to $d_1$ the list of all packets in $Q^1_{\{2\}}$ and $Q_\text{mix}$ of the vr-network.\footnote{One can see that both $d_1$ and $d_2$ need to receive the lists of packets in $Q^1_{\{2\}}$, $Q^2_{\{1\}}$, and $Q_\text{mix}$. Therefore, $s$ can simply {\em broadcast} (the changes) of the three lists to both $d_1$ and $d_2$.} After receiving the list, $d_1$ simply removes from its buffer any linear sum $[X_i+Y_j]$ that satisfies neither (a) nor (b).  

The above pruning mechanism ensures that only the packets useful for future decoding are kept in the buffer of $d_1$ and $d_2$. Furthermore, it also leads to the following lemma. 
\begin{lemma}\label{lemma:bound-receiver-buffer}Assume the lists of packets in $Q^1_{\{2\}}$, $Q^2_{\{1\}}$, and $Q_\text{mix}$ are sent to $d_1$ after every time slot. The number of packets in the buffer of $d_1$ is upper bounded by $|Q^1_{\{2\}}|+|Q^2_{\{1\}}|+|Q_\text{mix}|$. 
\end{lemma}

{\em Proof: }From our discussion, the total number of uncoded packets $X_i$ or $Y_j$ in the buffer of $d_1$ is upper bounded by $|Q^2_{\{1\}}|$. Also, the total number of linear sum $[X_i+Y_j]$ in the buffer of $d_1$ is upper bounded by $|Q_\text{mix}|$ plus the number of $Y_j$ packets in $Q^1_{\{2\}}$, which is further bounded by $|Q_\text{mix}|+|Q^1_{\{2\}}|$. As a result, the total number of packets in the buffer of $d_1$ is upper bounded by $|Q^1_{\{2\}}|+|Q^2_{\{1\}}|+|Q_\text{mix}|$. \qed

Lemma~\ref{lemma:bound-receiver-buffer} implies that as long as the queues in the vr-network are stabilized, the actual memory usage at both the source and the destinations can be stabilized simultaneously. Moreover, the combined memory usage of the source and 2 receivers will be upper bounded by $Q^1_\emptyset+ Q^2_\emptyset+ 3|Q^1_{\{2\}}|+3|Q^2_{\{1\}}|+3|Q_\text{mix}|$ in the vr-network. 

{\em Remark:} In addition to efficient decoding and buffer management, we notice that in the proposed INC scheme, only the binary XOR is used and each transmitted packet is either an uncoded packet or a linear sum of two packets. Therefore, during transmission we only need to store 1 or 2 packet sequence numbers in the header of the uncoded/coded packet, depending on whether we send an uncoded packet or a linear sum. As a result, the communication overhead of the proposed scheme is very small.

\section{The Proposed Scheduling Solution \label{sec:new_scheduling}}

In this section, we first formalize the model of SPNs with random departure and then we propose a new scheme that achieves the optimal throughput region for SPNs with random departure. We conclude this section by providing the corresponding stability/throughput analysis.

\subsection{A Simple SPN model with Random Departure} \label{subsec:01SPN}
Although our solution applies to general SPNs with random departure, for illustration purposes we describe our scheme by focusing on a simple SPN model with random departure, which we termed the (0,1) random SPN. The (0,1) random SPN includes the INC vr-network in Section~\ref{sec:new-coding} as a special example and is thus sufficient for our discussion.

Recall the definitions in Section~\ref{subsubsec:SPN-deterministic} for SPNs with deterministic departure (we use {deterministic SPNs} as shorthand). The differences between the (0,1) random SPN and the deterministic SPN are:

Difference 1: In a deterministic SPN, SA $n$ can be activated only if for all $k$ in the input queues $\SAin{n}$, queue $k$ has at least $\bin{k}{n}$ number of packets in the queue. For comparison, in a (0,1) random SPN, SA $n$ can be activated only if for all $k\in\SAin{n}$, queue $k$ has at least 1 packet in the queue. For easier future reference, we say SA $n$ is {\em feasible} at time $t$ if at time $t$ queue $k$ has at least 1 packet for all $k\in\SAin{n}$. Otherwise, we say SA $n$ is infeasible at time $t$.

Difference 2:  In a deterministic SPN, when SA $n$ is activated with the channel quality $c$, exactly $\bin{k}{n}(c)$ number of packets will leave queue $k$ for all $k\in\SAin{n}$. In a (0,1) random SPN, when SA $n$ is activated with the channel quality $c$ (assuming SA $n$ is feasible), the number of packets leaving queue $k$ is a binary random variable, $\bin{k}{n}(c)$, with mean $\overline{\bin{k}{n}(c)}$ for all $k\in\SAin{n}$. Namely, with probability $\overline{\bin{k}{n}(c)}$, 1 packet will leave queue $k$ and with probability $1-\overline{\bin{k}{n}(c)}$ no packet will leave queue $k$. Since the packet consumption is Bernoulli, in a (0,1) random SPN, it is possible that an SA consumes zero packet even after being activated. However, since we do not know how many packets will be consumed beforehand, the (0,1) random SPN imposes that all the input queues have at least 1 packet before we can activate an SA, even though when we actually activate the SA, it sometimes consumes zero packet. For comparison, in a deterministic SPN, an SA $n$ is feasible if all its input queues have at least $\bine{k}{n}(\cq(t))$ packets and it will always consume exactly $\bine{k}{n}(\cq(t))$ packets from its input queues once activated (see Difference 1).

Difference 3:
In a (0,1) random SPN, when SA $n$ is activated with the channel quality $c$ (assuming SA $n$ is feasible), the number of packets entering queue $k$ is a binary random variable with mean $\overline{\bout{k}{n}(c)}$ for all $k\in\SAout{n}$. 

\par We also use the following 3 technical assumptions for the (0,1) random SPN: Assumption 1: Given any channel quality $c\in\bigcq$, both the input and output service matrix $\Bin$ and $\Bout$ are independently distributed over time. Assumption 2: Each vector in the set of possible service vectors $\mathfrak{X}$ can have at most 1 non-zero coordinate. Namely, we can activate at most one service activity (out of totally $N$ SAs) at any given time. Assumption 3: For any $\cq(t)$, the expectation of $\bin{k}{n}(\cq(t))$ (resp. $\bout{k}{n}(\cq(t))$) with $k\in\SAin{n}$ (resp. $k\in\SAout{n}$) is always strictly in $(0,1]$. Namely, we do not consider the limiting case in which the Bernoulli random variables are always 0. Assumption 1 is related to the practical scenarios. Assumptions 2 and 3 are for rigorously proving the stability region.

\par One can easily verify that the three INC vr-networks in Figs.~\ref{fig:BPEC_eg}, \ref{fig:5-type}, and~\ref{fig:7-type} are special examples of the (0,1) random SPN and they satisfy the 3 technical assumptions as well.

\subsection{The Proposed Scheduler For (0,1) Random SPNs} \label{subsec:proposed-DMW}
Similar to the DMW algorithm, each queue $k$ maintains a real-valued counter $q_k(t)$, the virtual queue length. Initially, $q_k(1)$ is set to 0. For any time $t$, the realization of each entry in the input and output service matrices $\Bin$ and $\Bout$ takes values in either 0 or 1 since we are focusing on a (0,1) random SPN. We compute $\overline{\Bin(\cq(t))}\stdef\EE(\Bin|\cq(t))$ and $\overline{\Bout(\cq(t))}\stdef\EE(\Bout|\cq(t))$, the {\em expected} input and output service matrices, respectively, when the channel quality is $\cq(t)$. The entries of $\overline{\Bin(\cq(t))}$ and $\overline{\Bout(\cq(t))}$ are denoted by $\overline{\bin{k}{n}(\cq(t))}$ and $\overline{\bout{k}{n}(\cq(t))}$, respectively. Obviously, by definition, the expected input and output service rates are non-negative numbers. For each time $t$, we choose the preferred service vector by the back-pressure decision rule \eqref{eq:DMW-schedule} except for that the back-pressure vector $\ve{d}(t)$ is now computed by
\begin{align}
\ve{d}(t)=\left(\overline{\Bin(\cq(t))}-\overline{\Bout(\cq(t))}\right)^\tran \ve{q}(t).\label{eq:new-backpressure}
\end{align}
We use the new back-pressure vector $\ve{d}(t)$ plus \eqref{eq:DMW-schedule} to find the preferred SA $n^*$, i.e., all the coordinates of $\ve{x}^*$ are zero except for the $n^*$-th coordinate being one. We then check whether the preferred SA $n^*$ is feasible. If so, we officially schedule SA $n^*$. If not, we let the system to be idle,\footnote{The reason of letting the system idle is to facilitate rigorous stability analysis. In practice, we can choose arbitrarily any other feasible SA at that moment.} i.e., the actually scheduled service vector $\ve{x}(t)=\0$ is now all-zero.
 
\par Regardless of whether the preferred SA $n^*$ is feasible or not, we update $\ve{q}(t)$ by
\begin{align}
\ve{q}(t+1) =& \ve{q}(t)+\im\cdot \ve{a}(t)\nonumber \\
 &+ \left(\overline{\Bout(\cq(t))}-\overline{\Bin(\cq(t))}\right)\cdot \ve{x}^*(t). \label{eq:DMW-update-new}
\end{align}
Note that $\ve{q}(t)$ can sometimes take negative values since we do not project $\ve{q}(t)$ to positive reals. 

\par In short, we borrow the wisdom of DMW so that we can make scheduling decisions based on the virtual queue lengths $q_k(t)$ that can take negative values. But then we update $q_k(t)$ only by the expected service rates rather than the actual service rates since we are dealing with a random SPN instead of a deterministic SPN. For notation simplicity, we denote the proposed scheduler for (0,1) random SPNs by $\SCH_\text{avg}$.

\subsection{Performance Analysis\label{subsec:analysis}}

The example in Section~\ref{sec:SPN_random} shows that one challenge of the SPN with random departure is that $Q_k(t)$ may grow unboundedly (sublinearly) even when the expected flow-conservation law in \eqref{eq:balance} is satisfied. In this work, we prove that the sublinearly growing queues in the example of Section~\ref{sec:SPN_random} are actually the worst possible case that could happen. Namely, for SPNs with random departure, we can always find an algorithm such that all queue lengths grow sublinearly when the input rates are within the optimal stability region.

\par Note that from a throughput perspective, sublinear growth  means that the throughput penalty incurred by the growing queues is negligible since the throughput is the average number of the packet arrivals per second and only the linear terms matter in the long run. Moreover, for any scheme $A$ that achieves sublinearly growing queues, it is likely (without any rigorous proof) that we can convert it to a bounded queue scheme by (i) Run scheme $A$ until any of the sublinearly growing queue length hits some pre-defined threshold; (ii) Stop scheme $A$ and run a naive scheme $B$ that focuses on ``draining'' the queues of the network; (iii) When running scheme $B$, put any new arrival  packets into a separate buffer $Q$; (iv) After scheme $B$ successfully drains out all the queues, we start to run scheme $A$ again and we inject the packets collected in $Q$ gradually back to the system. The above 4 steps guarantee that the queue lengths are bounded. Heuristically, they also approach the optimal throughput since the queues grow sublinearly, the penalty of running the ``draining-stage scheme B" should also be negligible when choosing a sufficiently large threshold in Step~(i). 

From the above reasonings, we believe that sublinearly growing queues are as good as the bounded queues from a practical perspective. The following analysis is based on the concept of sublinearly growing queue lengths.
\begin{definition}
A queue length $q(t)$ {\em grows sublinearly} if for any $\epsilon>0$ and $\delta>0$, there exists $t_0$ such that
\begin{align}
\prob(|q(t)|>\epsilon t)< \delta,~\forall t>t_0.
\end{align}
Since we assume that the input activities $\ve{a}(t)$ have bounded support, an equivalent definition of sublinear growth is: $q(t)$ grows sublinearly if for any $\rho>0$ there exists $t_0$ such that 
\begin{align}
\EE\{|q(t)|\}<\rho  t, ~\forall t>t_0.
\end{align} 
An SPN is {\em sublinearly stable} if all the queues grow sublinearly.
\end{definition}

{\em Remark: }As a result of the above definition, one can observe that the summation of finitely many sublinearly-growing queues is still sublinearly-growing.

The following two propositions characterize the sublinear stability region of any (0,1) random SPN. Proposition~\ref{prop:outer} specifies the outer bound of the stability region, and Proposition~\ref{prop:inner} specifies an inner bound.
\begin{proposition}\label{prop:outer}Consider any (0,1) random SPN. A rate vector $\ve{R}$ can be sublinearly stabilized only if
there exist $\ve{s}_c\in \Lambda$ for all $c\in\bigcq$ such that
\begin{align}
\im\cdot \ve{R}+\sum_{c\in\bigcq}f_c\cdot \overline{\Bout(c)}\cdot \ve{s}_c=\sum_{c\in\bigcq}f_c\cdot \overline{\Bin(c)}\cdot \ve{s}_c.\label{eq:new-balance}
\end{align}
\end{proposition}

Proposition~\ref{prop:outer} can be derived by conventional flow conservation arguments as in \cite{jiang2009stable} and the proof is thus omitted. 

\begin{proposition}\label{prop:inner}
For any SPN that satisfies the three assumptions in Section~\ref{subsec:01SPN} and any rate vector $\ve{R}$, if there exist $\ve{s}_c\in \Lambda^\circ$ for all $c\in\bigcq$ such that \eqref{eq:new-balance} holds, then the proposed scheme $\SCH_\text{avg}$ in Section~\ref{subsec:proposed-DMW} can sublinearly stabilize the SPN with arrival rate $\ve{R}$.
\end{proposition}

{\em Outline of the proof of Proposition~\ref{prop:inner}: }Let each queue $k$ keep another two real-valued counters $\qkinter{k}(t)$ and $\Qkinter{k}(t)$, termed the {\em intermediate virtual queue length} and {\em intermediate actual queue length}. Recall that $q_k(t)$ is the virtual queue length and $Q_k(t)$ is the actual queue length. There are thus 4 different queue length values\footnote{$\qkinter{k}(t)$ and $\Qkinter{k}(t)$ are used only for the proof and are not needed when running the scheduling algorithm.} for each queue $k$. To prove $\ve{Q}(t)$ can be sublinearly stabilized by $\SCH_\text{avg}$, we will show that both $\Qkinter{k}(t)$ and the absolute difference $|Q_k(t)-\Qkinter{k}(t)|$ can be sublinearly stabilized by $\SCH_\text{avg}$ for all $k$. Since the summation of sublinearly-growing random processes is still sublinearly-growing, $\ve{Q}(t)$ can be sublinearly stabilized by $\SCH_\text{avg}$, and we have thus proven Proposition~\ref{prop:inner}.

To that end, we first specify the update rules for $\qkinter{k}(t)$ and $\Qkinter{k}(t)$. Initially, $\qkinter{k}(1)$ and $\Qkinter{k}(1)$ are set to 0 for all $k$. In the end of each time $t$, we compute $\qinter(t+1)$ using the preferred schedule $\ve{x}^*(t)$ chosen by $\SCH_{\text{avg}}$:
\begin{align}
\qinter(t+1) =& \qinter(t)+\im\cdot \ve{a}(t)\nonumber \\
&+ \left(\Bout(\cq(t))-\Bin(\cq(t))\right)\cdot \ve{x}^*(t). \label{eq:DMW-update-inter}
\end{align}
If we compare \eqref{eq:DMW-update-inter} with the computation of $\ve{q}(t)$ in \eqref{eq:DMW-update-new}, $\ve{q}^\text{inter}(t)$ is updated based on the {\em realization} of the input and output service matrices while $\ve{q}(t)$ is updated based on the {\em expected} input and output service matrices. Equivalently, we can rewrite \eqref{eq:DMW-update-inter} as
\begin{align}
\qkinter{k}(t+1)=\qkinter{k}(t)-\mu_{\text{out},k}(t)+\mu_{\text{in},k}(t),~\forall k, \label{eq:qk-inter-update}
\end{align}
where
\begin{align}
&\mu_{\text{out},k}(t) = \sum_{n=1}^N \left(\bine{k}{n}(\cq(t)) \cdot x_n^*(t) \right), \label{eq:mu_out_def} \\
&\mu_{\text{in},k}(t) = \sum_{m=1}^M\left( \alpha_{k,m}\cdot a_m(t) \right)+\sum_{n=1}^N \left(\bout{k}{n}(\cq(t))\cdot x_n^*(t)\right). \label{eq:mu_in_def}
\end{align}
Here, $\mu_{\text{out},k}$ is the amount of packets coming ``out of queue $k$", which is decided by the ``input rates of SA $n$". Similarly, 
$\mu_{\text{in},k}$ is the amount of packets ``entering queue $k$", which is decided by the ``output rates of SA $n$". We also update $\Qinter(t+1)$ by
\begin{align}
\Qkinter{k}(t+1)=\left(\Qkinter{k}(t)-\mu_{\text{out},k}(t)\right)^+ + \mu_{\text{in},k}(t),~\forall k, \label{eq:Qk-inter-update}
\end{align}
where $(v)^+ = \max\{0,v\}$.

The difference between $\qkinter{k}(t)$ and $\Qkinter{k}(t)$ is that the former can be still be strictly negative when updated via \eqref{eq:qk-inter-update} while we enforce the latter to be non-negative. 

\par To compare $\Qkinter{k}(t)$ and $Q_k(t)$, we observe that  by \eqref{eq:Qk-inter-update}, $\Qkinter{k}(t)$ is purely updated by the preferred service vector $\ve{x}^*(t)$ without considering whether the preferred SA $n^*$ is feasible or not (see Difference 1 in Section~\ref{subsec:01SPN}). That is, in the case that SA $n^*$ is infeasible, then SA $n^*$ cannot be carried out successfully. Therefore, the system remains idle and the actual queue length $Q_k(t+1)=Q_k(t)$ for all $k=1$ to $K$ or $Q_k(t)$ increases if there is external arrival at queue $k$. In contrast, even though SA $n^*$ cannot be carried out successfully, we still update $\Qkinter{k}(t+1)$ by \eqref{eq:mu_out_def} to \eqref{eq:Qk-inter-update} for all queue $k$. As a result, the $\Qkinter{k}(t)$ values will still change\footnote{In the original DMW algorithm for deterministic SPNs \cite{jiang2009stable}, the quantity ``actual queue length" is updated by \eqref{eq:Qk-inter-update}. The ``actual queue lengths in \cite{jiang2009stable}" thus refer to a conceptual register value $\Qkinter{k}(t)$ rather than the number of physical packets in the buffer/queue. In this work, we rectify this inconsistency by renaming ``the actual queue lengths in \cite{jiang2009stable}" the ``intermediate actual queue lengths $\Qkinter{k}(t)$."} for those $k\in \SAin{n}\cup \SAout{n}$.
 
To evaluate the absolute difference $|Q_k(t)-\Qkinter{k}(t)|$, for any time $t$ and any queue $k$, we first define an event, which is called the {\em null activity} of queue $k$ at time $t$. Since we assume at any time $t$, only one SA can be scheduled, we use $n(t)$ to denote the preferred SA suggested by the back-pressure scheduler in \eqref{eq:DMW-schedule} and \eqref{eq:new-backpressure}. As a result, at time $t$, we say the null activity occurs at queue $k$ if (i) $k\in\SAin{n(t)}$ and (ii) $\Qkinter{k}(t)<\bine{k}{n}(\cq(t))$. That is, the null activity describes the event that the preferred SA shall consume the packets in queue $k$ (since $k\in \SAin{n(t)}$) but the intermediate actual queue length $\Qkinter{k}(t)$ is less than the realization $\bine{k}{n}(\cq(t))$. Note that the null activity is defined based on the intermediate actual queue length $\Qkinter{k}(t)$ and does not distinguish whether the actual queue length $Q_k(t)$ is larger or less than 1. Therefore the null activities are not directly related to the event that SA $n$ is infeasible.

\par Let $N_{\mathsf{NA},k}(t)$ be the aggregate number of null activities occurred at queue $k$ up to time $t$. Then we can write $N_{\mathsf{NA},k}(t)$ as
\begin{align}
&\Nna{k}(t)\eqdef \nonumber \\
&\sum_{\tau=1}^{t} I(k\in\SAin{n(\tau)}) \cdot  I(\Qkinter{k}(\tau)<\bine{k}{n(\tau)}(\cq(\tau))), \nonumber
\end{align}
where $I(\cdot)$ is the indicator function.

\par The following lemma upper bounds the difference of $Q_k(t)$ and $\Qkinter{k}(t)$ by the aggregate numbers of null activities.

\begin{lemma}\label{lemma:bound-Qk}
For all $k=1,2,...,K$, there exist $K$ non-negative coefficients $\gamma_{1},...,\gamma_{K}$ such that
\begin{align}
&\EE(|Q_k(t)-\Qkinter{k}(t)|) \leq \sum_{\tilde{k}=1}^K \gamma_{\tilde{k}} \Nna{\tilde{k}}(t). \label{eq:bound-Qk}
\end{align}
for all $t=1$ to $\infty$.
\end{lemma}

The proof of Lemma~\ref{lemma:bound-Qk} is relegated to Appendix~\ref{app:bound-Qk}. In Appendix~\ref{app:sub_growth}, we prove that both $\Qkinter{k}(t)$ and $\Nna{k}(t)$ of (0,1) random SPN can be sublinearly stabilized by $\SCH_\text{avg}$ for all $k$.\footnote{The DMW algorithm for SPNs were first introduced in \cite{jiang2009stable}. However, in that paper, the authors rename the intermediate actual queue lengths defined in \eqref{eq:Qk-inter-update} of this paper as the actual queue length and prove that $\Qkinter{k}(t)$ can be stabilized for deterministic SPNs. However, proving $\Qkinter{k}(t)$ can be stabilized does not necessarily mean that $Q_k(t)$ can be stabilized, as discussed in the paragraphs after \eqref{eq:Qk-inter-update}. The critical part of proving $|Q_k(t)-\Qkinter{k}(t)|$ is stabilized is unfortunately missing in \cite{jiang2009stable}. One contribution of this work is to provide in Lemma~\ref{lemma:bound-Qk} the first rigorous proof showing that $|Q_k(t)-\Qkinter{k}(t)|$ can be stabilized as well.} Therefore, by Lemma~\ref{lemma:bound-Qk},  $\Qkinter{k}(t)$ and $|Q_k(t)-\Qkinter{k}(t)|$ can be sublinearly stabilized and so can $Q_k(t)$. Proposition~\ref{prop:inner} is thus proven.

\section{The Combined Solution} \label{sec:combined}

We are now ready to combine the discussions in Sections~\ref{sec:new-coding} and \ref{sec:new_scheduling}. As discussed in Section~\ref{sec:new-coding}, the 7 operations form a vr-network as described in Fig.~\ref{fig:7-type} and both the source and the two receivers perform encoding and decoding according to the packet movements in the vr-network, respectively. Specifically, there are $K=5$ queues, $M=2$ IAs, and $N=7$ SAs. The 5-by-2 input matrix $\mathcal{A}$ contains 2 ones, since the packets arrive at either $Q^1_\emptyset$ or $Q^2_\emptyset$. Given the channel quality $\cq(t)=c$, the expected input and output service matrices $\avg{\Bin(c)}$ and $\avg{\Bout(c)}$ can be derived from Table~\ref{tab:transition}. 

\par We use the following concrete example to illustrate our procedure. Suppose that the channel quality $\cq(t)$ is Bernoulli with parameter $1/2$  (i.e., flipping a perfect coin). Also suppose that when $\cq(t)=0$, with probability $0.5$ (resp.\ $0.7$) destination $d_1$ (resp.\ destination $d_2$) can successfully receive a packet transmitted by source $s$; and when $\cq(t)=1$, with probability $2/3$ (resp.\ $1/3$) destination $d_1$ (resp.\ destination $d_2$) can successfully receive a packet transmitted by source $s$. Further assume that all the success events of $d_1$ and $d_2$ are independent. Please also see Appendix~\ref{app:combine-capacity} for further details on the matrix construction. If we order the 5 queues as $\left[\ve{Q}_\emptyset^1,\ve{Q}_\emptyset^2,\ve{Q}_{\{2\}}^1,\ve{Q}_{\{1\}}^2,\ve{Q}_\text{mix}\right]$, the 7 service activities as $\left[\text{NC1},\text{NC2},\text{DX1},\text{DX2},\text{PM},\text{RC},\text{CX}\right]$, then the matrices of the SPN become
\begin{align}
&\im = \left[
           \begin{array}{ccccc}
             1 & 0 & 0 & 0 & 0 \\
             0 & 1 & 0 & 0 & 0 \\
           \end{array}
         \right]^\tran, \nonumber \\
&\overline{\Bin(0)} =\left[
        \begin{array}{ccccccc}
          0.85 & 0 & 0 & 0 & 0.85 & 0 & 0 \\
          0 & 0.85 & 0 & 0 & 0.85 & 0 & 0 \\
          0 & 0 & 0.5 & 0 & 0 & 0 & 0.5 \\
          0 & 0 & 0 & 0.7 & 0 & 0 & 0.7 \\
          0 & 0 & 0 & 0 &  0 &  0.85 & 0 \\
        \end{array}
      \right], \nonumber \\
&\overline{\Bin(1)} =\left[
         \begin{array}{ccccccc}
           7/9 & 0 & 0 & 0 & 7/9 & 0 & 0 \\
           0 & 7/9 & 0 & 0 & 7/9 & 0 & 0 \\
           0 & 0 & 2/3 & 0 & 0 & 0 & 2/3 \\
           0 & 0 & 0 & 1/3 & 0 & 0 & 1/3 \\
           0 & 0 & 0 & 0 &  0 &  7/9 & 0 \\
         \end{array}
       \right], \nonumber \\
&\overline{\Bout(0)} = \left[
        \begin{array}{ccccccc}
          0 & 0 & 0 & 0 & 0 & 0 & 0 \\
          0 & 0 & 0 & 0 & 0 & 0 & 0 \\
          0.35 & 0 & 0 & 0 & 0 & 0.35 & 0 \\
          0 & 0.15 & 0 & 0 & 0 & 0.15 & 0 \\
          0 & 0 & 0 & 0 &  0.85 &  0 & 0 \\
        \end{array}
      \right], \nonumber \\      
&\overline{\Bout(1)} = \left[
        \begin{array}{ccccccc}
          0 & 0 & 0 & 0 & 0 & 0 & 0 \\
          0 & 0 & 0 & 0 & 0 & 0 & 0 \\
          1/9 & 0 & 0 & 0 & 0 & 1/9 & 0 \\
          0 & 4/9 & 0 & 0 & 0 & 4/9 & 0 \\
          0 & 0 & 0 & 0 &  7/9 &  0 & 0 \\
        \end{array}
      \right], \nonumber \\  
&\ve{s}_c = \left[ \xncone~\xnctwo~\xdxone~\xdxtwo~\xpm~\xrc~\xcx \right]^\tran. \nonumber
\end{align}
For example, the seventh column of $\overline{\Bin(0)}$ indicates that when $\cq(t)=0$ and the {\sc Classic-XOR} is activated, with probability 0.5 (resp.\  0.7) 1 packet will be consumed from queue $\ve{Q}_{\{2\}}^1$ (resp.\ $\ve{Q}_{\{1\}}^2$). The third row of $\overline{\Bout(1)}$ indicates that when $\cq(t)=1$,  queue $\ve{Q}_{\{2\}}^1$ will increase by 1 with probability 1/9 (resp.\ 1/9) if coding choice {\sc Non-Coding-1} (resp.\ {\sc Reactive-Coding}) is activated since it corresponds to the event that $d_1$ receives the transmitted packet but $d_2$ does not.

\par Since there are 7 coding operations (SAs), each vector in $\mathfrak{X}$ is a 7-dimensional binary vector. Since we are allowed to choose any one of the 7 operations or choose to transmit nothing, 7 of the 8 vectors are the Dirac delta vectors and the rest is an all-zero vector. We can now use the proposed DMW scheduler in \eqref{eq:DMW-schedule}, \eqref{eq:new-backpressure}, and \eqref{eq:DMW-update-new} to compute the preferred scheduling decision. We activate the preferred decision if it is feasible. If not, then the system remains idle.

\par For general channel parameters (including but not limited to this simple example), after computing the $\overline{\Bin(c)}$ and $\overline{\Bout(c)}$ of the vr-network in Fig.~\ref{fig:7-type} with the help of Table~\ref{tab:transition}, we can explicitly compare the sublinear stability region in Propositions~\ref{prop:outer} and \ref{prop:inner} with the Shannon capacity region in \cite{WangHan14}. In the end, we have the following proposition.

\begin{proposition}\label{prop:combine-capacity} The sublinear stability region of the proposed INC-plus-SPN-scheduling scheme matches the block-code capacity of time-varying channels.
\end{proposition}
\noindent
The detailed proof of Proposition~\ref{prop:combine-capacity} is provided in Appendix~\ref{app:combine-capacity}.

{\em Remark:} During numerical simulations, we notice that we 
can further revise the proposed scheme to reduce the actual queue lengths $Q_k(t)$ by $\approx 50\%$ even though we do not have any rigorous proofs/performance guarantees for the revised scheme. That is, when making the scheduling decision by \eqref{eq:DMW-schedule}, we can compute $\ve{d}(t)$ by
\begin{align}
\ve{d}(t)=\left(\overline{\Bin(\cq(t))}-\overline{\Bout(\cq(t))}\right)^\tran \ve{q}^\text{inter}(t).\label{eq:inter-backpressure}
\end{align}
where $\ve{q}^\text{inter}(t)$ is the intermediate virtual queue length defined in \eqref{eq:qk-inter-update} of Section~\ref{subsec:analysis}. The intuition behind is that the new back-pressure in \eqref{eq:inter-backpressure} allows the scheme to directly control $q_k^\text{inter}(t)$, which, when compared to the virtual queue $\ve{q}(t)$ in \eqref{eq:DMW-update-new}, is more closely related to the actual queue length\footnote{There are four types of queue lengths in this work: $\ve{q}(t)$, $\qinter(t)$, $\Qinter(t)$, and $\ve{Q}(t)$ and they range from the most artificially-derived $\ve{q}(t)$ to the most realistic metric, the actual queue length $\ve{Q}(t)$.} $Q_k(t)$.

\subsection{Extensions For Rate Adaption} \label{subsec:extention_ACM}
We close this section by noting that the proposed solution can be naturally extended to the case of rate adaptation, which is also known as adaptive coding and modulation. For illustration purposes, we consider the following simple example of adaptive coding and modulation scheme.

\par Consider 2 possible error correcting rates (1/2 and 3/4); 2 possible modulation schemes QPSK and 16QAM; and jointly there are 4 possible combinations. The lowest throughput combination is rate-1/2 plus QPSK and the highest throughput combination is rate-3/4 plus 16QAM. Assuming the packet size is fixed. If the highest throughput combination takes 1-unit time to finish sending 1 packet, then the lowest throughput combination will take 3-unit time. For these 4 possible (rate,modulation) combinations, we denote the unit-time to finish transmitting 1 packet as $T_1$ to $T_4$, respectively. 

\par For the $i$-th (rate,modulation) combination, $i=1$ to $4$, source $s$ can measure the probability that $d_1$ and/or $d_2$ successfully hears the transmission, and denote the corresponding probability vector by $\vec{p}^{(i)}$. Source $s$ then uses $\vec{p}^{(i)}$ to compute the  $\overline{\Bini(\cq(t))}$ and $\overline{\Bouti(\cq(t))}$ for the vr network. Then it computes the backpressure by
\begin{align}
\ve{d}^{(i)}(t)=\left(\overline{\Bini(\cq(t))}-\overline{\Bouti(\cq(t))} \right)^\tran \ve{q}(t). \nonumber
\end{align}
We can now compute the preferred scheduling choice by 
\begin{align}
\argmax_{i\in\{1,2,3,4\},\ve{x}\in\mathfrak{X}} \frac{\ve{d}^{(i)}(t)^\tran \cdot \ve{x} }{T_i}\label{eq:new-sch-ACM}
\end{align} 
and update the virtual queue length $\ve{q}(t)$ by \eqref{eq:DMW-update-new}. Namely, the backpressure $\ve{d}^{(i)}(t)^\tran \cdot \ve{x}$ is scaled inverse proportionally with respect to $T_i$, the time it takes to finish the transmission of 1 packet. If the preferred SA $n^*$ is feasible, then we use the $i$-th (rate,modulation) combination plus the coding choice $n^*$ for the current transmission. If the preferred SA $n^*$ is infeasible, then we either choose another (rate,modulation) combination plus coding choice arbitrarily or simply let the system idle. 

\par One can see that the new scheduler \eqref{eq:new-sch-ACM} automatically balances the packet reception status (the $\ve{q}(t)$ terms), the success overhearing probability of different (rate,modulation) (the $\overline{\Bini(\cq(t))}$ and $\overline{\Bouti(\cq(t))}$ terms), and different amount of time it takes to finish transmission of a coded/uncoded packet  (the $T_i$ term). In all the numerical experiments of Section~\ref{sec:simulation}, the new scheduler \eqref{eq:new-sch-ACM} robustly achieves the optimal throughput with adaptive coding and modulation. 

\section{Simulation Results\label{sec:simulation}}

\begin{figure}
\centering
  \includegraphics[width=9cm]{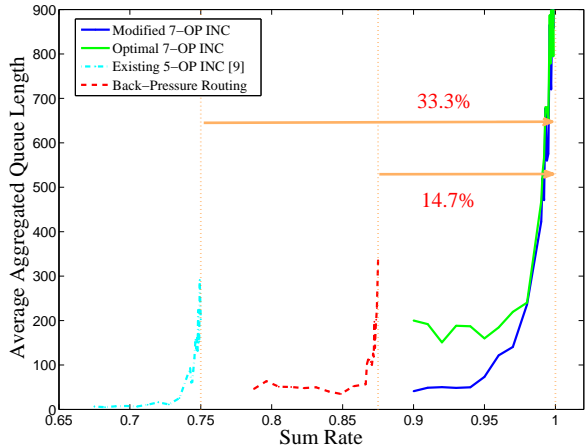}\\
  \caption{The backlog of four different schemes for a time-varying channel with $\cq(t)$ uniformly distributed on $\{1,2\}$, and the packet delivery probability being $\vec{p}=(0,0.5,0.5,0)$ if $\cq(t)=1$ and $\vec{p}=(0,0,0,1)$ if $\cq(t)=2$.}\label{fig:7vs5}
\end{figure}

\begin{figure}
\centering
\subfigure[$(f_1,f_2,f_3,f_4)=(0.15,0.15,0.35,0.35)$. \label{fig:random}]{
  \includegraphics[width=9cm]{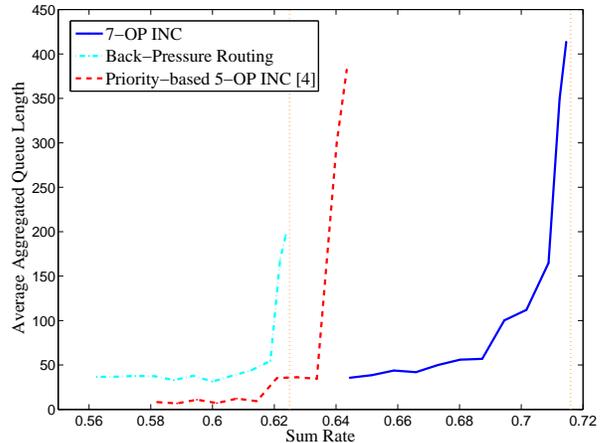}}
\subfigure[$(f_1,f_2,f_3,f_4)=(0.25,0.25,0.25,0.25)$. \label{fig:round}]{
  \includegraphics[width=9cm]{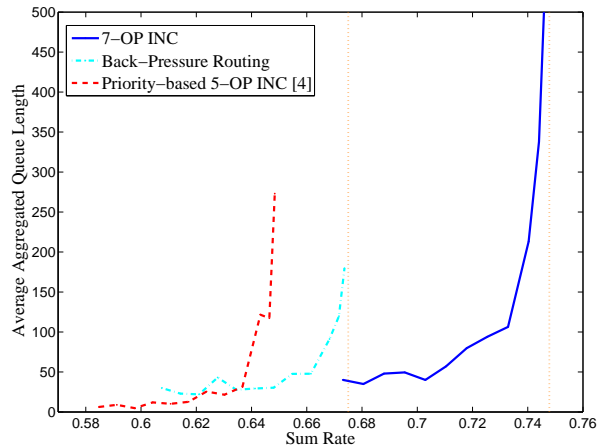}}  \caption{\label{fig:overall}The backlog comparison with $\cq(t)$ chosen from $\{1,2,3,4\}$ and $\vec{p}^{(1)}=(0.14,0.06,0.56,0.24)$, $\vec{p}^{(2)}=(0.14,0.56,0.06,0.24)$, $\vec{p}^{(3)}=(0.04,0.16,0.16,0.64)$, and $\vec{p}^{(4)}=(0.49,0.21,0.21,0.09)$.}
\end{figure}

In this section, we simulate the proposed optimal 7-operation INC + scheduling solution and compare the results  with the existing INC solutions and the (back-pressure) pure-routing solutions.

In Fig.~\ref{fig:7vs5}, we simulate a simple time-varying channel situation first described in Section~\ref{subsubsec:degenerate-XOR}. Specifically, the channel quality $\cq(t)$ is i.i.d.\ distributed and for any $t$,  $\cq(t)$ is uniformly distributed on $\{1,2\}$. When $\cq(t)=1$, the success probabilities are $\vec{p}^{(1)}=(0,0.5,0.5,0)$ and when $\cq(t)=2$, the success probabilities are $\vec{p}^{(2)}=(0,0,0,1)$, respectively. We consider four different schemes: (i) Back-pressure (BP) + pure routing; (ii) BP + INC with 5 operations \cite{georgiadis2006resource,AthanasiadouGeorgiadis13}; (iii) The proposed DMW+INC with 7 operations, and (iv) The modified DMW+INC with 7 operations that use $\qkinter{k}(t)$ to compute the back pressure, see \eqref{eq:inter-backpressure}, instead of $q_k(t)$ in \eqref{eq:new-backpressure}.

\par We choose perfectly fair $(R_1,R_2)=(\theta, \theta)$ and gradually increase the $\theta$ value and plot the stability region. For each experiment, i.e., each $\theta$, we run the schemes for $10^5$ time slots. The horizontal axis is the sum rate $R_1+R_2=2\theta$ and the vertical axis is the aggregate backlog  (averaged over 10 trials) in the end of $10^5$ slots. By the results in \cite{WangHan14}, the sum rate Shannon capacity is 1 packet/slot, the best possible rate for 5-OP INC is 0.875 packet/slot, and the best pure routing rate is 0.75 packet/slot, which are plotted as vertical lines in Fig.~\ref{fig:7vs5}.  The simulation results confirm our analysis. The proposed 7-operation dynamic INC has a stability region matching the Shannon block code capacity and provides $14.7\%$ throughput improvement over the 5-operation INC, and $33.3\%$ over the pure-routing solution.

\par Also, both our original proposed solution (using $q_k(t)$) and the modified solution (using $\qkinter{k}(t)$) can approach the stability region while the modified solution has smaller backlog. This phenomenon is observed throughout all our experiments. As a result, in the following experiments, we only report the results of the modified solution using $\qkinter{k}(t)$ to compute the backpressure.

\par Next we simulate the scenario of 4 different channel qualities: $\bigcq=\{1,2,3,4\}$. The varying channel qualities could model the situations like the different packet transmission rates and loss rates due to time-varying interference caused by the primary traffic in a cognitive radio environment. We assume four possible channel qualities with the corresponding probability distributions being $\vec{p}^{(1)}=(p^{(1)}_{\overline{d_1d_2}}, p^{(1)}_{d_1\overline{d_2}},p^{(1)}_{\overline{d_1}d_2},p^{(1)}_{d_1d_2})=(0.14,0.06,0.56,0.24)$, $\vec{p}^{(2)}=(0.14,0.56,0.06,0.24)$, $\vec{p}^{(3)}=(0.04,0.16,0.16,0.64)$, and $\vec{p}^{(4)}=(0.49,0.21,0.21,0.09)$ in both Figs.~\ref{fig:random} and \ref{fig:round}. The difference is that in Fig.~\ref{fig:random}, the channel quality $\cq(t)$ is i.i.d.\ distributed with probability (frequency) $(f_1,f_2,f_3,f_4)$ being $(0.15, 0.15, 0.35, 0.35)$. In Fig.~\ref{fig:round} the $\cq(t)$ is again i.i.d.\ but with different frequency $(f_1,f_2,f_3,f_4)= (0.25, 0.25, 0.25, 0.25)$. Again, we assume perfect fairness $(R_1,R_2)=(\theta,\theta)$ and gradually increase the $\theta$ value. The sum-rate Shannon block-code capacity is $R_1+R_2=0.716$ when $(f_1,f_2,f_3,f_4)=(0.15, 0.15, 0.35, 0.35)$ and $R_1+R_2=0.7478$ when $(f_1,f_2,f_3,f_4)= (0.25, 0.25, 0.25, 0.25)$, and the pure routing sum-rate capacity is $R_1+R_2=0.625$ when $(f_1,f_2,f_3,f_4)=(0.15, 0.15, 0.35, 0.35)$ and $R_1+R_2=0.675$ when $(f_1,f_2,f_3,f_4)= (0.25, 0.25, 0.25, 0.25)$. We simulate our modified 7-OP INC, the priority-based solution in \cite{GeorgiadisTassiulas13}, and a standard back-pressure routing scheme \cite{tassiulas1992stability}. Each point of the curves is the average of 10 trials and each trial lasts for $10^5$ slots. 

\par Although the priority-based scheduling solution is provably optimal for fixed channel quality, it is less robust and can sometimes be substantially suboptimal (see Fig.~\ref{fig:round}) due to the ad-hoc nature of the priority-based policy. For example, as depicted by Figs.~\ref{fig:random} and \ref{fig:round}, the pure-routing solution outperforms the 5-operation scheme for one set of frequency $(f_1,f_2,f_3,f_4)$ while the order is reversed for another set of frequency. On the other hand, the proposed 7-operation scheme consistently outperforms all the existing solutions and has a stabiliby region matching the Shannon block-code capacity. We have tried many other combinations of time-varying channels. In all our simulations, the proposed DMW scheme always achieves the block-code capacity in \cite{WangHan14} and outperforms routing and any existing solutions \cite{AthanasiadouGeorgiadis13,GeorgiadisTassiulas13}.

\begin{figure}
\centering
  \includegraphics[width=9cm]{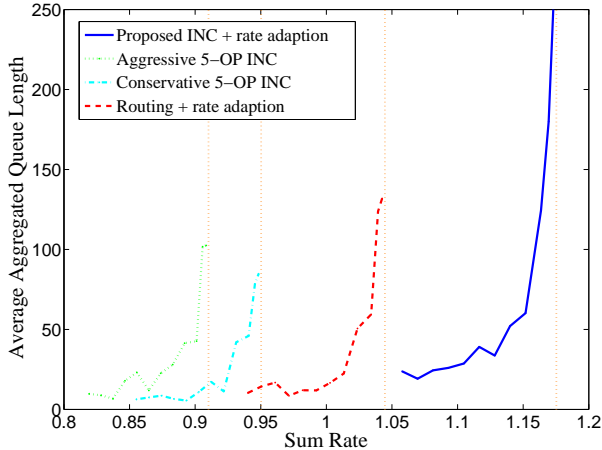}\\
  \caption{The backlog of four different schemes for rate adaptation with two possible (error-correcting-code rate,modulation) combinations. The back-pressure-based INC scheme in \cite{AthanasiadouGeorgiadis13} is used in both aggressive and conservative 5-OP INC, where the former always chooses the high-throughput (rate,modulation) combination while the latter always chooses the low-throughput (rate,modulation) combination.}\label{fig:acm}
\end{figure}

\par Our solution in Section~\ref{subsec:extention_ACM} is the first dynamic INC design that is guaranteed to achieve the optimal linear INC capacity with rate-adaptation (adaptive coding and modulation) \cite{WangHan14}. Fig.~\ref{fig:acm} compares its performance with existing routing-based rate-adaptation scheme and the existing INC schemes, the latter of which are designed without rate adaptation. We assume there are two available (error-correcting-code rate,modulation) combinations to be selected. We assume that the first combination takes 1 second to finish transmitting a single packet and the second combination takes 1/3 second to finish a single packet. That is, the transmission rate of the second combination is 3 times faster than the first combination.

\par We further assume the packet delivery probability is $\vec{p}=(0.1\cdot 0.05,0.95\cdot 0.1,0.05\cdot 0.9,0.95\cdot 0.9)$ if the first combination is selected and $\vec{p}=(0.6\cdot 0.8,0.8\cdot 0.4,0.2\cdot 0.6,0.2\cdot 0.4)$ if the second combination is selected. That is, the low-throughput combination is likely to be overheard by both destinations and the high-throughput combination has a much lower success transmission probability. We can compute the corresponding Shannon block-code capacity region by modifying the equations in \cite{WangHan14}. We then use the proportional fairness objective function $\xi(R_1,R_2)=\log(R_1)+\log(R_2)$ and find the maximizing  $R_1^*$ and $R_2^*$ over the Shannon capacity region, which are $R_1^*= 0.6508$  packets per second and $R_2^*=0.5245$ packets per second in this example.

After computing the optimal block code capacity, we assume the following dynamic packet arrivals. We define $(R_1,R_2)=\theta\cdot (R_1^*,R_2^*)$ for any given $\theta\in(0,1)$. For any experiment (i.e., for any given $\theta$), the arrivals of session-$i$ packets is a Poisson random process with rate $R_i$ packets per second for $i=1,2$. That is, if the low-throughput combination 1 is selected to transmit 1 packet, then during the 1 second it takes to finish, the number of arrivals of session-$i$ packets is a Possion random variable with mean $R_i\cdot 1$ packets. Similarly, if the high-throughput combination is selected to transmit 1 packet, then during the 1/3 second it takes to finish transmission, the number of arrivals of session-$i$ packets is a Possion random variable with mean $R_i\cdot 1/3$ packets.

\par Each point of the curves of Fig.~\ref{fig:acm} consists of 10 trials and each trial lasts for $10^5$ seconds. We compare the performance of our scheme in Section~\ref{subsec:extention_ACM} with (i) Pure-routing with rate-adaptation; (ii) aggressive 5-OP INC, i.e., use the scheme in \cite{AthanasiadouGeorgiadis13} and always choose combination 2; and (iii) conservative 5-OP INC, i.e., use the scheme in \cite{AthanasiadouGeorgiadis13} and always choose combination 1. We also plot the optimal routing-based rate-adaptation rate and the optimal Shannon-block-code capacity rate as vertical lines.

\par We can observe that since our proposed scheme jointly decides which (rate,modulation) combination to use and which INC operation to encode the packet in an optimal way, see \eqref{eq:new-sch-ACM}, the stability region of our scheme matches the block-code Shannon capacity with rate-adaptation. It provides $12.51\%$ throughput improvement over the pure routing-based rate-adaptation solution (which is represented by the red dash line in Fig.~\ref{fig:acm}).

\par Furthermore, we observe that if we perform INC but always choose the low-throughput (rate,modulation), as suggested in some existing works \cite{RayanchuSenWuBanerjeeSengupta08}, then the largest sum-rate $R_1+R_2=\theta_\text{cnsv.\ 5-OP}^*(R_1^*+R_2^*)=0.9503$, which is worse than pure routing with rate-adaptation $\theta_\text{routing,RA}^*(R_1^*+R_2^*)=1.0446$. Even if we always choose the high-throughput (rate,modulation) with 5-OP INC, then the largest sum-rate $R_1+R_2=\theta_\text{aggr.\ 5-OP}^*(R_1^*+R_2^*)=0.9102$ is even worse than the conservative 5-OP INC capacity. We have tried many other rate-adaptation scenarios. In all our simulations, the proposed DMW scheme always achieves the block-code capacity and outperforms pure-routing, conservative 5-OP INC, and aggressive 5-OP INC.

\par It is worth emphasizing that in our simulation, for any fixed (rate,modulation) combination, the channel quality is also fixed. Therefore since 5-OP scheme is throughput optimal for fixed channel quality \cite{GeorgiadisTassiulas09}, it is guaranteed that the 5-OP scheme is throughput optimal when using a fixed (rate,modulation) combination. Our results thus show that using a fixed (rate,modulation) combination is the main reason of the suboptimal performance and the proposed scheme in \eqref{eq:DMW-schedule}, \eqref{eq:DMW-update-new}, and \eqref{eq:new-sch-ACM} can dynamically decide which (rate,modulation) combination to use for each transmission and achieve the largest possible stability region.

\section{Conclusion\label{sec:conclusion}}

We have proposed a new 7-operation INC scheme together with the corresponding scheduling algorithm to achieve the optimal downlink throughput of the 2-flow access point network with time varying channels. Based on binary XOR operations, the proposed solution  admits ultra-low encoding/decoding complexity with efficient buffer management and minimal communication and control overhead. The proposed algorithm has also been generalized for rate adaptation and it again robustly achieves the optimal throughput in all the numerical experiments. The proposed algorithm has also been generalized for rate adaptation and it again robustly achieves the optimal throughput in all the numerical experiments. A byproduct of this paper is a throughput-optimal scheduling solution for SPNs with random departure, which could further broaden the applications of SPNs to other real-world applications.

\appendix

\subsection{Proof of Lemma~\ref{lemma:bound-Qk}} \label{app:bound-Qk}
Recall that we assume the SPN under consideration is acyclic, and hence we could arrange the queues from the upstream to the downstream and index them from $1$ (the most upstream) to $K$ (the most dowsntream). Recall that we use $n(t)$ to denote the preferred SA chosen by the back-pressure scheduler. The proof of Lemma~\ref{lemma:bound-Qk} consists of proving the following lemmas.

We first prove the following lemma.

\begin{lemma} \label{lemma:bound-Qk-Qkinter}
For any $k=1,2,...,K$ and any time $t$, 
\begin{align}
&|Q_k(t)-\Qkinter{k}(t)| \nonumber \\
&\leq \sum_{\tau=1}^{t-1}I(\exists k'\in\SAin{n(\tau)}: Q_{k'}(\tau)=0). \label{eq:bound-Qk-Qkinter}
\end{align}
\end{lemma}

\begin{lemma} \label{lemma:divide_to_4_cases}
For any $k=1,2,...,K$ and any time $\tau$,
\begin{align}
& I(k\in\SAin{n(\tau)})\nonumber \\
\leq& I(k\in\SAin{n(\tau)})I(\Qkinter{k}(\tau)<\bine{k}{n(\tau)}(\cq(\tau))) \nonumber\\
&+ I(k\in\SAin{n(\tau)})I(Q_{k}(\tau)=0)\nonumber \\
&\cdot I(\bine{k}{n(\tau)}(\cq(\tau))=0) I(1\leq \Qkinter{k}(\tau))\nonumber \\
&+ I(k\in\SAin{n(\tau)})I(Q_{k}(\tau)=0)\nonumber \\
&\cdot I(\bine{k}{n(\tau)}(\cq(\tau))=0) I(0\leq \Qkinter{k}(\tau)<1)\nonumber \\
&+ I(k\in\SAin{n(\tau)})I(Q_{k}(\tau)=0)\nonumber \\
&\cdot I(\bine{k}{n(\tau)}(\cq(\tau))=1) I(1\leq \Qkinter{k}(\tau)). \label{eq:divide_to_4_cases}
\end{align}
\end{lemma}

By combining Lemma~\ref{lemma:bound-Qk-Qkinter} and Lemma~\ref{lemma:divide_to_4_cases} and by the union bound, we can upper bound the value of $|Q_k(t)-\Qkinter{k}(t)|$ as follows. 
\begin{align}
&|Q_k(t)-\Qkinter{k}(t)| \nonumber \\
\leq&\sum_{k'=1}^K\sum_{\tau=1}^{t-1}I(k'\in\SAin{n(\tau)})I(\Qkinter{k'}(\tau)<\bine{k'}{n(\tau)}(\cq(\tau))) \nonumber\\
&+\sum_{k'=1}^K\sum_{\tau=1}^{t-1}I(k'\in\SAin{n(\tau)})I(Q_{k'}(\tau)=0)\nonumber \\
&\cdot I(\bine{k'}{n(\tau)}(\cq(\tau))=0) I(1\leq \Qkinter{k'}(\tau))\nonumber \\
&+\sum_{k'=1}^K\sum_{\tau=1}^{t-1}I(k'\in\SAin{n(\tau)})I(Q_{k'}(\tau)=0)\nonumber \\
&\cdot I(\bine{k'}{n(\tau)}(\cq(\tau))=0) I(0\leq \Qkinter{k'}(\tau)<1)\nonumber \\
&+\sum_{k'=1}^K\sum_{\tau=1}^{t-1}I(k'\in\SAin{n(\tau)})I(Q_{k'}(\tau)=0)\nonumber \\
&\cdot I(\bine{k'}{n(\tau)}(\cq(\tau))=1) I(1\leq \Qkinter{k'}(\tau)). \label{eq:bound-Qk-Qkinter-final}
\end{align}

Recall that the goal of Lemma~\ref{lemma:bound-Qk} is to upper bound the expectation of $|Q_k(t)-\Qkinter{k}(t)|$ as a weighted sum of $\EE\{\Nna{k'}(t)\}$. We then observe that the expectation of the first term of the RHS of \eqref{eq:bound-Qk-Qkinter-final} is indeed the sum of $\EE\{\Nna{k'}(t)\}$. Therefore, to complete the proof of Lemma~\ref{lemma:bound-Qk}, we only need to upper bound the expectation of the second to the fourth terms of the RHS of \eqref{eq:bound-Qk-Qkinter-final} by some weighted sum of $\EE\{\Nna{k'}(t)\}$. The following Lemmas~\ref{lemma:new_lemma_6} to \ref{lemma:new_lemma_8} upper bound the expectation of the second to the fourth terms, respectively.

\begin{lemma} \label{lemma:new_lemma_6}
For any $k=1,..., K$, there exists a constant $\gamma_k$ such that
\begin{align}
&\EE (\sum_{\tau=1}^{t}I(k\in\SAin{n(\tau)})I(Q_{k}(\tau)=0)\nonumber \\
& \cdot I(\bine{k}{n(\tau)}(\cq(\tau))=0) I(1\leq \Qkinter{k}(\tau)))\nonumber \\
&\leq \gamma_k \EE(\sum_{\tau=1}^{t}I(k\in\SAin{n(\tau)})I(Q_{k}(\tau)=0)\nonumber \\
&\cdot I(\bine{k}{n(\tau)}(\cq(\tau))=1) I(1\leq \Qkinter{k}(\tau)) \label{eq:new_lemma_6}
\end{align}
for all $t=1$ to $\infty$. 
\end{lemma}

Namely, the expectation of the second term of the RHS of \eqref{eq:bound-Qk-Qkinter-final} is upper bounded by $\gamma_k$ times the expectation of the fourth term of the RHS of \eqref{eq:bound-Qk-Qkinter-final}.

\begin{lemma} \label{lemma:new_lemma_7}
For any $k=1,..., K,$ there exists a constant $\gamma_k$ such that
\begin{align}
&\EE (\sum_{\tau=1}^{t}I(k\in\SAin{n(\tau)})I(Q_{k}(\tau)=0)\nonumber \\
&\cdot I(\bine{k}{n(\tau)}(\cq(\tau))=0) I(0\leq \Qkinter{k}(\tau)<1))\nonumber \\
&\leq \gamma_k \EE(\sum_{\tau=1}^{t}I(k\in\SAin{n(\tau)})I(\Qkinter{k}(\tau)<\bine{k}{n(\tau)}(\cq(\tau))) \nonumber
\end{align}
for all $t=1$ to $\infty$.
\end{lemma}

\par Lemma~\ref{lemma:new_lemma_8} upper bounds the expectation of the fourth term of \eqref{eq:bound-Qk-Qkinter-final}, which is also used to upper bound the second term of \eqref{eq:bound-Qk-Qkinter-final} through Lemma~\ref{lemma:new_lemma_6}.
\begin{lemma} \label{lemma:new_lemma_8}
For any $k$ value, there exists $\gamma_1$ to $\gamma_{k-1}$ such that
\begin{align}
&\EE(\sum_{\tau=1}^t I(k\in\SAin{n(\tau)})I(\Qkinter{k}(\tau)\geq 1)\nonumber\\
&\cdot I(\bine{k}{n(\tau)}(\cq(\tau))=1)I(Q_k(\tau)=0))\nonumber\\
&\leq \sum_{k'=1}^{k-1} \gamma_{k'}\EE (\sum_{\tau=1}^{t}I(k'\in\SAin{n(\tau)})I(\Qkinter{k'}(\tau)<\bine{k'}{n(\tau)}(\cq(\tau)))) \label{eq:lemma-8-ineq}
\end{align}
for all $t=1$ to $\infty$.
\end{lemma}

Finally by applying Lemma~\ref{lemma:new_lemma_6} and Lemma~\ref{lemma:new_lemma_8} to the second term of the RHS of \eqref{eq:bound-Qk-Qkinter-final}, applying Lemma~\ref{lemma:new_lemma_7} to the third term of the RHS of \eqref{eq:bound-Qk-Qkinter-final}, and applying Lemma~\ref{lemma:new_lemma_8} to the fourth term of the RHS of \eqref{eq:bound-Qk-Qkinter-final}, we have proven the following statement: for any $k$, there exist $\gamma_1$ to $\gamma_K$ such that
\begin{align}
&\EE(|Q_k(t)-\Qkinter{k}(t)|)\nonumber\\
&\leq \sum_{k'=1}^K \gamma_{k'} \EE(\sum_{\tau=1}^{t-1}I(k'\in\SAin{n(\tau)})I(\Qkinter{k'}(\tau)<\bine{k'}{n(\tau)}(\cq(\tau)))) \nonumber
\end{align}
for all $t=1$ to $\infty$. The proof of Lemma~\ref{lemma:bound-Qk-Qkinter} to Lemma~\ref{lemma:new_lemma_8} are relegated to Appedix~\ref{app:5_lemma_proofs}. Lemma \ref{lemma:bound-Qk} is thus proven. \qed

\subsection{Proofs of Lemma~\ref{lemma:bound-Qk-Qkinter} to Lemma~\ref{lemma:new_lemma_8}} \label{app:5_lemma_proofs}
{\em Proof of Lemma~\ref{lemma:bound-Qk-Qkinter}: }Before proving Lemma~\ref{lemma:bound-Qk-Qkinter}, we first rewrite the LHS of \eqref{eq:bound-Qk-Qkinter} as
\begin{align}
&|Q_k(t)-\Qkinter{k}(t)| \nonumber \\
&=\sum_{\tau=1}^{t-1}\left(|Q_k(\tau+1)-\Qkinter{k}(\tau+1)|-|Q_k(\tau)-\Qkinter{k}(\tau)|\right). \nonumber
\end{align}
So to prove \eqref{eq:bound-Qk-Qkinter}, it suffices to show that the following inequality holds for all $k$ and $\tau<t$.
\begin{align}
&\left(|Q_k(\tau+1)-\Qkinter{k}(\tau+1)|-|Q_k(\tau)-\Qkinter{k}(\tau)|\right) \nonumber\\
&\leq I(\exists k'\in\SAin{n(\tau)}: Q_{k'}(\tau)=0). \label{eq:bound-Qk-Qkinter-2}
\end{align}

We now prove \eqref{eq:bound-Qk-Qkinter-2}. By set relationship, one can easily verify that one and only one of the following 3 possible cases is true at each time $\tau$.
\begin{enumerate}
\item $k\not\in \SAin{n(\tau)}\cup \SAout{n(\tau)}$.
\item $k\in \SAin{n(\tau)}\cup \SAout{n(\tau)}$ and SA $n(\tau)$ is feasible.
\item $k\in \SAin{n(\tau)}\cup \SAout{n(\tau)}$ and SA $n(\tau)$ is not feasible.
\end{enumerate}

In the case of 1), the LHS of \eqref{eq:bound-Qk-Qkinter-2} at time $\tau$ is zero since $Q_k(\tau+1)-Q_k(\tau)=\Qkinter{k}(\tau+1)-\Qkinter{k}(\tau)=\sum_{m=1}^M \alpha_{k,m}a_m(\tau)$. Inequality \eqref{eq:bound-Qk-Qkinter-2} thus holds obviously.

\par In the case of 2), the scheduled SA $n(\tau)$ is feasible. Suppose $k\in\SAout{n(\tau)}$. Then the LHS of \eqref{eq:bound-Qk-Qkinter-2} at time $\tau$ is always 0 since $Q_k(\tau+1)-Q_k(\tau)=\Qkinter{k}(\tau+1)-\Qkinter{k}(\tau)=\bout{k}{n(\tau)}(\cq(\tau))+\sum_{m=1}^M \alpha_{k,m}a_m(\tau)$. Suppose $k\in\SAin{n(\tau)}$. Then $Q_k(\tau+1) = Q_k(\tau)-\bine{k}{n(\tau)}(\cq(\tau))+\sum_{m=1}^M \alpha_{k,m}a_m(\tau)$. There are now two sub-cases: $\Qkinter{k}(\tau)\geq Q_k(\tau)\geq 1$ or $\Qkinter{k}(\tau)<Q_k(\tau)$. (The case that $Q_k(\tau)=0$ is not possible since we now consider the scenario SA $n(\tau)$ is feasible.) In the first sub-case, since $\Qkinter{k}(\tau)\geq Q_k(\tau)\geq 1$ and since SA $n(\tau)$ is feasible, we must have $\Qkinter{k}(\tau+1)-\Qkinter{k}(\tau)=Q_k(\tau+1)-Q_k(\tau)=-\bine{k}{n(\tau)}(\cq(\tau))+\sum_{m=1}^M \alpha_{k,m}a_m(\tau)$. As a result, the LHS of \eqref{eq:bound-Qk-Qkinter-2} at time $\tau$ is again 0. In the second sub-case, $\Qkinter{k}(\tau+1) = (\Qkinter{k}(\tau)-\bine{k}{n(\tau)}(\cq(\tau)))^+ +\sum_{m=1}^M \alpha_{k,m}a_m(\tau)$, and $Q_k(\tau+1) = \Qkinter{k}(\tau)-\bine{k}{n(\tau)}(\cq(\tau)) +\sum_{m=1}^M \alpha_{k,m}a_m(\tau)$ since in this case we assume SA $n(\tau)$ is feasible and thus $Q_k(\tau)\geq 1$. Recall that $\Qkinter{k}(\tau)< Q_k(\tau)$ in this sub-case. Therefore, $(\Qkinter{k}(\tau)-\bine{k}{n(\tau)}(\cq(\tau)) )^+\leq  (Q_k(\tau)-\bine{k}{n(\tau)}(\cq(\tau)) )^+ = (Q_k(\tau)-\bine{k}{n(\tau)}(\cq(\tau)))$. We thus have $\Qkinter{k}(\tau+1)\leq Q_k(\tau+1)$. As a result, the LHS of \eqref{eq:bound-Qk-Qkinter-2} at time $\tau$ becomes
\begin{align}
&(|Q_k(\tau+1)-\Qkinter{k}(\tau+1)|-|Q_k(\tau)-\Qkinter{k}(\tau)|)\nonumber\\
=&(Q_k(\tau+1) -Q_k(\tau)+\Qkinter{k}(\tau)-\Qkinter{k}(\tau+1))\nonumber\\
=&-\bine{k}{n(\tau)}(\cq(\tau)) \nonumber \\ &+(\Qkinter{k}(\tau)-\max\{\Qkinter{k}(\tau)-\bine{k}{n(\tau)}(\cq(\tau)), 0\})\nonumber\\
=&-\bine{k}{n(\tau)}(\cq(\tau)) +\min\{\bine{k}{n(\tau)}(\cq(\tau)),\Qkinter{k}(\tau)\}\nonumber\\
\leq &0\nonumber.
\end{align}
Since the RHS of \eqref{eq:bound-Qk-Qkinter-2} is always non-negative, \eqref{eq:bound-Qk-Qkinter-2} holds in the case of 2).

In the case of 3), the preferred SA $n(\tau)$ is not feasible. Without loss of generality, we assume that there is no external arrival at queue $k$ in time $\tau$ since any external arrival will change $Q_k(\tau)$ and $\Qkinter{k}(\tau)$ by the same amount. Since SA $n(\tau)$ is not feasible, we have $Q_k(\tau+1)=Q_k(\tau)$. On the other hand, $\Qkinter{k}(\tau)$ might still increase or decrease at most by 1 since the update rule of $\Qkinter{k}(\tau)$ \eqref{eq:Qk-inter-update} does not depend on whether SA $n(\tau)$ is feasible or not. Since $\Qkinter{k}(\tau)$ changes by at most 1, the LHS of \eqref{eq:bound-Qk-Qkinter-2} at time $\tau$ is upper bounded by 1 in this case while the RHS of \eqref{eq:bound-Qk-Qkinter-2} is always 1 since SA $n(\tau)$ is no feasible. Thus \eqref{eq:bound-Qk-Qkinter-2} holds in the case of 3).

In summary, for all $k$ and $\tau<t$, \eqref{eq:bound-Qk-Qkinter-2} holds in all 3 possible cases. Lemma~\ref{lemma:bound-Qk-Qkinter} is proven. \qed

{\em Proof of Lemma~\ref{lemma:divide_to_4_cases}: }
Suppose $k\in\SAin{n(\tau)}$. We claim that one and only one of the following 4 possible cases is true:
\begin{enumerate}
\item $\Qkinter{k}(\tau)<\bine{k}{n(\tau)}(\cq(\tau))$.
\item $Q_{k}(\tau)=0$, $\bine{k}{n(\tau)}(\cq(\tau))=0$, and $1\leq \Qkinter{k}(\tau)$.
\item $Q_{k}(\tau)=0$, $\bine{k}{n(\tau)}(\cq(\tau))=0$, and $0\leq \Qkinter{k}(\tau)<1$.
\item $Q_{k}(\tau)=0$, $\bine{k}{n(\tau)}(\cq(\tau))=1$, and $1\leq \Qkinter{k}(\tau)$.
\end{enumerate}

The reason is as follows. For any fixed $k$, we either have $\Qkinter{k}(\tau)<\bine{k}{n(\tau)}(\cq(\tau))$;
or $Q_{k}(\tau)=0$ and $\Qkinter{k}(\tau)\geq \bine{k}{n(\tau)}(\cq(\tau))$. In the former scenario, we have 1). In the latter scenario, we can further partition the event based on the values of $\bine{k}{n(\tau)}(\cq(\tau))$ and $\Qkinter{k}(\tau)$ and we thus have 2) to 4). The four cases correspond to the four terms in the RHS of \eqref{eq:divide_to_4_cases}. The proof of Lemma~\ref{lemma:divide_to_4_cases} is complete. \qed

{\em Proof of Lemma~\ref{lemma:new_lemma_6}: }Obviously we have
\begin{align}
&\EE (\sum_{\tau=1}^{t}I(k\in\SAin{n(\tau)})I(Q_{k}(\tau)=0)\nonumber \\
& \cdot I(\bine{k}{n(\tau)}(\cq(\tau))=0) I(1\leq \Qkinter{k}(\tau)))\nonumber \\
&\leq \EE(\sum_{\tau=1}^{t}I(k\in\SAin{n(\tau)})I(Q_{k}(\tau)=0)\nonumber \\
&\cdot I(1\leq \Qkinter{k}(\tau)).\nonumber
\end{align}

We then observe that $\bine{k}{n(\tau)}(\cq(\tau))$, the channel realization from queue $k$ to SA $n(\tau)$ during time $\tau$, is independent of $n(\tau)$, $Q_k(\tau)$, and $\Qkinter{k}(\tau)$, which depend only on the history from time 1 to $(\tau-1)$, not on the realization of $\bine{k}{n(\tau)}(\cq(\tau))$ in time $\tau$.

Furthermore, recall that $\bine{k}{n(\tau)}(\cq(\tau))$ is a Bernoulli random variable with $\EE(I(\bine{k}{n(\tau)}(\cq(\tau))=1))=\avg{\bine{k}{n(\tau)}(\cq(\tau))}$. Define
\begin{align}
\gamma_k=\frac{1}{\min_{c\in\bigcq, n\in [1,N], k\in \SAin{n}} \avg{\bine{k}{n}(c)}},\nonumber
\end{align}
which always exists since we assume $\min_{c\in\bigcq, n\in [1,N], k\in \SAin{n}} \avg{\bine{k}{n}(c)}>0$ in the SPN of interest (Assumption~3). Since whether $\bine{k}{n}(\cq(\tau))=1$ is independent of $n(\tau)$, $Q_k(\tau)$, and $\Qkinter{k}(\tau)$, we have
\begin{align}
&\EE (\sum_{\tau=1}^{t}I(k\in\SAin{n(\tau)})I(Q_{k}(\tau)=0)\nonumber \\
& \cdot I(1\leq \Qkinter{k}(\tau)))\nonumber \\
&\leq \gamma_k \EE(\sum_{\tau=1}^{t}I(k\in\SAin{n(\tau)})I(Q_{k}(\tau)=0)\nonumber \\
&\cdot I(\bine{k}{n(\tau)}(\cq(\tau))=1) I(1\leq \Qkinter{k}(\tau)),\nonumber
\end{align}
which completes the proof of Lemma~\ref{lemma:new_lemma_6}. \qed

{\em Proof of Lemma~\ref{lemma:new_lemma_7}: }We have
\begin{align}
&\EE(\sum_{\tau=1}^{t}I(k\in\SAin{n(\tau)})I(Q_{k}(\tau)=0)\nonumber\\
&\cdot I(\bine{k}{n(\tau)}(\cq(\tau))=0) I(0\leq \Qkinter{k}(\tau)<1))\nonumber\\
\leq& \gamma_k \EE (\sum_{\tau=1}^{t}I(k\in\SAin{n(\tau)})I(Q_{k}(\tau)=0)\nonumber\\
&\cdot I(\bine{k}{n(\tau)}(\cq(\tau))=1) I(0\leq \Qkinter{k}(\tau)<1))\label{eq:new_lemma_7-1} \\
\leq& \gamma_k \EE (\sum_{\tau=1}^{t}I(k\in\SAin{n(\tau)})I(\Qkinter{k}(\tau)<\bine{k}{n(\tau)}(\cq(\tau))), \label{eq:new_lemma_7}
\end{align}
where \eqref{eq:new_lemma_7-1} follows from the same argument as used in the proof of Lemma~\ref{lemma:new_lemma_6}; \eqref{eq:new_lemma_7} follows from the fact that if $\bine{k}{n(\tau)}(\cq(\tau))=1$ and $0\leq \Qkinter{k}(\tau)<1$, then $\Qkinter{k}(\tau)<\bine{k}{n(\tau)}(\cq(\tau))$. Thus Lemma~\ref{lemma:new_lemma_7} is proven. \qed

{\em Proof of Lemma~\ref{lemma:new_lemma_8}: }Define $\Delta Q_k(\tau)\eqdef \Qkinter{k}(\tau)-Q_k(\tau)$. We first state the following four claims and use these claims to prove Lemma~\ref{lemma:new_lemma_8}. The proof of these four claims are relegated to Appendix~\ref{app:claims_proof_lemma8}.

\begin{claim} \label{claim:Q1_larger_Qinter1}
For the most upstream queue ($k=1$) we have $Q_1(\tau)\geq \Qkinter{1}(\tau)$ for all $\tau$.
\end{claim} 

\begin{claim}\label{claim:new_claim_1}
For any $k=1,2,...,K$ and any time $t$, we have
\begin{align}
&\sum_{\tau=1}^t I(k\in\SAin{n(\tau)})I(\Qkinter{k}(\tau)\geq 1)\nonumber\\
&\cdot I(\bine{k}{n(\tau)}(\cq(\tau))=1)I(Q_k(\tau)=0)\nonumber\\
&\leq \sum_{\tau=1}^t I(k\in \SAout{n(\tau)})I(\Delta Q_k(\tau+1)>\Delta Q_k(\tau)). \label{eq:new_claim_1}
\end{align}
\end{claim}

\begin{claim} \label{claim:Qk-Qkinter-2}
For any $\tau=1$ to $\infty$, we have
\begin{align}
& I\left(\Delta Q_k(\tau+1)>\Delta Q_k(\tau) \right)I(k\in \SAout{n(\tau)}) \nonumber\\
&\leq \sum_{k'=1}^{k-1} I(k'\in\SAin{n(\tau)})I(\Qkinter{k'}(\tau)<\bine{k'}{n(\tau)} (\cq(\tau))) \nonumber \\
&+ \sum_{k'=1}^{k-1} I(k'\in\SAin{n(\tau)})I(Q_{k'}(\tau)=0)\nonumber\\
&\cdot I(\bine{k'}{n(\tau)} (\cq(\tau)))=0)I(1\leq \Qkinter{k'}(\tau))\nonumber\\
&+ \sum_{k'=1}^{k-1} I(k'\in\SAin{n(\tau)})I(Q_{k'}(\tau)=0)\nonumber\\
&\cdot I(\bine{k'}{n(\tau)} (\cq(\tau)))=0)I(0\leq \Qkinter{k'}(\tau)<1)\nonumber\\
&+ \sum_{k'=1}^{k-1} I(k'\in\SAin{n(\tau)})I(Q_{k'}(\tau)=0)\nonumber\\
&\cdot I(\bine{k'}{n(\tau)} (\cq(\tau)))=1)I(1\leq \Qkinter{k'}(\tau)). \label{eq:Qk-Qkinter-2}
\end{align}
\end{claim}

\begin{claim} \label{lemma:Qk-Qkinter}
For any $k=1,2,...,K$ and any time $t$, we have
\begin{align}
&\sum_{\tau=1}^t I(k\in \SAin{n(\tau)})I(\Qkinter{k}(\tau)\geq 1)\nonumber \\ 
&\cdot I(\bine{k}{n(\tau)}(\cq(\tau))=1) I(Q_k(\tau)=0) \nonumber \\
&\leq \sum_{k'=1}^{k-1} \sum_{\tau=1}^t I(k'\in \SAin{n(\tau)}) I(\Qkinter{k}(\tau)<\bine{k'}{n(\tau)}(\cq(\tau)))\nonumber \\
&+ \sum_{k'=1}^{k-1} \sum_{\tau=1}^t I(k'\in \SAin{n(\tau)}) I(\Qkinter{k'}(\tau)\geq 1) \nonumber \\
&\cdot I(\bine{k'}{n(\tau)}(\cq(\tau))=1) I(Q_{k'}(\tau)=0) \nonumber \\
&+ \sum_{k'=1}^{k-1} \sum_{\tau=1}^t I(k'\in \SAin{n(\tau)}) I(\Qkinter{k'}(\tau)\geq 1) \nonumber \\
&\cdot I(\bine{k'}{n(\tau)}(\cq(\tau))=0) I(Q_{k'}(\tau)=0)\nonumber \\
&+ \sum_{k'=1}^{k-1} \sum_{\tau=1}^t I(k'\in \SAin{n(\tau)}) I(1>\Qkinter{k'}(\tau)\geq 0) \nonumber \\
&\cdot I(\bine{k'}{n(\tau)}(\cq(\tau))=0)I(Q_{k'}(\tau)=0). \label{eq:Qk-Qkinter}
\end{align}
\end{claim}

With the above four claims, we are now ready to prove Lemma~\ref{lemma:new_lemma_8}. We prove Lemma~\ref{lemma:new_lemma_8} by induction on the value of $k$. Consider the case of $k=1$ first. By Claim~\ref{claim:Q1_larger_Qinter1}, we have $Q_1(\tau)\geq \Qkinter{1}(\tau)$ for all $\tau$. Therefore, whenever $Q_1(\tau)=0$, we must have $\Qkinter{1}(\tau)=0$. As a result, the LHS of \eqref{eq:lemma-8-ineq} is always 0 for $k=1$. Lemma~\ref{lemma:new_lemma_8} thus holds for $k=1$.

Now consider general $k$. By Lemma~\ref{lemma:Qk-Qkinter} and taking the expectation on both sides, we have
\begin{align}
&\EE(\sum_{\tau=1}^t I(k\in \SAin{n(\tau)})I(\Qkinter{k}(\tau)\geq 1)\nonumber\\ &\cdot I(\bine{k}{n(\tau)}(\cq(\tau))=1) I(Q_k(\tau)=0))\nonumber\\
\leq& \sum_{k'=1}^{k-1} \EE(\sum_{\tau=1}^t I(k'\in \SAin{n(\tau)}) I(\Qkinter{k}(\tau)<\bine{k'}{n(\tau)}(\cq(\tau))))\nonumber\\
&+ \sum_{k'=1}^{k-1} \EE(\sum_{\tau=1}^t I(k'\in \SAin{n(\tau)}) I(\Qkinter{k'}(\tau)\geq 1)\nonumber\\
&\quad\cdot I(\bine{k'}{n(\tau)}(\cq(\tau))=1) I(Q_{k'}(\tau)=0))\nonumber\\
&+ \sum_{k'=1}^{k-1} \EE(\sum_{\tau=1}^t I(k'\in \SAin{n(\tau)}) I(\Qkinter{k'}(\tau)\geq 1)\nonumber\\
&\quad\cdot I(\bine{k'}{n(\tau)}(\cq(\tau))=0) I(Q_{k'}(\tau)=0))\nonumber\\
&+ \sum_{k'=1}^{k-1} \EE(\sum_{\tau=1}^t I(k'\in \SAin{n(\tau)}) I(1>\Qkinter{k'}(\tau)\geq 0)\nonumber\\
&\quad\cdot I(\bine{k'}{n(\tau)}(\cq(\tau))=0)I(Q_{k'}(\tau)=0)). \label{eq:new_lemma_8-1} 
\end{align}

We look at the second term of the RHS of \eqref{eq:new_lemma_8-1} first. Notice that by induction hypothesis, for $k'=1,...,k-1$, there exists $\gamma_{k',1}$ to $\gamma_{k',k'-1}$ such that
\begin{align}
&\EE(\sum_{\tau=1}^t I(k'\in \SAin{n(\tau)}) I(\Qkinter{k'}(\tau)\geq 1)\nonumber\\
&\quad\cdot I(\bine{k}{n(\tau)}(\cq(\tau))=1) I(Q_{k'}=0))\nonumber\\
&\leq \sum_{k''=1}^{k'-1} \gamma_{k',k''}\nonumber\\
&\cdot \EE (\sum_{\tau=1}^{t}I(k''\in\SAin{n(\tau)})I(\Qkinter{k''}(\tau)<\bine{k''}{n(\tau)}(\cq(\tau)))).\nonumber
\end{align}
The above inequality shows that the second term of the RHS of \eqref{eq:new_lemma_8-1} can be bounded by a weighted sum of $\EE (\sum_{\tau=1}^{t}I(k'\in\SAin{n(\tau)})I(\Qkinter{k'}(\tau)<\bine{k'}{n(\tau)}(\cq(\tau))))$ for $k'=1,...,k-1$.

We now look at the third term of the RHS of \eqref{eq:new_lemma_8-1}. By Lemma~\ref{lemma:new_lemma_6}, for $k'=1,...,k-1$, there exists a constant $\gamma_{k'}'$ such that
\begin{align}
&\EE(\sum_{\tau=1}^t I(k'\in \SAin{n(\tau)}) I(\Qkinter{k'}(\tau)\geq 1)\nonumber\\
&\quad\cdot I(\bine{k'}{n(\tau)}(\cq(\tau))=0) I(Q_{k'}(\tau)=0))\nonumber\\
&\leq \gamma_{k'}'\EE(\sum_{\tau=1}^t I(k'\in \SAin{n(\tau)}) I(\Qkinter{k'}(\tau)\geq 1)\nonumber\\
&\quad\cdot I(\bine{k'}{n(\tau)}(\cq(\tau))=1) I(Q_{k'}(\tau)=0)).\nonumber
\end{align}
Again by the same argument for the second term of the RHS of \eqref{eq:new_lemma_8-1}, the third term of the RHS of \eqref{eq:new_lemma_8-1} can also be bounded by a weighted sum of $\EE (\sum_{\tau=1}^{t}I(k'\in\SAin{n(\tau)})I(\Qkinter{k'}(\tau)<\bine{k'}{n(\tau)}(\cq(\tau))))$ for $k'=1,...,k-1$.

By Lemma~\ref{lemma:new_lemma_7}, the fourth term of the RHS of \eqref{eq:new_lemma_8-1} can also be bounded by a weighted sum of $\EE (\sum_{\tau=1}^{t}I(k'\in\SAin{n(\tau)})I(\Qkinter{k'}(\tau)<\bine{k'}{n(\tau)}(\cq(\tau))))$ for $k'=1,...,k-1$.

Since all the 4 terms in the RHS of \eqref{eq:new_lemma_8-1} can be upper bounded by a weighted sum of $\EE (\sum_{\tau=1}^{t}I(k'\in\SAin{n(\tau)})I(\Qkinter{k'}(\tau)<\bine{k'}{n(\tau)}(\cq(\tau))))$ for $k'=1,...,k-1$, we have thus proven Lemma~\ref{lemma:new_lemma_8}. \qed

\subsection{Proofs of Claims for Lemma~\ref{lemma:new_lemma_8}} \label{app:claims_proof_lemma8}
{\em Proof of Claim~\ref{claim:Q1_larger_Qinter1}: }By the definition of $Q_k(t)$ and $\Qkinter{k}(t)$, we have $Q_k(1)=0=\Qkinter{k}(1)$. The desired inequality holds when $\tau=1$. Suppose the inequality holds for some $\tau$. We now prove that the inequality also holds for $\tau+1$. To that end, we first notice that any external arrival at time $\tau$ will increase the $\Qkinter{1}(\tau)$ and $Q_1(\tau)$ by the same amount. Therefore, the external arrivals will not affect the order between $\Qkinter{1}(\tau)$ and $Q_1(\tau)$ and we can thus assume there is no external arrival in time $t$ without loss of generality. Consider the first scenario in which $1\notin \SAin{n(\tau)}$. Since $1\notin \SAin{n(\tau)}$, no packets will leave queue $1$. Since we have $\Qkinter{1}(\tau)\leq Q_1(\tau)$ to begin with, we will still have $\Qkinter{1}(\tau+1)\leq Q_1(\tau+1)$.

\par Now consider the scenario of $1\in \SAin{n(\tau)}$ and the following two cases: Case 1: $\Qkinter{1}(\tau)<\bine{1}{n(\tau)}(\cq(\tau))$. In this case, at the beginning of time $\tau+1$, $\Qkinter{1}(\tau+1)=0$ due to the update rule \eqref{eq:Qk-inter-update}. Since the actual queue length $Q_k(\tau)$ is non-negative, we must have $\Qkinter{1}(\tau+1)\leq Q_1(\tau+1)$. Case 2: $\Qkinter{1}(\tau)\geq \bine{1}{n(\tau)}(\cq(\tau))$. In this case, we have $\Qkinter{1}(\tau+1)= \Qkinter{1}(\tau)-\bine{1}{n(\tau)}(\cq(\tau))$ (recall that we assume no external). We observe that the actual queue length $Q_1$ either decreases by $\bine{1}{n(\tau)}(\cq(\tau))$ or remain the same, depending on whether SA $n(\tau)$ can be carried out successfully or not (see Difference 1 in the Section~\ref{subsec:01SPN}). Therefore, the decrease amount of $\Qkinter{1}(\tau)$ is no less than the decrease amount of $Q_1(\tau)$, which together with the fact that $\Qkinter{1}(\tau)\leq Q_1(\tau)$ imply $\Qkinter{1}(\tau+1)\leq Q_1(\tau+1)$. By induction, we have proven that $Q_1(\tau)\geq \Qkinter{1}(\tau)$ for all $\tau$. \qed

{\em Proof of Claim~\ref{claim:new_claim_1}: }Since both $Q_k(\tau)$ and $\Qkinter{k}(\tau)$ are integer-valued random processes, $\Delta Q_k(\tau)$ is also an integer-valued random process. Furthermore, we observe that the changes of $Q_k(\tau)$ and $\Qkinter{k}(\tau)$ is always in the same direction. Namely, if $\Qkinter{k}(\tau)$ increases,\footnote{ For ease of exposition, we do not count the external arrivals since any external arrival will increase $Q_k$ and $\Qkinter{k}$ by the same amount.} then it means that $k$ is one of the output queues of SA $n(\tau)$, which means that $Q_k(\tau)$ can either increase or remain the same (the latter is due to the fact that the preferred SA $n(\tau)$ may be infeasible). Similarly, if $\Qkinter{k}(\tau)$ decreases then $Q_k(\tau)$ can decrease or remain the same. Since the largest change of $Q_k$ (resp.\ $\Qkinter{k}$) is at most $1$ and they move in the same direction, it can be easily shown that the change of $\Delta Q_k(\tau)$ is also at most $1$. To simplify the expression, for the time being, we sometimes ignore the queue index $k$ in $\Delta Q_k(\tau)$. That is, we will write $\Delta Q_k(\tau)$ as $\Delta Q(\tau)$ in the remaining of this proof.

In the following, we iteratively define two sequences of time instants, $\{s_i:\forall i\}$ and $\{t_i:\forall i\}$. The first time instant is $s_1=1$. Then for any $i$, define $t_i\in (s_i,t+1]$ as the largest time instant such that for all time instant $\tilde{\tau}\in (s_i,t_i)$, we have $\Delta Q(\tilde{\tau})> 0$. Note that for all $\tilde{\tau}\in (s_i,s_i+1)$ we have $\Delta Q(\tilde{\tau})> 0$ since $(s_i,s_i+1)$ is an empty interval. As a result, $t_i$ always exists and is uniquely defined as long as we have $s_i\leq t$ to begin with. Furthermore, since $\Delta Q(\tau)$ is an integer-valued random process with change at most 1 at each time slot, we can observe that $\Delta Q(t_i)=0$ if $t_i\leq t$ and $\Delta Q(t_i)>0$ if $t_i=t+1$. In summary, $\Delta Q(t_i)\geq 0$.

\par After defining $t_i$, we define $s_{i+1}\in [t_i,t]$ as the time instant such that for all time instant $\tilde{\tau}\in (t_i,s_{i+1}]$, we have $\Delta Q(\tilde{\tau})\leq 0$ and $\Delta Q(s_{i+1}+1)>0$. This time, such $s_{i+1}$ may or may not exist. For example, if $\Delta Q(\tilde{\tau} )\leq  0$ for all $\tilde{\tau}\in [t_i, t+1]$, then $s_{i+1}$ does not exist since even the largest possible choice of $s_{i+1}=t$ still does not satisfy the requirement $\Delta Q(s_{i+1}+1)>0$. However, one may observe that we must have $\Delta Q(s_{i+1})=0$ whenever $s_{i+1}$ exists. The reason is that $\Delta Q(\tau)$ changes by at most one in any two consecutive time slots. Therefore, the facts that $\Delta Q(\tilde{\tau})\leq 0$ and $\Delta Q(s_{i+1}+1)>0$ jointly imply $\Delta Q(s_{i+1})=0$. In summary, $[s_i,t_i)$ is the $i$-th ``continuous interval" such that all $\tau\in (s_i,t_i)$ satisfy $\Delta Q(\tau)>0$.

\par Define $M_s$ as the number of $(s_i,t_i)$ pairs that do exist. Since $s_1=1$ is clearly defined, we have $M_s\geq 1$. We will now argue that for any $i=1$ to $M_s$, we have
\begin{align}
&\sum_{\tau=s_i}^{t_i-1}I(\Delta Q(\tau+1)<\Delta Q(\tau))\nonumber\\
\leq& \sum_{\tau=s_i}^{t_i-1}I(\Delta Q(\tau+1)>\Delta Q(\tau)). \label{eq:new_23}
\end{align}

To see the correctness of \eqref{eq:new_23}, we first observe that
\begin{align}
\Delta Q(t_i)=\Delta Q(s_i)+\sum_{\tau=s_i}^{t_i-1}(\Delta Q(\tau+1)-\Delta Q(\tau)). \nonumber
\end{align}

Since $\Delta Q(s_i)=0$ and $\Delta Q(t_i)\geq 0$, we have
\begin{align}
&\sum_{\tau=s_i}^{t_i-1} (\Delta Q(\tau+1)-\Delta Q(\tau))^+\nonumber\\
\geq& \sum_{\tau=s_i}^{t_i-1} (\Delta Q(\tau+1)-\Delta Q(\tau))^-,\nonumber
\end{align}
where $(v)^+=\max\{0,v\}$ and $(v)^-=\max\{0,-v\}$. Since $\Delta Q(\tau)$ moves by at most 1, we thus have \eqref{eq:new_23}.

Now we turn our focus back to proving Claim~\ref{claim:new_claim_1}. We notice that when
\begin{align}
&I(k\in\SAin{n(\tau)})I(\Qkinter{k}(\tau)\geq 1)\nonumber\\
&\cdot I(\bine{k}{n(\tau)}(\cq(\tau))=1)I(Q_k(\tau)=0)=1, \label{new-tau-break-1}
\end{align}
we have $\Delta Q(\tau)=\Qkinter{k}(\tau)-Q_k(\tau)\geq 1$. Moreover, we argue that $\Delta Q(\tau+1)=\Delta Q(\tau)-1$. The reason is as follows. Since queue $k$ is one of the input queues of the preferred SA at $\tau$ and the queue lengths satisfy $\Qkinter{k}(\tau)\geq 1$ and $Q_k(\tau)=0$, $\Qkinter{k}$ will decrease by one according to the update rule \eqref{eq:Qk-inter-update} while $Q_k$ remains zero since the preferred SA $n(\tau)$ is infeasible. As a result, $\Delta Q(\tau+1)=\Delta Q(\tau)-1$.

We thus have the following,
\begin{align}
&\sum_{\tau=1}^t I(k\in\SAin{n(\tau)})I(\Qkinter{k}(\tau)\geq 1)\nonumber\\
&\cdot I(\bine{k}{n(\tau)}(\cq(\tau))=1)I(Q_k(\tau)=0)\nonumber\\
&\leq \sum_{i=1}^{M_s} \sum_{\tau=s_i}^{t_i-1} I(\Delta Q(\tau+1)<\Delta Q(\tau))\label{eq:new_29}
\end{align}
The reason is that any $\tau$ that satisfies \eqref{new-tau-break-1} will have $\Delta Q(\tau)\geq 1$, and by our construction of $s_i$ and $t_i$, for all $i=1$ to $M_s$, such $\tau$ must fall into one of the intervals $[s_i,t_i)$. Also, any $\tau$ that satisfies \eqref{new-tau-break-1} will have $\Delta Q(\tau+1)<\Delta Q(\tau)$. As a result, we have \eqref{eq:new_29}. We can continue upper bounding \eqref{eq:new_29} by
\begin{align}
\eqref{eq:new_29}&\leq \sum_{i=1}^{M_s} \sum_{\tau=s_i}^{t_i-1} I(\Delta Q(\tau+1)>\Delta Q(\tau))\label{eq:new_30}\\
&= \sum_{i=1}^{M_s} \sum_{\tau=s_i}^{t_i-1} I(\Delta Q(\tau+1)>\Delta Q(\tau))I(k\in\SAout{n(\tau)}) \label{eq:new_31}\\
&\leq \sum_{\tau=1}^{t} I(\Delta Q(\tau+1)>\Delta Q(\tau))I(k\in\SAout{n(\tau)}),\label{eq:new_32}
\end{align}
where \eqref{eq:new_30} follows from \eqref{eq:new_23}; and \eqref{eq:new_32} follows from including additionally those $\tau$ not in any of the interval $[s_i,t_i)$. Except for proving \eqref{eq:new_31}, the proof of Claim~\ref{claim:new_claim_1} is complete.

\par In the remaining part of this proof, we will rigorously prove \eqref{eq:new_31}. To that end, we first notice that
\begin{align}
&I(\Delta Q(\tau+1)-\Delta Q(\tau) >0) \nonumber \\
=&I(\Delta Q(\tau+1)-\Delta Q(\tau) >0)\cdot I(k\in\SAout{n(\tau)})\nonumber \\
& +I(\Delta Q(\tau+1)-\Delta Q(\tau) >0)\cdot I(k\not\in\SAout{n(\tau)}) \label{eq:new-proof-CCW8}
\end{align}

In the next paragraph, we will prove that when $k\not\in \SAout{n(\tau)}$, we always have either ``$\Delta Q(\tau+1)-\Delta Q(\tau) \leq 0$'' or ``$\Delta Q(\tau)< 0$.'' It means that the term $I(\Delta Q(\tau+1)-\Delta Q(\tau) >0)\cdot I(k\not\in\SAout{n(\tau)})$ is either $0$ or the $\tau$ value is not counted in any of the $[s_i,t_i)$ intervals since by our construction we always have $\Delta Q(s_i)= 0$ and any $\tilde{\tau}\in (s_i,t_i)$ satisfying $\Delta Q(\tilde{\tau})>0$. As a result, \eqref{eq:new_31} is true.

\par Consider the situation when $k\not\in \SAout{n(\tau)}$ and consider two sub-cases: If SA $n(\tau)$ turns out to be infeasible, then $Q_k(\tau+1)=Q_k(\tau)+\sum_{m=1}^M \alpha_{k,m} a_m(\tau)$. Also, we always have $\Qkinter{k}(\tau+1)\leq \Qkinter{k}(\tau)+\sum_{m=1}^M \alpha_{k,m}a_m(\tau)$ since $k\not\in \SAout{n(\tau)}$ implies that the intermediate queue length $\Qkinter{k}$ can only decrease or remain the same (except when there is external arrival $\sum_{m=1}^M \alpha_{k,m}a_m(t)$). As a result, in this sub-case, we have $\Delta Q(\tau+1)-\Delta Q(\tau) \leq 0$.

\par In the second sub-case: SA $n(\tau)$ is feasible, we have $Q_k(\tau+1)=Q_k(\tau)+\sum_{m=1}^M \alpha_{k,m}a_m(\tau)-I(k\in \SAin{n(\tau)})\bine{k}{n(\tau)}(\cq(\tau))$. Namely, when not counting the external arrival, $Q_k$ can now possibly decrease if $k\in \SAin{n(\tau)}$ or it will remain the same if $k\not\in \SAin{n(\tau)}$. Our goal is to show that either ``$\Delta Q(\tau+1)-\Delta Q(\tau) \leq 0$'' or ``$\Delta Q(\tau)< 0$.'' To that end, we prove the equivalent statement that $\Delta Q(\tau)\geq 0$ implies $\Delta Q(\tau+1)=\Delta Q(\tau)$. Since SA $n(\tau)$ is feasible, we have $Q_k(\tau)\geq 1$. Since $\Delta Q(\tau)=\Qkinter{k}(\tau)-Q_k(\tau)\geq 0$, we have $\Qkinter{k}(\tau)\geq Q_k(\tau)\geq 1$. Therefore, if $k\in \SAin{n(\tau)}$, then both $\Qkinter{k}(\tau)$ and $Q_k(\tau)$ will decrease by the same amount $\bine{k}{n(\tau)}(\cq(\tau))$; and if $k\not\in \SAin{n(\tau)}$, both $\Qkinter{k}(\tau)$ and $Q_k(\tau)$ will remain the same except for the external arrival. We thus have $\Qkinter{k}(\tau+1)=\Qkinter{k}(\tau)+\sum_{m=1}^M \alpha_{k,m}a_m(t)-I(k\in \SAin{n(\tau)})\bine{k}{n(\tau)}(\cq(\tau))$. Namely, $\Qkinter{k}(\tau)$ will experience the same change as $Q_k(\tau)$. As a result $\Delta Q(\tau+1)=\Delta Q(\tau)$. The proof of \eqref{eq:new_31} is complete and the proof of Claim~\ref{claim:new_claim_1} is thus also complete. \qed

{\em Proof of Claim~\ref{claim:Qk-Qkinter-2}: }If $k\not\in\SAout{n(\tau)}$, the LHS of \eqref{eq:Qk-Qkinter-2} is zero and the inequality always holds. If $k\in\SAout{n(\tau)}$, we claim that at least of the following 5 possible cases is true:
\begin{enumerate}
\item For all queues $k'\in\SAin{n(\tau)}$, $Q_{k'}(\tau)\geq 1$ and $\Qkinter{k'}(\tau)\geq \bine{k'}{n(\tau)}(\cq(\tau))$.
\item There exists a queue $k'\in\SAin{n(\tau)}$ with $\Qkinter{k'}(\tau)<\bine{k'}{n(\tau)}(\cq(\tau))$.
\item There exists a queue $k'\in\SAin{n(\tau)}$ with $Q_{k'}(\tau)=0$; $\bine{k'}{n(\tau)}(\cq(\tau))=0$; and $1\leq \Qkinter{k'}(\tau)$.
\item There exists a queue $k'\in\SAin{n(\tau)}$  with $Q_{k'}(\tau)=0$; $\bine{k'}{n(\tau)}(\cq(\tau))=0$; and $0\leq \Qkinter{k'}(\tau)<1$.
\item There exists a queue $k'\in\SAin{n(\tau)}$  with $Q_{k'}(\tau)=0$ and $\bine{k'}{n(\tau)}(\cq(\tau))=1$, and $1\leq \Qkinter{k'}(\tau)$.
\end{enumerate}

The reason is as follows. If 1) does not hold, either there exists a $k'$ such that $\Qkinter{k'}(\tau)<\bine{k'}{n(\tau)}(\cq(\tau))$;
or there exists a $k'$ such that $Q_{k'}(\tau)=0$ and $\Qkinter{k'}(\tau)\geq \bine{k'}{n(\tau)}(\cq(\tau))$. In the former scenario, we have 2). In the latter scenario, we can further partition the event based on the values of $\bine{k'}{n(\tau)}(\cq(\tau))$ and $\Qkinter{k'}(\tau)$, which leads to 3) to 5).

In the case of 1), SA $n(\tau)$ is feasible in the beginning of time $\tau$. Hence SA $n(\tau)$ will be activated. Since we now consider the scenario of $k\in \SAout{n(\tau)}$, both $Q_k(\tau)$ and $\Qkinter{k}(\tau)$ increase by the same amount, $\bout{k}{n(\tau)}(\cq(\tau))+\sum_{m=1}^M \alpha_{k,m}a_m(\tau)$. As a result, $\Delta Q(\tau+1)=\Delta Q(\tau)$. The LHS of \eqref{eq:Qk-Qkinter-2} equals to zero and the inequality \eqref{eq:Qk-Qkinter-2} holds.

\par In the case of 2), the first term of the RHS of \eqref{eq:Qk-Qkinter-2} is at least 1 because there exists a queue $k'\in\SAin{n(\tau)}$ such that $I(k'\in\SAin{n(\tau)})I(\Qkinter{k'}(\tau)<\bine{k'}{n(\tau)} (\cq(\tau)))=1$. Since the LHS of \eqref{eq:Qk-Qkinter-2} is at most 1, the inequality \eqref{eq:Qk-Qkinter-2} holds. We can observe the same relationship between 3) and the second term, 4) and the third term, and 5) and the fourth term of the RHS of \eqref{eq:Qk-Qkinter-2}. Since \eqref{eq:Qk-Qkinter-2} holds for all 5 cases, the proof of Claim~\ref{claim:Qk-Qkinter-2} is complete. \qed

{\em Proof of Claim~\ref{lemma:Qk-Qkinter}:} Notice that joinly Claims~\ref{claim:new_claim_1} and~\ref{claim:Qk-Qkinter-2} immediately give us Claim 4. \qed

\subsection{Proof of Sublinearly Growing $\Qkinter{k}(t)$ and $\Nna{k}(t)$} \label{app:sub_growth}
In the next lemma, we will shows that $\SCH_\text{avg}$ can sublinearly stabilize $\Qkinter{k}(t)$ and $\Nna{k}(t)$ for all $k$.
\begin{lemma} \label{lemma:sub-growth}
Consider any rate vector $\ve{R}$ such that there exist $\ve{s}_c\in \Lambda^\circ$ for all $c\in\bigcq$ satisfying \eqref{eq:new-balance}. The proposed $\SCH_\text{avg}$ can sublinearly stabilize $\qkinter{k}(t)$, $\Nna{k}(t)$, and $\Qkinter{k}(t)$ for all $k$.
\end{lemma}

We will prove the sublinear growth of the four quantities separately.

{\em Proof of sublinearly growing $q_k(t)$ and $\qkinter{k}(t)$: }First, we provide the conventional stability definition.
\begin{definition}
A queue length $q(t)$ is stable if
\begin{align}
\limsup_{t\to\infty} \frac{1}{t}\sum_{\tau=1}^t\EE\{|q(t)|\}<\infty. \label{eq:traditional-stab}
\end{align}
And the network is stable if all the queues are stable.
\end{definition}

As discussed in Section~\ref{subsec:proposed-DMW}, the back-pressure vector computation \eqref{eq:new-backpressure} and the update rule \eqref{eq:DMW-update-new} are only based on the expected input and output service rate matrix $\overline{\Bine(\cq(t))}$ and $\avg{\Bout(\cq(t))}$, which are deterministic matrices. As a result, they can be viewed as the virtual queue lengths of a deterministic SPN. In the existing proof in \cite{jiang2009stable}, it has been shown that the virtual queue length $\ve{q}(t)$ of a deterministic SPN can be stabilized by $\SCH_\text{avg}$. As a result, $\SCH_\text{avg}$ can also stabilize the virtual queue length $q_k(t)$ for all $k$ in the given (0,1) random SPN.

\par Notice that given the past arrival vectors and the past and current channel quality, i.e., given $\cq(t)$ and $\{\ve{a},\cq\}_1^{t-1}$, the quantity $\ve{q}(t)$ and $\ve{x}^*(t)$ is no longer random and is of deterministic value, see the update rules of \eqref{eq:new-backpressure} and \eqref{eq:DMW-update-new}. The following lemma establishes the connection between $\ve{q}(t)$ and $\ve{q}^{\text{inter}}(t)$.
\begin{lemma} \label{lemma:VQ1_VQ2}
$\ve{q}(t)$ is the expectation of $\two{\ve{q}}(t)$ conditioned on $\{\ve{a},\cq\}_1^{t-1}$. That is, $\one{\ve{q}}(t) = \EE\{\two{\ve{q}}(t)| \{\ve{a},\cq\}_1^{t-1}\}$.
\end{lemma}

{\em Proof of Lemma~\ref{lemma:VQ1_VQ2}: }This lemma can be proven iteratively. When $t=1$, since $\ve{q}(t)=\qinter(t)=\ve{0}$, the zero vector, Lemma~\ref{lemma:VQ1_VQ2} holds automatically. Suppose Lemma~\ref{lemma:VQ1_VQ2} holds for some $t$. By comparing \eqref{eq:DMW-update-new} and \eqref{eq:DMW-update-inter}, we can see that Lemma~\ref{lemma:VQ1_VQ2} holds for $t+1$ as well. \qed

For any $k\in \{1,2,...,K\}$, we square both sides of \eqref{eq:qk-inter-update} and we thus have
\begin{align}
&q_k^\text{inter}(t+1)^2 - q_k^\text{inter}(t)^2 \nonumber \\
=& (\mu_{\out,k}(t)-\mu_{\iin,k}(t))^2 - 2q_k^\text{inter}(t)\left(\mu_{\out,k}(t)-\mu_{\iin,k}(t) \right). \nonumber
\end{align}

\par Similar to \eqref{eq:mu_out_def} and \eqref{eq:mu_in_def} we can define the average arrival rate and departure rate of queue $k$ as follows.
\begin{align}
&\overline{\mu_{\out,k}}(t) = \sum_{n=1}^N \left(\overline{\bine{k}{n}(\cq(t))} x_n^*(t) \right), \nonumber \\
&\overline{\mu_{\iin,k}}(t) = \sum_{m=1}^M \left( \alpha_{k,m} a_m(t) \right)+\sum_{n=1}^N \left(\overline{\bout{k}{n}(\cq(t))} x_n^*(t)\right). \label{eq:avg_mu}
\end{align}
By taking the expectation conditioned on the past and current arrival vectors and past channel quality on both sides until time $t$, we have
\begin{align}
&\EE\{q_k^\text{inter}(t+1)^2|\{\ve{a},\cq\}_1^t\} - \EE\{q_k^\text{inter}(t)^2|\{\ve{a},\cq\}_1^t\} \nonumber \\
=& \EE\{(\mu_{\out,k}(t)-\mu_{\iin,k}(t))^2|\{\ve{a},\cq\}_1^t\} \nonumber \\
&- 2 \EE\{q_k^\text{inter}(t)\left(\mu_{\out,k}(t)-\mu_{\iin,k}(t) \right)|\{\ve{a},\cq\}_1^t\} \nonumber \\
=& \EE\{(\mu_{\out,k}(t)-\mu_{\iin,k}(t))^2|\{\ve{a},\cq\}_1^t\} \nonumber \\
&- 2 q_k(t)\left(\avg{\mu_{\out,k}}(t)-\avg{\mu_{\iin,k}}(t) \right) \label{eq:suf_NM2_1} \\
\leq& C^2 + 2 |q_k(t)| U, \label{eq:suf_NM2_2}
\end{align}
where \eqref{eq:suf_NM2_1} follows from the observation that $\qkinter{k}(t)$ is a constant given $\{\ve{a},\cq\}_1^{t-1}$ and $\overline{\mu_{\out,k}}(t)$ and $\overline{\mu_{\iin,k}}(t)$ are the conditional expectation of $\mu_{\out,k}(t)$ and $\mu_{\iin,k}(t)$ \eqref{eq:avg_mu} given $\{\ve{a},\cq\}_1^{t}$; and \eqref{eq:suf_NM2_2} follows from defining $C$ to be the upper bound of $|\mu_{\out,k}(t)-\mu_{\iin,k}(t)|$ and $U$ to be the upper bound\footnote{$C$ and $U$ exist because $\mu_{\out,k}(t)$ and  $\mu_{\iin,k}(t)$ have bounded support by our definition.} of $|\overline{\mu_{\out,k}}(t)-\overline{\mu_{\iin,k}}(t)|$. Now we take the expectation over all possible past arrival vectors and past channel quality.
\begin{align}
\EE\{q_k^\text{inter}(t+1)^2\} - \EE\{q_k^\text{inter}(t)^2\} \leq C^2 + 2 U \EE\{|q_k(t)|\}. \label{eq:new-tau-t}
\end{align}

Eq.~\eqref{eq:new-tau-t} also holds if we replace the time index $t$ by $\tau$.  By summing up \eqref{eq:new-tau-t} (with time index $\tau)$ for $\tau=1$ to $\tau=t-1$ and by noticing $\qkinter{k}(1)=0$, we have
\begin{align}
&\EE\{q_k^\text{inter}(t)^2\} - \EE\{q_k^\text{inter}(1)^2\} \nonumber \\
=&\EE\{q_k^\text{inter}(t)^2\}\leq (t-1) C^2 + 2U \sum_{\tau=1}^{t-1}\EE\{|q_k(\tau)|\}. \nonumber
\end{align}

Since $q_k(t)$ is stable and thus satisfies $\limsup_{t\to\infty}\frac{1}{t} \sum_{\tau=1}^t\EE\{|q_k(\tau)|\} < \infty$, there exists an $L$ value such that $\frac{1}{t} \sum_{\tau=1}^t\EE\{|q_k(\tau)|\}\leq L$ for all possible $t$ values. We then have 
\begin{align}
&\frac{1}{t-1}\EE\{q_k^\text{inter}(t)^2\}\nonumber\\
\leq& C^2 + 2U \frac{1}{t-1}\sum_{\tau=1}^{t-1}\EE\{|q_k(\tau)|\}\nonumber\\
\leq& C^2 + 2UL. \nonumber
\end{align}
for arbitrary $t$ values.

For any arbitrarily given $\epsilon'>0$, we now apply Markov inequality with the second moment expression to derive
\begin{align}
\prob(|q_k^\text{inter}(t)|\geq \epsilon' t) \leq \frac{1}{\epsilon'^2 t^2} \EE\{q_k^\text{inter}(t)^2\} \leq \frac{C+2UL}{\epsilon'^2 t}. \nonumber
\end{align}
For any arbitrarily given $\delta>0$, let $t_0$ be the first $t$ such that $\frac{C+2UL}{\epsilon'^2 t} < \delta$. Then we have
\begin{align}
\prob(|q_k^\text{inter}(t)|\geq \epsilon' t) < \delta,~\forall t>t_0. \nonumber
\end{align}
Thus we have proven the sublinear growth of $\two{\ve{q}}(t)$. \qed

Before we continue our proofs of sublinearly growing $\Nna{k}(t)$ and $\Qkinter{k}(t)$, we state the following claim first. Define {\em the deficit}, $D_k$, for all $k$ as the difference between $\Qkinter{k}$ and $\qkinter{k}$. That is, at any time $t$,
\begin{align}
&D_k(t) = \Qkinter{k}(t)-\qkinter{k}(t),~\forall k. \label{eq:deficit-update}
\end{align}

\begin{claim} \label{claim:dk_nondecreasing}
For all $k$, the function $D_k(t)$ is non-decreasing and it grows sublinearly.
\end{claim}

The proof of Claim~\ref{claim:dk_nondecreasing} is relegated to Appendix~\ref{app:proof_dk_nondecreasing}. We now continue our proofs.

{\em Proof of sublinearly growing $\Nna{k}(t)$: } Recall the definition of the null activity at queue $k$ ($k\in\SAin{n(t)}$, and $\Qkinter{k}(t)<\mu_{\out,k}(t)$). In the proof of Claim~\ref{claim:dk_nondecreasing}, in particular \eqref{eq:D_nondecreasing}, we can see that the null activity occurs at queue $k$ at time $t$ if and only if $D_k(t+1) > D_k(t)$. As a result,
\begin{align}
\Nna{k}(t) = \sum_{\tau = 1}^t I(D_k(\tau+1)>D_k(\tau)). \nonumber
\end{align}

Recall that $\Qkinter{k}(t)$ is an integer-valued random process and so is $\mu_{\text{out},k}(t) = \sum_{n=1}^N \left(\bine{k}{n}(\cq(t)) \cdot x_n^*(t) \right)$. As a result, whenever $\mu_{\text{out},k}(t)-\Qkinter{k}(t)>0$, we must have $\mu_{\text{out},k}(t)-\Qkinter{k}(t)\geq 1$. Using this observation and the fact that $D_k(t)$ is non-decreasing, we have
\begin{align}
\sum_{\tau = 1} ^t I(D_k(\tau+1)>D_k(\tau)) \leq D_k(t+1). \nonumber
\end{align}
The above argument implies $\Nna{k}(t)\leq D_k(t+1)$. Since $D_k(t)$ grows sublinearly as proven in Claim~\ref{claim:dk_nondecreasing}, we have proven that $N_{\mathsf{NA},k}(t)$ also grows sublinearly. \qed

{\em Proof of sublinearly growing $\Qkinter{k}(t)$: }By \eqref{eq:deficit-update}, 
\begin{align}
\Qkinter{k}(t)=\qkinter{k}(t)+D_k(t). \nonumber
\end{align}
We have shown that both $\qkinter{k}$ and $D_k(t)$ grow sublinearly, and hence $\Qkinter{k}$ also grows sublinearly. \qed

The above discussion on $q_k(t)$, $\qkinter{k}(t)$, $\Nna{k}(t)$, and $\Qkinter{k}(t)$ completes the proof of Lemma~\ref{lemma:sub-growth}.

\subsection{Proof of Proposition~\ref{prop:combine-capacity}}\label{app:combine-capacity}
To compare polytopes in Proposition~\ref{prop:WangISIT12} and Proposition~\ref{prop:inner}, we first list all the linear constraints describing each region separately. For Proposition~\ref{prop:inner}, the region can be described by \eqref{eq:new-balance}. Following from Table~\ref{tab:transition}, we can explicitly write $\im$ and $\om$ as follows. To facilitate matrix labeling, we order the 7 operations as $[\text{NC1},\text{NC2},\text{DX1},\text{DX2},\text{PM},\text{RC},\text{CX}]$, and order the 5 queues as $\left[\ve{Q}_\emptyset^1,\ve{Q}_\emptyset^2,\ve{Q}_{\{2\}}^1,\ve{Q}_{\{1\}}^2,\ve{Q}_\text{mix}\right]$. Let $\vec{p}^{[c]}\eqdef \vec{p}(c)$ for all $c\in\bigcq$ be the probability vector which represents the reception status probabilities when the channel quality is $c$. Given the above definitions, we can write $\im$, $\avg{\Bin}$, $\avg{\Bout}$, and the average service vector, $\ve{s}_c$, under channel quality $c$ for any $c\in\bigcq$ as
\begin{align}
&\im = \left[
           \begin{array}{ccccc}
             1 & 0 & 0 & 0 & 0 \\
             0 & 1 & 0 & 0 & 0 \\
           \end{array}
         \right]^\tran, \nonumber \\
&\avg{\Bin(c)} \nonumber\\&=\left[
        \begin{array}{ccccccc}
          p_{d_1\vee d_2}^{[c]} & 0 & 0 & 0 & p_{d_1\vee d_2}^{[c]} & 0 & 0 \\
          0 & p_{d_1\vee d_2}^{[c]} & 0 & 0 & p_{d_1\vee d_2}^{[c]} & 0 & 0 \\
          0 & 0 & p_{d_1}^{[c]} & 0 & 0 & 0 & p_{d_1}^{[c]} \\
          0 & 0 & 0 & p_{d_2}^{[c]} & 0 & 0 & p_{d_2}^{[c]} \\
          0 & 0 & 0 & 0 &  0 &  p_{d_1\vee d_2}^{[c]} & 0 \\
        \end{array}
      \right], \nonumber \\
&\avg{\Bout(c)} = \left[
        \begin{array}{ccccccc}
          0 & 0 & 0 & 0 & 0 & 0 & 0 \\
          0 & 0 & 0 & 0 & 0 & 0 & 0 \\
          p_{\overline{d_1}d_2}^{[c]} & 0 & 0 & 0 & 0 & p_{\overline{d_1}d_2}^{[c]} & 0 \\
          0 & p_{d_1\overline{d_2}}^{[c]} & 0 & 0 & 0 & p_{d_1\overline{d_2}}^{[c]} & 0 \\
          0 & 0 & 0 & 0 &  p_{d_1\vee d_2}^{[c]} &  0 & 0 \\
        \end{array}
      \right], \nonumber \\      
&\ve{s}_c = \left[ \xncone~\xnctwo~\xdxone~\xdxtwo~\xpm~\xrc~\xcx \right]^\tran. \nonumber
\end{align}

As a result, the throughput region in Proposition~\ref{prop:inner} can be expressed by a collection of 5+1 linear (in)equalities, where the first 5 equalities correspond to the flow-conservation law of queues 1 to 5 and the 6-th inequalities follows from $\ve{s}_c$ being drawn from the convex hull $\Lambda$: That is,
\begin{align}
&\sum_{\forall c\in\bigcq}f_c\left(\xncone + \xpm\right)p_{d_1\vee d_2}^{[c]} = R_1, \label{eq:poly-inner-1} \\
&\sum_{\forall c\in\bigcq}f_c\left(\xnctwo + \xpm\right)p_{d_1\vee d_2}^{[c]} = R_2, \label{eq:poly-inner-2} \\
&\sum_{\forall c\in\bigcq}f_c\left(\xcx + \xdxone\right)p_{d_1}^{[c]} = \sum_{\forall c\in\bigcq}f_c\left(\xncone + \xrc\right)p_{\overline{d_1}d_2}^{[c]}, \label{eq:poly-inner-3} \\
&\sum_{\forall c\in\bigcq}f_c\left(\xcx + \xdxtwo\right)p_{d_2}^{[c]} = \sum_{\forall c\in\bigcq}f_c\left(\xnctwo + \xrc\right)p_{d_1\overline{d_2}}^{[c]}, \label{eq:poly-inner-4} \\
&\sum_{\forall c\in\bigcq}f_c\xrc p_{d_1\vee d_2}^{[c]} = \sum_{\forall c\in\bigcq}f_c\xpm p_{d_1\vee d_2}^{[c]}, \label{eq:poly-inner-5} \\
&\xncone + \xnctwo +\xdxone +\xdxtwo +\xpm +\xrc +\xcx \leq 1, \nonumber \\
& \quad\quad\quad\quad\quad\quad\quad\quad\quad\quad\quad\quad\quad\quad\quad\quad\quad\quad \forall c\in\bigcq. \label{eq:poly-inner-6}
\end{align}

On the other hand, by Lemma~8 of \cite{WangHan14}, the polytype in Proposition~\ref{prop:WangISIT12} can also be expressed by another collection of linear (in)equalities:
\begin{align}
&x_0^{[c]}+x_9^{[c]}+x_{18}^{[c]}+x_{27}^{[c]}+x_{31}^{[c]}+x_{63}^{[c]}+x_{95}^{[c]}\leq 1,\forall c\in\bigcq, \label{eq:poly-wang-1}\\
&y_1=\sum_{\forall c\in\bigcq} f_c \left(x_0^{[c]}+x_9^{[c]}+x_{18}^{[c]}+x_{27}^{[c]}+x_{31}^{[c]}+x_{63}^{[c]}\right)p_{d_1}^{[c]}, \label{eq:poly-wang-2}\\
&y_2=\sum_{\forall c\in\bigcq} f_c \left(x_0^{[c]}+x_9^{[c]}+x_{18}^{[c]}+x_{27}^{[c]}+x_{31}^{[c]}+x_{95}^{[c]}\right)p_{d_2}^{[c]}, \label{eq:poly-wang-3}\\
&y_3=R_1+\sum_{\forall c\in\bigcq} f_c \left(x_0^{[c]}+x_9^{[c]}\right)p_{d_1}^{[c]}, \label{eq:poly-wang-4}\\
&y_4=R_1+\sum_{\forall c\in\bigcq} f_c \left(x_0^{[c]}+x_{18}^{[c]}+x_{27}^{[c]}\right)p_{d_2}^{[c]}, \label{eq:poly-wang-5}\\
&y_5=\sum_{\forall c\in\bigcq} f_c \left(x_0^{[c]}+x_9^{[c]}+x_{18}^{[c]}\right)p_{d_1\vee d_2}^{[c]}, \label{eq:poly-wang-6}\\
&y_6=R_1+\sum_{\forall c\in\bigcq} f_c \left(x_0^{[c]}+x_9^{[c]}\right)p_{d_1\vee d_2}^{[c]}, \label{eq:poly-wang-7}\\
&y_7=R_2+\sum_{\forall c\in\bigcq} f_c \left(x_0^{[c]}+x_{18}^{[c]}\right)p_{d_1\vee d_2}^{[c]}, \label{eq:poly-wang-8}
\end{align}
and
\begin{align}
&y_1=y_3; \quad y_2=y_4;\label{eq:poly-wang-17}\\
&y_5=y_6=y_7=R_1+R_2.\label{eq:poly-wang-16}
\end{align}

To prove that the dynamic-arrival stability region in \eqref{eq:poly-inner-1}-\eqref{eq:poly-inner-6} matches the block-coding capacity in \eqref{eq:poly-wang-1}--\eqref{eq:poly-wang-16}, we need to prove that for any $(R_1,R_2)$ and the accompanying $x_{(\cdot)}^{[c]}$ and $y_{(\cdot)}$ variables satisfying \eqref{eq:poly-wang-1} to \eqref{eq:poly-wang-16}, we can always find out another set of  $\ve{s}_c=[\xncone,\xnctwo,\xdxone,\xdxtwo,\xpm,\xrc,\xcx]$ variables such that $(R_1,R_2)$ and $\ve{s}_c$ jointly satisfying \eqref{eq:poly-inner-1} to \eqref{eq:poly-inner-6}. To do so, we will verify that the following one-to-one mapping $\ve{x}_{(\cdot)}^{[c]}$ satisfies \eqref{eq:poly-inner-1} to \eqref{eq:poly-inner-6}.
\begin{align}
&\xncone = x_{18}^{[c]},~\xnctwo = x_{9}^{[c]},~\xdxone=x_{63}^{[c]},\xdxtwo=x_{95}^{[c]}, \nonumber \\
&\xpm=x_{0}^{[c]},~\xrc=x_{27}^{[c]},~\xcx=x_{31}^{[c]}. \label{eq:poly-assign}
\end{align}

Ineq.~\eqref{eq:poly-inner-6} is true as a direct result of \eqref{eq:poly-wang-1}. We now prove that \eqref{eq:poly-inner-1} holds. By \eqref{eq:poly-wang-16}, we have $y_7=R_1+R_2$. By \eqref{eq:poly-wang-8}, we then have
\begin{align}
&y_7=R_2+\sum_{\forall c\in\bigcq} f_c \left(\xpm+\xncone\right)p_{d_1\vee d_2}^{[c]} = R_1+R_2 \nonumber \\
\Rightarrow &\sum_{\forall c\in\bigcq} f_c \left(\xpm+\xncone\right)p_{d_1\vee d_2}^{[c]} = R_1, \nonumber
\end{align}
which implies \eqref{eq:poly-inner-1}. \eqref{eq:poly-inner-2} can be proven by symmetric arguments. Next we check \eqref{eq:poly-inner-5}. Again by the fact that $y_5=R_1+R_2$, we have
\begin{align}
&y_5=\sum_{\forall c\in\bigcq} f_c \left(\xpm+\xncone+\xnctwo+\xrc\right)p_{d_1\vee d_2}^{[c]}\nonumber\\
&\quad = R_1+R_2 \nonumber \\
\Rightarrow &\sum_{\forall c\in\bigcq} f_c \left(\xncone+\xrc\right)p_{d_1\vee d_2}^{[c]} = R_1, \label{eq:poly-match-1}
\end{align} 
where \eqref{eq:poly-match-1} follows from substituting \eqref{eq:poly-inner-2} into $y_5=R_1+R_2$. Combining \eqref{eq:poly-match-1} with \eqref{eq:poly-inner-1}, we have
\begin{align}
&R_1 = \sum_{\forall c\in\bigcq} f_c \left(\xncone+\xrc\right)p_{d_1\vee d_2}^{[c]} \nonumber \\
&\quad =\sum_{\forall c\in\bigcq}f_c\left(\xncone + \xpm\right)p_{d_1\vee d_2}^{[c]} \nonumber \\
&\Rightarrow  \sum_{\forall c\in\bigcq}f_c\xrc p_{d_1\vee d_2}^{[c]} = \sum_{\forall c\in\bigcq}f_c\xpm p_{d_1\vee d_2}^{[c]}, \nonumber
\end{align}
which implies \eqref{eq:poly-inner-5}. Finally we check \eqref{eq:poly-inner-3} and \eqref{eq:poly-inner-4}. By \eqref{eq:poly-wang-17},  we have
\begin{align}
&y_3 = y_1 \nonumber \\
\Rightarrow& \sum_{\forall c\in\bigcq} f_c \left(\xpm+\xncone+\xnctwo+\xrc+\xcx+\xdxone\right)p_{d_1}^{[c]} \nonumber \\
&=R_1+\sum_{\forall c\in\bigcq} f_c \left(\xpm+\xnctwo\right)p_{d_1}^{[c]} \nonumber \\\
\Rightarrow& \sum_{\forall c\in\bigcq} f_c \left(\xncone+\xrc+\xcx+\xdxone\right)p_{d_1}^{[c]}=R_1. \label{eq:poly-match-2}
\end{align}

Combining \eqref{eq:poly-match-2} and \eqref{eq:poly-match-1}, we have
\begin{align}
&R_1=\sum_{\forall c\in\bigcq} f_c \left(\xncone+\xrc+\xcx+\xdxone\right)p_{d_1}^{[c]} \nonumber \\
&\quad = \sum_{\forall c\in\bigcq} f_c \left(\xncone+\xrc\right)p_{d_1\vee d_2}^{[c]}. \label{eq:poly-match-3}
\end{align}

Following from the fact that $p_{d_1\vee d_2}^{[c]}=p_{d_1}^{[c]}+p_{\overline{d_1}d_2}^{[c]}$, we can rewrite \eqref{eq:poly-match-3} as
\begin{align}
\sum_{\forall c\in\bigcq}f_c\left(\xcx + \xdxone\right)p_{d_1}^{[c]} = \sum_{\forall c\in\bigcq}f_c\left(\xncone + \xrc\right)p_{\overline{d_1}d_2}^{[c]}, \nonumber
\end{align}
which implies\eqref{eq:poly-inner-3}. \eqref{eq:poly-inner-4} can be derived by symmetric arguments. Thus we complete the proof of Proposition~\ref{prop:combine-capacity}. 

\subsection{Lemma~\ref{lemma:new-lemma}} \label{app:new-lemma}
We use $\mathcal{P}$ to denote a finite collection of probability distributions and each distribution is of zero mean and finite support. For simplicity, we say $\mathcal{P}=\{P_1,P_2,\cdots, P_{K}\}$ where $K=|\mathcal{P}|$.  
\begin{lemma}\label{lemma:new-lemma}
There exists a fixed constant $C>0$ such that for any arbitrary $K$ non-negative integers $L_1,L_2,...,L_K$, the following inequality always holds. 
\begin{align}
\prob(\sum_{k=1}^K \sum_{i=1}^{L_k} X_i^{(k)}\geq 0)>C
\end{align}
where for any $k$, the random variables $X_i^{(k)} \sim P_k$ are i.i.d. for different $i$ values and the random processes $\{X_i^{(k)}:i\}$ are independently distributed for different $k$ values.  
\end{lemma}

{\em Proof: } We prove this lemma by induction on the size of $\mathcal{P}$. When $K=|\mathcal{P}|=1$, the  probability of interest becomes $\prob(\sum_{i}^{L} X_i\geq 0)$ where we drop the index $k$ for simplicity. By the central limit theorem, there exists an $l_0$ such that when $L>l_0$, the probability of interest is $>1/4$ (which can be made arbitrarily close to $1/2$ but we choose $1/4$ for simplicity). Choose $C=\min(   \min\{\prob(\sum_{l=1}^L X_i\geq 0): l\leq l_0\}, 1/4)$. We claim that such a $C$ value is strictly positive. The reason is that $ \min\{\prob(\sum_{l=1}^L X_i\geq 0): l\leq l_0\}\neq 0$ because of the assumptions that $X_i$ is zero mean and i.i.d. From the above construction we have
\begin{align}
\prob(\sum_{i=1}^L X_i\geq 0)>C,\quad \forall L. 
\end{align}

We now consider the case of $K=|\mathcal{P}|\geq 2$. For any arbitrarily given $L_1$ to $L_K$, the probability of interest satisfies
\begin{align}
&\prob(\sum_{k=1}^K\sum_{i=1}^{L_1} X_{i}^{(k)} \geq 0) \nonumber \\
\geq& \prob(\sum_{i=1}^{L_k} X_{i}^{(k)} \geq 0,\forall k) \nonumber \\
=& \prod_{k=1}^K \prob(\sum_{i=1}^{L_k} X_{i}^{(k)}\geq 0). \label{eq:claim_Tk_2}
\end{align}
We have shown that for each $k$, there exists a constant $C_k>0$ such that $\prob(\sum_{i=1}^{L_k} X_{i}^{(k)}\geq 0)> C_k$ for any arbitrary $L_k$. Hence the product in \eqref{eq:claim_Tk_2} is larger than $C  \stackrel{\triangle}{=}\prod_{k=1}^K C_k$ for any arbitrary $L_1$ to $L_K$. Lemma~\ref{lemma:new-lemma} is thus proven. 

\subsection{Proof of Claim~\ref{claim:dk_nondecreasing}}\label{app:proof_dk_nondecreasing}
For all $k$, the reason why $D_k(t)$ is non-decreasing is because
\begin{align}
D_k(t+1) &=  \Qkinter{k}(t+1)-\qkinter{k}(t+1) \nonumber \\
         &= \left(\Qkinter{k}(t)-\mu_{\out,k}(t) \right)^+ - \left(\qkinter{k}(t)-\mu_{\out,k}(t) \right) \nonumber \\
         &= \Qkinter{k}(t)-\mu_{\out,k}(t) + \left( \mu_{\out,k}(t)-\Qkinter{k}(t) \right)^+ \nonumber \\
         &\quad - \left( \qkinter{k}(t)-\mu_{\out,k}(t) \right) \nonumber \\
         &= D_k(t)+\left( \mu_{\out,k}(t)-\Qkinter{k}(t) \right)^+. \label{eq:D_nondecreasing}
\end{align}

We now prove that $D_k(t)$ grows sublinearly for all $k$. Define $p_k(t) \stackrel{\triangle}{=} -\qkinter{k}(t-1)+\mu_{\out,k}(t-1)$ for all $t\geq 2$ and $p_k(1)=-\qkinter{k}(1)=0$. Notice that $p_k(t)$ grows sublinearly because $\qkinter{k}(t)$ grows sublinearly and $\mu_{\out,k}(t-1)$ is bounded. We notice that $D_k(t)$ is the running maximum of $p_k(t)$ since by \eqref{eq:D_nondecreasing},
\begin{align}
D_k(t) &= D_k(t-1)+\left(\mu_{\out,k}(t-1)-\Qkinter{k}(t-1) \right)^+ \nonumber \\
        &= D_k(t-1)+\max\{0, \mu_{\out,k}(t-1)-\Qkinter{k}(t-1)\} \nonumber \\
        &= \max\{D_k(t-1), -\qkinter{k}(t-1)+\mu_{\out,k} (t-1)\} \label{eq:newDk1} \\
        &= \max\{D_k(t-1), p_k(t)\}\nonumber\\
        &=\max_{1\leq \tau\leq t} p_k(\tau),\nonumber
\end{align}
where \eqref{eq:newDk1} follows from \eqref{eq:deficit-update}.

Recall $\overline{\mu_{\out,k}}(t)$ and $\overline{\mu_{\iin,k}}(t)$ are the expectation of $\mu_{\out,k}(t)$ and $\mu_{\iin,k}(t)$, respectively, conditioned on the arrival vectors and the channel quality until time $t$. And by \eqref{eq:DMW-update-new}, we can update $q_k(t)$ as
\begin{align}
q_k(t+1) = q_k(t)-\overline{\mu_{\out,k}}(t)+\overline{\mu_{\iin,k}}(t),~\forall k. \label{eq:qk-update}
\end{align}

Define $\overline{p_k}(t) \stackrel{\triangle}{=} -q_k(t-1)+\overline{\mu_{\out,k}}(t-1)=-q_k(t)+ \overline{\mu_{\iin,k}}(t-1)$. That is, $\overline{p_k}(t) $ is the conditional expectation of $p_k(t)$ given $\{\ve{a},\cq\}_1^{t-1}$. Define $p_k'(t) \stackrel{\triangle}{=} p_k(t)-\overline{p_k}(t)$. That is, $p_k'(t)$ is the difference between the random variable $p_k(t)$ and its conditional expectation $\overline{p_k}(t)$. Thus far, we have decompose
\begin{align}
p_k(t) = \overline{p_k}(t)+p_k'(t) \nonumber
\end{align}
as the summation of the average term $\overline{p_k}(t)$ and the random variation term $p_k'(t)$, where the latter has zero mean. We now define $D_k'(t)$ to be the running maximum of the $p_k'(t)$ and $\overline{D_k}(t)$ to be the running maximum of $\overline{p_k}(t)$. That is,
\begin{align}
& D_k'(t) = \max_{1\leq\tau\leq t} p_k'(\tau), \nonumber \\
&\overline{D_k}(t) = \max_{1\leq\tau\leq t} \overline{p_k}(\tau). \nonumber
\end{align}
In the following, we will prove:  Step 1: $\overline{p_k}(t)$ is stable and $p_k'(t)$ grows sublinearly; Step 2: $D_k'(t)$ grows sublinearly; and Step 3: $\overline{D_k}(t)$ grows sublinearly. Note that by definition, we always have $0\leq D_k(t)\leq D_k'(t)+\overline{D_k}(t)$. As a result, Steps 2 and 3 imply $D_k(t)$ also grows sublinearly. The proof is complete. 

Step 1: $\overline{p_k}(t)$ is stable because $q_k(t)$ is stable and $\overline{\mu_{\iin,k}}(t-1)$ is bounded. Furthermore, $p_k'(t)$ grows sublinearly from the fact that the summation/difference of one stable queue and one sublinearly stable queue is sublinearly stable\footnote{One can easily verify that with bounded initial value, stability implies sublinear stability.}. The proof of Step 1 is complete. \qed

Step 2: We now show that $D_k’(t)$ grows sublinearly. Recall that $p_k’(t)$ is the random variation term with mean zero and $D_k’(t)$ is the running maximum of the random variation. As a result, in essence, the $D_k’(t)$ is similar to the running maximum of a random walk with zero drift. The following proof is adapted from the standard proof that the running maximum of a zero-drift random walk is sublinearly growing [Chapter 4, \cite{Durrett:2010:PTE:1869916}].

\par Let $T_k'(b) \stackrel{\triangle}{=}\min\{t\geq 1: p_k'(t)\geq b\}$ be the hitting time of $p_k'(t)$ exceeding the threshold $b$.
\begin{claim} \label{claim:Tk}
There exists $C>0$ such that for all $t\geq 1$, all $b>0$, and all possible past arrival vector realizations and channel quality realizations $\{\a,\cq\}_1^{t-1}$, we have
\begin{align}
\prob(p_k'(t)\geq b| T_k'(b)\leq t,\{\ve{a},\cq\}_{1}^{t-1}) > C.
\end{align}
\end{claim}

{\it Proof of Claim~\ref{claim:Tk}: }
Let $\Delta \mu_{\iin,k}(t) \stackrel{\triangle}{=}\mu_{\iin,k}(t) - \overline{\mu_{\iin,k}}(t)$, $\Delta \mu_{\out,k}(t) \stackrel{\triangle}{=}\mu_{\out,k}(t) - \overline{\mu_{\out,k}}(t)$, and $\Delta \mu_k(t) \stackrel{\triangle}{=} \Delta \mu_{\iin,k}(t)-\Delta \mu_{\out,k}(t)$. By \eqref{eq:qk-update} and \eqref{eq:qk-inter-update},
\begin{align}
&\qkinter{k}(t) = \sum_{\tau=1}^{t-1} (\mu_{\iin,k}(\tau)-\mu_{\out,k}(\tau)), \nonumber \\
&q_k(t) = \sum_{\tau=1}^{t-1} (\overline{\mu_{\iin,k}}(\tau)-\overline{\mu_{\out,k}}(\tau)), \nonumber \\
\text{and}~&\qkinter{k}(t)-q_k(t) = \sum_{\tau=1}^{t-1} \Delta\mu_{k}(\tau). \nonumber
\end{align}
Then by the definitions of $p_k(t)$, $\overline{p_k}(t)$, and $p_k'(t)$, we have
\begin{align}
&p_k(t) = -\sum_{\tau=1}^{t-2} (\mu_{\iin,k}(\tau)-\mu_{\out,k}(\tau)) + \mu_{\out,k}(t-1), \nonumber \\
&\overline{p_k}(t) = -\sum_{\tau=1}^{t-2} (\overline{\mu_{\iin,k}}(\tau)-\overline{\mu_{\out,k}}(\tau)) + \overline{\mu_{\out,k}}(t-1), \nonumber \\
&p_k'(t) = -\sum_{\tau=1}^{t-2} \Delta\mu_{k}(\tau) + \Delta \mu_{\out,k}(t-1). \label{eq:new-pk-prime1}
\end{align}

By \eqref{eq:new-pk-prime1} we have
\begin{align}
&p_k'(t) - p_k'(T_k'(b)) \nonumber\\
=& \left(\sum_{\tau=1}^{t-2} \Delta\mu_{k}(\tau) + \Delta \mu_{\out,k}(t-1)\right) \nonumber \\
&\quad-\left(\sum_{\tau=1}^{T_k'(b)-2} \Delta\mu_{k}(\tau) + \Delta \mu_{\out,k}(T_k'(b)-1)\right) \nonumber \\
=&\sum_{\tau=T_k'(b)-1}^{t-2} \Delta\mu_{k}(\tau) + \Delta \mu_{\out,k}(t-1) -\Delta \mu_{\out,k}(T_k'(b)-1) \nonumber\\
=&\sum_{\tau=T_k'(b)}^{t-2} \Delta\mu_{k}(\tau) + \Delta \mu_{\out,k}(t-1)\nonumber\\
&\quad + (\Delta \mu_{\iin,k}(T_k'(b)-1) - 2\Delta\mu_{\out,k}(T_k'(b)-1))\nonumber
\end{align}
Define $\Delta \hat{\mu}_k (T_k'(b)-1)\eqdef \Delta \mu_{\iin,k}(T_k'(b)-1)- 2\Delta \mu_{\out,k}(T_k'(b)-1)$. Thus, we have
\begin{align}
&\prob(p_k'(t)\geq b| T_k'(b)\leq t,\{\ve{a},\cq\}_{1}^t) \nonumber \\
\geq  &\prob\bigg(\Delta \hat{\mu}_k(T_k'(b)-1)+\sum_{\tau=T_k'(b)}^{t-2} \Delta\mu_{k}(\tau) \nonumber \\
&\quad + \Delta \mu_{\out,k}(t-1) \geq 0| T_k'(b)\leq t,\{\ve{a},\cq\}_{1}^t\bigg) \label{eq:claim_Tk_1}
\end{align}
We now notice that in the RHS of \eqref{eq:claim_Tk_1}, there are $(t-T_k'(b)+1)$ summands in the probability expression, one for each $\tau\in[T_k'(b)-1,t-1]$. One can easily verify that conditioning on the past arrival vectors and past channel quality $\{\ve{a},\cq\}_1^{t}$, each summand is independently distributed. The reason is that when conditioning on $\{\ve{a},\cq\}_1^{t}$, both the virtual queue length vector $\ve{q}(\tau)$ and back-pressure scheduler become deterministic for all $\tau=1$ to $t$, see \eqref{eq:DMW-schedule}, \eqref{eq:new-backpressure}, and \eqref{eq:DMW-update-new}. As a result, the randomness of each summand depends only on the realization of $\bin{k}{n}(\tau)$ and $\bout{k}{n}(\tau)$ and they are independently distributed in our SPN model. Moreover, each summand is also of zero mean and bounded support. The reason is that the definitions of $\Delta \mu_{\iin,k}(\tau)$, $\Delta \mu_{\out,k}(\tau)$, and $\Delta \mu_k(\tau)$ ensure that these random variables are of zero mean. Also, since $\Bin(\tau)$ and $\Bout(\tau)$ are of bounded support, so are $\Delta\mu_{\iin,}(\tau)$, $\Delta \mu_{\out,k}(\tau)$, and $\Delta \mu_k(\tau)$.

\par Obviously the conditional distribution of each of the $t-T_k'(b)+1$ summands given $\{\ve{a},\cq\}_1^{t}$ depends on the values of $T_k'(b)$ and $t$ and the realization $\{\ve{a},\cq\}_1^{t}$. However, we further argue that there is a bounded number of distributions, denoted by $\mathcal{P}$, and each of the conditional distribution must be of a distribution $P\in \mathcal{P}$ regardless what are the values of $t$, $T_k'(b)$, and the realization $\{\ve{a},\cq\}_1^{t}$. Namely, even though there are infinitely many ways of having the $t$, $T_k'(b)$,  and the realization $\{\ve{a},\cq\}_1^{t}$ values, the number of possible distributions for all the summands is bounded. The reason is that the distributions of $\Delta\mu_{\iin,}(\tau)$, $\Delta \mu_{\out,k}(\tau)$, and $\Delta \mu_k(\tau)$ depend only on what is the actual schedule at time $\tau$. Since there is only a bounded number of possible scheduling decisions, the number of possible distributions for all the summands is bounded. 

By Lemma~\ref{lemma:new-lemma} in Appendix~\ref{app:new-lemma}, there exists a $C>0$ such that 
\begin{align}
\eqref{eq:claim_Tk_1}>C
\end{align}
for all $t$ and all possible past arrival vector realizations and channel quality realizations $\{\ve{a},\cq\}_1^{t-1}$. The proof of Claim~\ref{claim:Tk} is complete. \qed

Notice that by Claim \ref{claim:Tk}, there exists $C$ such that for all possible past arrival vector and channel quality realizations
\begin{align}
&\prob(p_k'(t)\geq b| T_k'(b)\leq t,\{\ve{a},\cq\}_{1}^{t-1}) \nonumber \\
=& \frac{\prob(p_k'(t)\geq b,T_k'(b)\leq t|\{\ve{a},\cq\}_{1}^t)}{\prob(T_k'(b)\leq t|\{\ve{a},\cq\}_{1}^{t-1})} \nonumber \\
=& \frac{\prob(p_k'(t)\geq b|\{\ve{a},\cq\}_{1}^{t-1})}{\prob(T_k'(b)\leq t|\{\ve{a},\cq\}_{1}^{t-1})} > C.
\end{align}
Meanwhile, since $D_k'(t)$ is the running maximum of $p_k'(t)$, we have
\begin{align}
&\prob(D_k'(t)\geq b|\{\ve{a},\cq\}_{1}^{t-1}) = \prob(T_k'(b)\leq t|\{\ve{a},\cq\}_{1}^{t-1}) \nonumber \\
<& \frac{1}{C}\prob(p_k'(t)\geq b|\{\ve{a},\cq\}_{1}^{t-1}).
\end{align}
Taking the expectation on both sides over all possible past arrival vectors and past channel quality, we have
\begin{align}
\prob(D_k'(t)\geq b) < \frac{1}{C}\prob(p_k'(t)\geq b).
\end{align}
Substituting $b$ by $\epsilon t$ in the above equation and using the fact that $p_k'(t)$ grows sublinearly, we have proven that $D_k'(t)$ grows sublinearly. The proof of Step 2 is complete. \qed

Step 3: We now prove the following claim.
\begin{claim} \label{claim:avg_Dk}
The following two inequalities are true for all possible realizations.
\begin{enumerate}
\item $\overline{D_k}(t+1)^2- \overline{D_k}(t)^2 \leq \max\{\overline{p_k}(t+1)^2-\overline{p_k}(t)^2,0\}+U^2$, where $U$ is the supremum over all possible $|\overline{\mu_{\out,k}}(t)-\overline{\mu_{\iin,k}}(t-1)|$. Note that $U$ always exists since in the random external arrivals and the random movements of the packets all have bounded support and $\overline{\mu_{\iin,k}}(t)$ and $\overline{\mu_{\out,k}}(t)$ are computed from the expected values of the random packets arrival and departures.
\item $\max\{\overline{p_k}(t+1)^2-\overline{p_k}(t)^2,0\}+U^2\leq 2|\overline{p_k}(t)|U+2U^2$.
\end{enumerate}
\end{claim}

{\it Proof of Claim~\ref{claim:avg_Dk}: }
We first prove 1). There are three possible cases. \\
Case 1: $\overline{D_k}(t)\geq \overline{p_k}(t+1)$. Since $\overline{D_k}(t)$ is the running maximum of $\overline{p_k}(t)$, $\overline{D_k}(t+1) = \overline{D_k}(t)$ in this case. Thus the left hand side of (i) is zero and the inequality holds. \\
Case 2: $\overline{D_k}(t)< \overline{p_k}(t+1)$ and $\overline{p_k}(t)\geq 0$. By the definition of $\overline{D_k}(t)$, we have $\overline{D_k}(t+1) = \overline{p_k}(t+1)$. Also, since $\overline{D_k}(t)$ is the running maximum of $\overline{p_k}(t)$, we have $0\leq \overline{p_k}(t)\leq \overline{D_k}(t)$, which implies $(\overline{p_k}(t))^2\leq (\overline{D_k}(t))^2$. Jointly, we thus have $\overline{D_k}(t+1)^2- \overline{D_k}(t)^2\leq \overline{p_k}(t+1)^2-\overline{p_k}(t)^2\leq \max\{\overline{p_k}(t+1)^2-\overline{p_k}(t)^2,0\}+U^2$.\\
Case 3: $\overline{D_k}(t)< \overline{p_k}(t+1)$ and $\overline{p_k}(t)< 0$. By the definition of $U$ and by \eqref{eq:new-pk-prime1}, we have $\overline{p_k}(t+1)\leq \overline{p_k}(t)+U$, which, together with the inequality $D_k(t)< \overline{p_k}(t+1)$ and the definition that $\overline{D_k}(t)$ being the running maximum of $\overline{p_k}(t)$, implies 
\begin{align}
\overline{D_k}(t)-U\leq \overline{p_k}(t)\leq \overline{D_k}(t). \nonumber
\end{align}

Since $\overline{D_k}(t)$ is always no less than zero, we thus have $-U\leq\overline{p_k}(t)< 0$, which in turn implies $U^2-\overline{p_k}(t)^2 \geq 0$. Since $\overline{D_k}(t+1)=\overline{p_k}(t+1)$, we now have, $\overline{D_k}(t+1)^2- \overline{D_k}(t)^2\leq\overline{p_k}(t+1)^2\leq \max\{\overline{p_k}(t+1)^2-\overline{p_k}(t)^2,0\}+U^2$. The proof of 1) is complete.

We now prove 2). Define $\Delta \overline{p_k}(t+1) \stackrel{\triangle}{=} \overline{p_k}(t+1) - \overline{p_k}(t)$. Then
\begin{align}
&\max\{\overline{p_k}(t+1)^2-\overline{p_k}(t)^2,0\}+U^2 \nonumber \\
=&\max\{(\overline{p_k}(t)+\Delta \overline{p_k}(t+1))^2-\overline{p_k}(t)^2,0\}+U^2 \nonumber \\
=&\max\{2\overline{p_k}(t)\Delta \overline{p_k}(t+1)+ \Delta\overline{p_k}(t+1)^2,0\}+U^2 \nonumber \\
\leq & 2|\overline{p_k}(t)\Delta \overline{p_k}(t+1)|+ |\Delta\overline{p_k}(t+1)^2|+U^2\nonumber\\
\leq & 2|\overline{p_k}(t)|U + 2U^2, \nonumber
\end{align}
where the last inequality follows from rewriting $\overline{\mu_{\iin,k}}(\tau)$ and $\overline{\mu_{\out,k}}(\tau)$ based on \eqref{eq:new-pk-prime1} and by the definition of $U$. \qed

Following from Claim \ref{claim:avg_Dk} and taking the expectation on both sides over all possible arrival vectors,
\begin{align}
\EE\{\overline{D_k}(t+1)^2\}-\EE\{\overline{D_k}(t)^2\} \leq 2 \EE\{|\overline{p_k}(t)|\} U +2U^2. \nonumber
\end{align}
Replacing the time index $t$ by $\tau$ and then summing up the above inequality with time index $\tau)$ from $\tau=1$ to $\tau=t-1$, we then have
\begin{align}
&\EE\{\overline{D_k}(t)^2\}\leq 2U\sum_{\tau=1}^{t-1}\EE\{|\overline{p_k}(\tau)|\}+2U^2 (t-1) \nonumber \\
\Rightarrow &\frac{1}{t}\EE\{\overline{D_k}(t)^2\}\leq 2U\frac{1}{t-1}\sum_{\tau=1}^{t-1}\EE\{|\overline{p_k}(\tau)|\}+2U^2. \nonumber
\end{align}

The fact that $\overline{p_k}(t)$ is stable implies that there exists an $L$ value such that $\frac{1}{t-1}\sum_{\tau=1}^{t-1}\EE\{|\overline{p_k}(\tau)|\}\leq L$ for all $t$. For any $\epsilon'>0$, $\delta>0$, we then apply the Markov inequality,
\begin{align}
\prob(\overline{D_k}(t)>\epsilon' t)\leq \frac{1}{\epsilon'^2 t^2}\EE\{\overline{D_k}(t)^2\} \leq \frac{1}{\epsilon'^2 t} \left(2UL + 2U^2\right). \nonumber
\end{align}
Let $t_0$ be the smallest $t$ such that $\frac{1}{\epsilon'^2 t} \left(2UL + 2U^2\right)<\delta$. Then $\prob(\overline{D_k}(t)>\epsilon' t)<\delta$ for all $t>t_0$, which completes the proof of Step 3. \qed

\bibliographystyle{IEEEtran}
\bibliography{WCK}

\end{document}